\documentclass[12pt]{iopart}

\usepackage{graphicx}
\usepackage{epstopdf}

\begin{document}

\newcommand{\bo}[1]{\mbox{$\mathbf{#1}$}}
\newcommand{\eqref}[1]{(\ref{#1})}
\newcommand{\qav}[1]{\mbox{$\left\langle\left\langle \, #1 \, \right\rangle\right\rangle$}}
\newcommand{\sqav}[1]{\mbox{$\langle\langle \, #1 \, \rangle\rangle$}}
\newcommand{\bqav}[1]{\mbox{$\bigg\langle\bigg\langle \, #1 \, \bigg\rangle\bigg\rangle$}}
\newcommand{\tav}[1]{\mbox{$\langle #1 \rangle$}}
\newcommand{\tgav}[2]{\mbox{$\langle #1 \rangle^{\text{TG}}_{#2}$}}
\newcommand{\MI}[2]{\mbox{$I [ #1 \, ;\,  #2 ] $}}
\newcommand{\N}[3]{\mbox{$N^{#1}_{#2 #3}$}}
\newcommand{\bk}[1]{\backslash {#1}}

\newcommand{\mess}[3]{\mbox{$M^{#1}_{#2 \rightarrow #3}$}}

\def\A{\bo{A}}
\def\a{\bo{a}}
\def\uai{u^a_i}
\def\ua{u^a}
\def\bamu{b^a_\mu}
\def\xoi{x^0_i}
\def\x{\bo{x}}
\def\D{\bo{D}}
\def\y{\bo{y}}
\def\s{\bo{s}}
\def\J{\bo{J}}
\def\w{\bo{w}}
\def\u{\bo{u}}
\def\W{\bo{W}}
\def\v{\bo{v}}
\def\r{\bo{r}}

\def\Qab{Q_{a b}}
\def\Qhab{\hat{Q}_{a b}}
\def\mz{\mathcal{Z}}
\def\Dq{\Delta Q}
\def\Du{\mathcal{D}u}
\def\Dz{\mathcal{D}z\,}
\def\Dqb{\Delta q}
\def\dqb{\Delta q}
\def\qob{q_0}
\def\ub{\overline{u}}

\def\H1{H_{\backslash 1}}

\def\pnz{P}

\title{Statistical mechanics of complex neural systems and high dimensional data}

\author{Madhu Advani, Subhaneil Lahiri, and Surya Ganguli}

\address{
Dept. of Applied Physics, Stanford University, Stanford, CA
}
\ead{sganguli@stanford.edu}

\begin{abstract}

Recent experimental advances in neuroscience have opened new vistas into the immense complexity of  neuronal networks.  This proliferation of data challenges us on two parallel fronts.  First, how can we form adequate theoretical frameworks for understanding how dynamical network processes cooperate across widely disparate  spatiotemporal scales to solve important computational problems?  And second, how can we extract meaningful models of neuronal systems from high dimensional datasets?  To aid in these challenges, we give a pedagogical review of a collection of ideas and theoretical methods arising at the intersection of statistical physics, computer science and neurobiology.   We introduce the interrelated replica and cavity methods, which originated in statistical physics as powerful ways to quantitatively analyze large highly heterogeneous systems of many interacting degrees of freedom.   We also introduce the closely related notion of message passing in graphical models, which originated in computer science as a distributed algorithm capable of solving large inference and optimization problems involving many coupled variables.  We then show how both the statistical physics and computer science perspectives can be applied in a wide diversity of contexts to problems arising in theoretical neuroscience and data analysis.   Along the way we discuss spin glasses, learning theory, illusions of structure in noise, random matrices, dimensionality reduction, and compressed sensing, all within the unified formalism of the replica method.  Moreover, we review recent conceptual connections between message passing in graphical models, and neural computation and learning.  Overall, these ideas illustrate how statistical physics and computer science might provide a lens through which we can uncover emergent computational functions buried deep within the dynamical complexities of neuronal networks.

\end{abstract}

\vspace{2pc}

\maketitle

\newpage 

\tableofcontents

\newpage

\section{Introduction}

\par
Neuronal networks are highly complex dynamical systems consisting of large numbers of neurons interacting through synapses \cite{kandel1991principles, dayan2001theoretical, ReikeWarlandvanSteveninckEtAl96}.  Such networks subserve dynamics over multiple time-scales.  For example, on fast time scales, on the order of milliseconds, synaptic connectivity is approximately constant, and this connectivity directs the flow of electrical activity through neurons.  On slower timescales, on the order of seconds to minutes and beyond, the synaptic connectivity itself can change through synaptic plasticity induced by the statistical structure of experience, which itself can stay constant over even longer timescales.   These synaptic changes are thought to underly our ability to learn from experience.  To the extent that such separations of timescale hold, one can exploit powerful tools from the statistical physics of disordered systems to obtain a remarkably precise understanding of neuronal dynamics and synaptic learning in basic models.   For example, the replica method and the cavity method, which we introduce and review below, become relevant because they allow us to understand the statistical properties of many interacting degrees of freedom that are coupled to each other through some fixed, or quenched, interactions that may be highly heterogenous, or disordered.

\par
However, such networks of neurons and synapses, as well as the dynamical processes that occur on them,  are not simply tangled webs of complexity that exist for their own sake.  Instead they have been sculpted over time, through the processes of evolution, learning and adaptation, to solve important computational problems necessary for survival.   Thus biological neuronal networks serve a {\it function} that is useful for an organism in terms of improving its evolutionary fitness.  The very concept of function does not of-course arise in statistical physics, as large disordered statistical mechanical systems, like glasses or non-biological polymers do not arise through evolutionary processes.   In general, the function that a biological network performs (which may not always be a-priori obvious) can provide a powerful way to understand both its structure and the details of its complex dynamics \cite{Lau:2007:Phys-Rev-E-Stat-Nonlin-Soft-Matter-Phys:17677098}.   As the functions performed by neuronal networks are often computational in nature, it can be useful to turn to ideas from distributed computing algorithms in computer science for sources of insight into how networks of neurons may learn and compute in a distributed manner.  In this review we also focus on distributed message passing algorithms whose goal is to compute the marginal probability distribution of a single degree of freedom in a large interacting system.   Many problems in computer science, including error correcting codes and constraint satisfaction, can be formulated as message passing problems \cite{mezard2009information}. As we shall review below, message passing is intimately related to the replica and cavity methods of statistical physics, and can serve as a framework for thinking about how specific dynamical processes of neuronal plasticity and network dynamics may solve computational problems like learning and inference.

\par
This combination of ideas from statistical physics and computer science are not only useful in thinking about how network dynamics and plasticity may mediate computation, but also for thinking about ways to analyze large scale datasets arising from high throughput experiments in neuroscience.  Consider a data set consisting of $P$ points in an $N$ dimensional feature space.   Much of the edifice of classical statistics and machine learning has been tailored to the situation in which $P$ is large and $N$ is small.   This is the low dimensional data scenario in which we have large amounts of data.  In such situations, many classical unsupervised machine learning algorithms can easily find structures or patterns in data, when they exist.  However, the advent of high throughput techniques in neuroscience has pushed us into a high dimensional data scenario in which both $P$ and $N$ are large, but their ratio is $O(1)$.  For example, we can simultaneously measure the activity of $O(100)$ neurons but often only under a limited number of trials (i.e. also $O(100)$) for any given experimental condition.  Also, we can measure the single cell gene expression levels of $O(100)$ genes but only in a limited number of cells.   In such a high dimensional scenario, it can be difficult to find statistically significant patterns in the data, as often classical unsupervised machine learning algorithms yield illusory structures.  The statistical physics of disordered systems again provides a powerful tool to understand high dimensional data, because many machine learning algorithms can be formulated as the minimization of a data dependent energy function on a set of parameters.  We review below how statistical physics plays a useful role in understanding possible illusions of structure in high dimensional data, as well as approaches like random projections and compressed sensing, which are tailored to the high dimensional data limit.

\par We give an outline and summary of this review as follows.  In section \ref{sec:spinglassmodel} we introduce the fundamental techniques of the replica method and cavity method within the context of a paradigmatic example, the Sherrington-Kirkpatrick (SK) model \cite{sherrington1975solvable} of a spin glass \cite{mezard1987spin, fischer1993spin, nishimori2001statistical}. In a neuronal network interpretation, such a system qualitatively models a large network in which the heterogenous synaptic connectivity is fixed and plays the role of quenched disorder.  On the otherhand, neuronal activity can fluctuate and we are interested in understanding the statistical properties of the neuronal activity.   We will find that certain statistical properties, termed self-averaging properties, do not depend on the detailed realization of the disordered connectivity matrix.   This is a recurring theme in these notes; in large random systems with microscopic heterogeneity, striking levels of almost deterministic macroscopic order can arise in ways that do not depend on the details of the heterogeneity.   Such order can govern dynamics and learning in neuronal networks, as well as the performance of machine learning algorithms in analyzing data, and moreover, this order can be understood theoretically through the replica and cavity methods. 

\par We end section \ref{sec:spinglassmodel} by introducing message passing which provides an algorithmic perspective on the replica and cavity methods.  Many models in equilibrium statistical physics are essentially equivalent to joint probability distributions over many variables, which are equivalently known and described as graphical models in computer science \cite{koller2009probabilistic}.  Moreover, many computations in statistical physics involve computing marginal probabilities of a single variable in such graphical models.  Message passing, also known in special cases as belief propagation \cite{pearl1988probabilistic}, involves a class of algorithms that yield dynamical systems whose fixed points are designed to approximate marginal probabilities in graphical models.  Another recurring theme in these notes is that certain aspects of neuronal dynamics may profitably be viewed through the lens of message passing; in essence, these neuronal (and also synaptic) dynamics can be viewed as approximate versions of message passing in a suitably defined graphical model.   This correspondence between neuronal dynamics and message passing allows for the possibility of both understanding the computational significance of existing neuronal dynamics, as well as deriving hypotheses for new forms of neuronal dynamics from a computational perspective.  

\par In section \ref{sec:smoflearning} we apply the ideas of replicas, cavities and messages introduced in section \ref{sec:spinglassmodel} to the problem of learning in neuronal networks as well as machine learning (see \cite{engel01} for a beautiful book length review of this topic).   In this context, training examples, or data play the role of quenched disorder, and the synaptic weights of a network, or the learning parameters of a machine learning algorithm, play the role of fluctuating statistical mechanical degrees of freedom.   In the zero temperature limit, these degrees of freedom are optimized, or learned, by minimizing an energy function.  The learning error, as well as aspects of the learned structure, can be described by macroscopic order parameters that do not depend on the detailed realization of the training examples, or data.  We show how to compute these order parameters for the classical perceptron \cite{rosenblatt1958perceptron, block1962perceptron}, thereby computing its storage capacity.  Also we compute these order parameters for classical learning algorithms, including Hebbian learning, principal components analysis (PCA), and K-means clustering, revealing that all of these algorithms are prone to discovering illusory structures that reliably arise in random realizations of high dimensional noise.  Finally, we end section \ref{sec:smoflearning} by discussing an application of message passing to learning with binary valued synapses, known to be an NP-complete problem \cite{blum1992training, amaldi1991complexity}.  The authors of \cite{braunstein2006learning, baldassi2007efficient} derived a biologically plausible learning algorithm capable of solving random instantiations of this problem by approximating message passing in a joint probability distribution over synaptic weights determined by the training examples. 
 
 \par In section \ref{sec:rmt}, we discuss the eigenvalue spectrum of random matrices.   Matrices from many random matrix ensembles have eigenvalue spectra whose probability distributions display fascinating macroscopic structures that do not depend on the detailed realization of the matrix elements.   These spectral distributions play a central role in a wide variety of fields \cite{mehta2004random, akemann2011oxford}; within the context of neural networks for example, they play a role in understanding the stability of linear neural networks, the transition to chaos in nonlinear networks \cite{sompolinsky1988chaos}, and the analysis of high dimensional data.  We begin section \ref{sec:rmt} by showing how replica theory can also provide a general framework for computing the typical eigenvalue distribution of a variety of random matrix ensembles.   Then we focus on understanding an ensemble of random empirical covariance matrices (the Wishart ensemble \cite{wishart1928generalised}) whose eigenvalue distribution, known as the Marcenko-Pastur distribution \cite{marchenko1967distribution}, provides a null model for the outcome of PCA applied to high dimensional data.  Moreover, we review how the eigenvalues of many random matrix ensembles can be thought of as Coulomb charges living in the complex plane, and the distribution of these eigenvalues can be thought of as the thermally equilibriated charge density of this Coulomb gas, which is stabilized via the competing effects of a repulsive two dimensional Coulomb interaction and an attractive confining external potential.   Moreover we review how the statistics of the largest eigenvalue, which obeys the Tracy-Widom distribution \cite{tracy1994level, tracy2002distribution}, can be understood simply in terms of thermal fluctuations of this Coulomb gas \cite{vivo2007large, majumdar2009large}.  The statistics of this largest eigenvalue will make an appearance later in section \ref{rdimred} when we discuss how random projections distort the geometry of manifolds.   Overall, section \ref{sec:rmt} illustrates the power of the replica formalism, and plays a role in connecting the statistical physics of two dimensional Coulomb gases to PCA in section \ref{secunsupill} and geometric distortions induced by dimensionality reduction in section \ref{sec:corrextreme}.     

\par
In section \ref{rdimred} we discuss the notion of random dimensionality reduction.   High dimensional data can be difficult to both model and process.  One approach to circumvent such difficulties is to reduce the dimensionality of the data; indeed many machine learning algorithms search for optimal directions on which to project the data.  As discussed in section \ref{secunsupill}, such algorithms yield projected data distributions that reveal low dimensional, illusory structures that do not exist in the data.  An alternate approach is to simply project the data onto a random subspace.  As the dimensionality of this subspace is lower than the ambient dimensionality of the feature space in which the data resides, features of the data will necessarily be lost.  However, it is often the case that interesting data sets lie along low dimensional submanifolds in their ambient feature space.  In such situations, a random projection above a critical dimension, that is more closely related to the dimensionality of the submanifold than to the dimensionality of the ambient feature space,  often preserves a surprising amount of structure of the submanifold.  In section \ref{rdimred} we review the theory of random projections and their ability to preserve the geometry of data submanifolds.  We end section \ref{rdimred} by introducing a statistical mechanics approach to random dimensionality reduction of simple random submanifolds, like point clouds and hyperplanes.  This analysis connects random dimensionality reduction to extremal fluctuations of 2D Coulomb gases discussed in sections \ref{sec:cg} and \ref{sec:tw}.

\par The manifold of sparse signals forms a ubiquitous and interesting low dimensional structure that accurately captures many types of data.  The field of compressed sensing \cite{bruckstein2009sparse,Candes:2008la}, discussed in section \ref{sec:compsense}, rests upon the central observation that a sparse high dimensional signal can be recovered from a random projection down to a surprisingly low dimension by solving a computationally tractable convex optimization problem, known as $L_1$ minimization.   In section \ref{sec:compsense} we focus mainly on the analysis of $L_1$ minimization based on statistical mechanics and message passing.    For readers who are more interested in applications of random projections, compressed sensing and $L_1$ minimization to neuronal information processing and data analysis, we refer them to \cite{ganguli2012annrevs}.  There, diverse applications of how the techniques in sections \ref{rdimred} and \ref{sec:compsense} can be used to acquire and analyze high dimensional neuronal data are discussed, including, magnetic resonance imaging \cite{lustig2008compressed, lustig2007sparse, parrish1995continuous},  compressed gene expression arrays \cite{dai2009compressive},  compressed connectomics \cite{hu2009reconstruction, mishchenko2011reconstruction}, receptive field measurements, and fluorescence microscopy \cite{wilt2009advances, taraska2010fluorescence} of multiple molecular species at high spatiotemporal resolution \cite{coskun2010lensless} using single pixel camera \cite{takhar2006new, duarte2008single} technology.
Also diverse applications of these same techniques to neuronal information processing are discussed in \cite{ganguli2012annrevs}, including 
semantic information processing \cite{rogers2004semantic, kiani2007object, kriegeskorte2008matching}, 
short-term memory \cite{ganguli2008memory, ganguli2010short},
neural circuits for $L_1$ minimization \cite{rozell2008sparse}, learning sparse representations \cite{olshausen1996emergence, perrinet2010role}, 
regularized learning of high dimensional synaptic weights from limited examples \cite{lage2009statistical}, and
axonally efficient long range brain communication through random projections \cite{coulter2010adaptive, Isely:2010fk,  Hillar:2011uq, kim2012spatial}.
\par
After introducing CS in \ref{sec:csintro}, we show how replica theory can be used to analyze its performance in section \ref{sec:csrepl}.   Remarkably, the performance of CS, unlike other algorithms discussed in section \ref{secunsupill}, displays a phase transition. For any given level of signal sparsity, there is a critical lower bound on the dimensionality of a random projection which is required to accurately recover the signal; this critical dimension decreases with increasing sparsity.  Also, in section \ref{sec:csmpass} we review how the $L_1$ minimization problem can be formulated as a message passing problem \cite{donoho2009message}.  This formulation yields a message passing dynamical system that qualitatively mimics neural network dynamics with a crucial history dependence terms. $L_1$ minimization via gradient descent has been proposed as a framework for neuronal dynamics underlying sparse coding in both vision \cite{hu11early} and olfaction \cite{koulakov11rindberg}.  On the otherhand, the efficiency of message passing in solving $L_1$ minimization, demonstrated in \cite{donoho2009message} may motivate revisiting the issue of sparse coding in neuroscience, and the role of history dependence in sparse coding network dynamics. 

\par  Finally, the appendix in section \ref{sec:repapp} provides an overview of the replica method, in a general form that is immediately applicable to spin glasses, perceptron learning, unsupervised learning, random matrices and compressed sensing.  Overall, the replica method is a powerful, if non-rigorous, method for analyzing the statistical mechanics of systems with quenched disorder.  We hope that this exposition of the replica method, combined with the cavity and message passing methods discussed in this review within a wide variety of disparate contexts, will help enable students and researchers in both theoretical neuroscience and physics to learn about exciting interdisciplinary advances made in the last few decades at the intersection of statistical physics, computer science, and neurobiology.

\section{Spin Glass Models of Neural Networks}
\label{sec:spinglassmodel}

The SK Model \cite{sherrington1975solvable} is a prototypical example of a disordered statistical mechanical system.  It has been employed as a simple model of spin glasses \cite{mezard1987spin, fischer1993spin}, as well as neural networks \cite{amit1985spin}, and has made a recent resurgence in neuroscience within the context of maximum entropy modeling of spike trains \cite{schneidman2006weak, shlens2006structure}.    It is defined by the energy function 
\begin{equation}
H(\s, \J) = - \frac{1}{2}  \sum_{ij} \J_{ij} s_i s_j,
\label{eq:skham}
\end{equation}
where the $s_i$ are $N$ spin degrees of freedom taking the values $\pm 1$.  In a neural network interpretation, $\s_i$ represents the activity state of a neuron and $\J$ is the synaptic connectivity matrix of the network.   This Hamiltonian yields an equilibrium Gibbs distribution of neural activity given by
\begin{equation}
P_{\mathbf J}(\s) = \frac{1}{Z[\J]}  e^{- \beta H(\s, {\mathbf J})}
\label{eq:spingibbs}
\end{equation}
where
\begin{equation}
 Z[{\mathbf J}] = \sum_{\s}  e^{- \beta  H(\s, {\mathbf J})}
 \end{equation}
is the partition function, and $\beta$ is an inverse temperature reflecting sources of noise.  The connectivity matrix is chosen to be random, where each $\J_{ij}$ is an independent, identically distributed (i.i.d) zero mean Gaussian with variance $1/N$. 
 
 \par The main property of interest is the statistical structure of high probability (low energy) activity patterns.   Much progress in spin glass theory \cite{mezard1987spin} has revealed a physical picture in which the Gibbs distribution in (\ref{eq:spingibbs}) decomposes at low temperature (large $\beta$) into many ``lumps'' of probability mass (more rigorously, pure states \cite{krzakala2007gibbs}) concentrated on subsets of activity patterns.  Equivalently, these lumps can be thought of as concentrated on the minima of a free energy landscape with many valleys.   Each lump, indexed by $a$, is characterized by a mean activity pattern $m_i^a = \tav{s_i}_a$, where $\tav{\cdot}_a$ is an average over configurations belonging to the free energy valley $a$, and a probability mass $P_a$ (the probability that a random activity pattern belongs to valley $a$).     In the large $N$ limit, free energy barriers between valleys diverge, so that in dynamical versions of this model, if an activity pattern starts in one valley, it will stay in that valley for infinite time.  Thus ergodicity is broken, as time average activity patterns are not equal to the full Gibbs average activity pattern. The network can thus maintain multiple steady states, and we are interested in understanding the structure of these steady states.  
 
 \par Now the detailed activity pattern in any free energy minimum $a$ (i.e the mean pattern $m^a_i$) depends on the detailed realization of the connectivity $\J$, and is hard to compute.   However, many interesting quantities, that involve averages over all neurons, are self-averaging, which by definition means that their fluctuations across different realizations of $\J$ vanish in the large $N$ limit.  As we see below, typical values of such quantities, for any given realization of $\J$, can be computed theoretically by computing their average over all $\J$.    One interesting quantity that probes the geometry of free energy minima is the distribution of overlaps between all pairs of activity patterns.   If the activity patterns belong to two valleys, $a$ and $b$, then the overlap is  
 \begin{equation}
 q_{ab} = \frac{1}{N} \sum_i m^a_i m^b_i.
 \label{eq:valleyovlap}  
 \end{equation}
 Now since $P_a$ is the probability a randomly chosen activity pattern belongs to valley $a$, the distribution of overlaps between any two pairs of activity patterns independently chosen from \eqref{eq:spingibbs} is given by
 \begin{equation}
 P_{\mathbf J}(q) = \sum_{ab} P_a \, P_b \, \delta (q - q_{ab}).
 \label{eq:maindaod}
 \end{equation}
 This distribution turns out not to be self-averaging (it fluctuates across realizations of $\J$), unless there is only one valley, or state (modulo the reflection symmetry $\s_i \rightarrow -\s_i)$, in which case the distribution becomes concentrated at a single number $q$, which is the self-overlap of the state,  $q = \frac{1}{N} \sum_i m^2_i$.  If there is indeed one state, then $q$  does not depend on the detailed realization of $\J$ and provides a measure of the variability of mean activity across neurons due to the quenched disorder in the connectivity.  In the case of multiple valleys, one can also compute the disorder averaged overlap distribution $\qav{P_{\mathbf J}(q)}_{\mathbf J}$; despite the fact that the overlap distribution $P_{\mathbf J}(q)$ may not be self-averaging, its average over $\J$ can  still yield a wealth of information about the geometric organization of free energy minima in neural activity space.  This can be done using the replica method, which we now introduce.   

\subsection{Replica Solution}
To understand the statistical properties of the Gibbs distribution in (\ref{eq:spingibbs}), it is useful to compute its free energy $-\beta F[{\mathbf J}] = \ln Z[\J]$.  Correlations between neurons can then be computed via suitable derivatives of the free energy.  Fortunately, the free energy is self-averaging, which means that to understand the free energy for any realization of $\J$, it suffices to compute its average over all $\J$:
\begin{equation}
\qav{-\beta F[{\mathbf J]}}_{\mathbf J} = \qav{\ln Z[{\mathbf J}]}_{\mathbf J} ,
\end{equation} 
where $\qav{\cdot}_{\mathbf J}$ denotes an average over the disorder $\J$.  This average is difficult to do because the logarithm appears inside the average.  The replica trick circumvents this difficulty by exploiting the identity 
\begin{equation}
\ln Z = \lim_{n \rightarrow 0} \frac{Z^n-1}{n} = \lim_{n \rightarrow 0} \frac{\partial}{\partial n} Z^n.
\end{equation}
This identity is useful because it allows us to first average over an integer power of $Z[\J]$, which can be done more easily, and then take the $n \rightarrow 0$ limit.  Appendix \ref{sec:repapp} provides a general outline of the replica approach that can be used for many problems.  Basically to compute the average over $Z^n$ it is useful to introduce $n$ replicated neuronal activity patterns $\s^a$, for  $a = 1,\dots,n$, yielding
\begin{equation}
\qav{Z^n}_{\mathbf J} = \bqav{\sum_{ \{ \s^a \}} \,  e^{\beta \sum_{a=1}^n \sum_{ij}  J_{ij} s^a_i s^a_j}}_{\mathbf J}.
\label{eq:skrepl1}
\end{equation}
Now the average over $\J$ can be performed because it is reduced to a set of Gaussian integrals.  To do so, we use the fundamental identity 
\begin{equation}
\tav{e^{zx}}_z = e^{\frac{1}{2} \sigma^2 x^2},
\label{eq:hubstrat}
\end{equation}
where $z$ is a zero mean Gaussian random variable with variance $\sigma^2$.  Applying this to (\ref{eq:skrepl1}) with $z=J_{ij}$, $\sigma^2 = \frac{1}{N}$, and $x = \beta \sum_a s^a_i s^a_j$ yields 
\begin{equation}
\qav{Z^n}_{\mathbf J} = \sum_{ \{ \s^a \}} \,  e^{\frac{1}{4N} \sum_{ij} (\sum_{a=1}^n  s^a_i s^a_j)^2} = \sum_{ \{ \s^a \}} \,  e^{N \frac{\beta^2}{4} \sum_{ab} \Qab^2},
\label{eq:disorderavsk}
\end{equation}
where 
\begin{equation}
\Qab = \frac{1}{N} \sum_{i=1}^N s^a_i s^b_i
\label{eq:ovlapspin}
\end{equation}
is the overlap matrix between replicated activity patterns.   

\par Thus although for any fixed realization of the quenched disorder $\J$, the replicated activity patterns $\s^a$ were independent, marginalizing over, or integrating out the disorder introduces attractive interactions between the replicas.  Consistent with the general framework, presented in section \ref{sec:repapp}, the interaction between replicas depends only on the overlap matrix $Q$, and we have in (\ref{eq:Eqabdef}),  $E(Q) = - \frac{\beta^2}{4} \sum_{ab} \Qab^2$. Thus minimization of this energy 
\begin{figure}[htbp]
   \centering
   \includegraphics[width=6in]{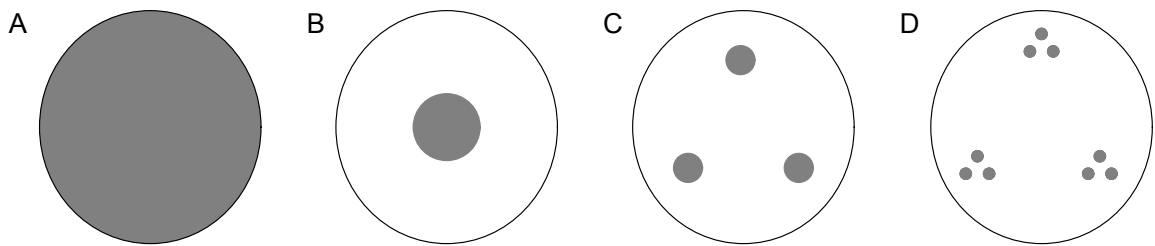} % requires the graphicx package
   \caption{Probability lumps in free energy valleys.  Schematic figures of the space of all possible neuronal or spin configurations (large circle) and the space of spin configurations with non-negligible probability under the Gibbs distribution in \eqref{eq:spingibbs} (shaded areas).  (A) At high temperature all spin configurations are explored by the Gibbs distribution.  Thus the inner product between two random spins drawn from the Gibbs distribution will typically have $0$ inner product, and so the replica order parameter $q$ is $0$.  (B) The replica symmetric ansatz for a low temperature phase: the spins freeze into a small set of configurations (free energy valley), which can differ from realization to realization of the connectivity $\J$.  However, the inner product between two random spins, and therefore also the replica order parameter, takes a nonzero value $q$ that does not depend on the realization of $\J$. (C) One possible ansatz for replica symmetry breaking (RSB) in which the replica overlap matrix $Q$ is characterized by two order parameters, $q_1 > q_2$.  This ansatz, known as 1-step RSB, corresponds to a scenario in which the Gibbs distribution breaks into multiple lumps,  with $q_1$ describing the typical inner product between two configurations chosen from the same lump, and $q_2$ describing the typical inner product between configurations from different lumps. (D) There exists a series of k-step RSB schemes describing scenarios in which the Gibbs distribution decomposes into a nested hierarchy of lumps of depth $k$.  This figure describes a possible scenario for $k=2$.  The true low temperature phase of the SK model is thought to be described by a particular $k=\infty$ RSB ansatz \cite{mezard1987spin}.}
\label{FigLandscape}
\end{figure}
function promotes alignment of the replicas.  The intuition is that for any fixed realization of $\J$, the replicas will prefer certain patterns.   Which patterns are preferred will vary across realizations of $\J$.  However, for any fixed realization of $\J$, the preferred set of patterns will be similar across replicas since the fluctuations of each replicated neuronal activity pattern are controlled by the same quenched connectivity $\J$.   Thus even after averaging over $\J$, we expect this similarity to survive, and hence we expect average overlaps between replicas to be nonzero. 

\par However, minimization of the energy $E(Q)$ alone does not determine the overlap matrix $Q_{ab}$.   One must still sum over $\s^a$ in (\ref{eq:disorderavsk}) which yields an entropic term corresponding to the number of replicated activity patterns with a given set of overlaps.  While energy minimization drives overlaps to be large, entropy maximization drives overlaps to be small, since there are many more replicated configurations with small, rather than large overlaps.   This competition between energy and entropy leads to a potentially nontrivial overlap matrix.  After computing this entropic term, the most likely value of the overlap matrix can be computed via the saddle point method, yielding a set of self-consistent equations for Q (a special case of  (\ref{eq:repsadd1}),(\ref{eq:repsadd2})):
\begin{equation}
\Qab = \langle s^a s^b \rangle_n,
\label{eq:mainskrep}
\end{equation}
where $\langle \cdot \rangle_n$ denotes an average with respect to the Gibbs distribution $P(s^1,\dots,s^n) = \frac{1}{Z} e^{-\beta H_{\rm eff}}$, with $H_{\rm eff} = -\beta \sum_{ab} s^a \Qab s^b$.   
\par  Now the physical meaning of the saddle point replica overlap matrix is explained in \ref{sec:ovlapmeaning}; it is simply related to the disorder averaged overlap distribution:
\begin{equation}
\qav{P_{\mathbf J}(q)}_{\mathbf J} = \lim_{n \rightarrow 0} \frac{1}{n(n-1)} \sum_{a \neq b} \delta ( q - Q_{ab} ),
\end{equation}
where $P_{\mathbf J}(q)$ is given by  (\ref{eq:maindaod}).  So the distribution of overlaps between pairs of free energy minima $m^a_i$ (weighted by their probability), is simply the distribution of off-diagonal matrix elements of the replica overlap matrix.   Thus, in searching for solutions to (\ref{eq:mainskrep}), any ansatz about the structure of $\Qab$ is implicitly an ansatz about the geometry  and multiplicity of free energy valleys in (\ref{eq:spingibbs}), averaged over $\J$. 
\par
Now the effective Hamiltonian yielding the average in (\ref{eq:mainskrep}) is symmetric with respect to permutations of the replica indices $a$ (i.e. permuting the rows and columns of $\Qab$).  Therefore it is natural to search for a replica  symmetric saddle point in which $Q_{ab} = q$ for all $a \neq b$.   This is equivalent to an assumption that there is only one free energy valley, and $q$ measures its heterogeneity.  Taking the $n \rightarrow 0$ limit with this replica symmetric ansatz yields a saddle point equation for $q$ (see \eqref{eq:saddlesk} for the derivation):
\begin{equation}
q = \qav{ \tanh^2 ( \beta \sqrt q z)}_z.
\label{eq:skrseq}
\end{equation}
At high temperature ($\beta < 1$), $q=0$ is the only solution, representing a ``paramagnetic" state (Fig. \ref{FigLandscape}A) in which activity patterns fluctuate over all possible configurations, and average neural activity $m_i$  is $0$ for all $i$ (Fig. \ref{FigLandscape}A).  At lower temperature ($\beta > 1$), a nonzero solution rises continuously from $0$, suggesting a phase transition to a ``frozen" state corresponding to a single valley (Fig. \ref{FigLandscape}B) in which each neuron has a different mean activity $m_i$.   
\par While this scenario seems plausible,  a further analysis of this solution \cite{sherrington1975solvable, kirkpatrick1978infinite} yields inconsistent physical predictions (like negative entropy for the system).   Within the replica framework, this inconsistency can be detected by showing that the replica symmetric  saddle point for $Q_{ab}$ is unstable \cite{de2001stability}, and so one must search for solutions in which $\Qab$ breaks the replica symmetry.  This corresponds to a physical picture in which there are many free energy minima.   A great deal of work has lead to a remarkably rich ansatz which predicts a nested hierarchical, tree like organization on the space of free energy minima (see Fig.\ \ref{FigLandscape}CD), known as an ultrametric structure \cite{rammal1986ultrametricity}.   It is striking that this highly symmetric and ordered low temperature hierarchical structure emerges generically from purely random, disordered couplings $\J_{ij}$.   Unfortunately, we will not explore this phenomenon further here, since for most of the applications of replica theory to neuronal processing and data analysis discussed below, a replica symmetric analysis turns out to be correct.
\subsection{Chaos in the SK Model and the Hopfield Solution}
\par
So far, in order to introduce the replica method, we have analyzed a toy neuronal network with a random symmetric connectivity matrix $\J$, and found that such a network exhibits broken replica symmetry corresponding to a hierarchy of low energy states that are stable with respect to thermal or noise induced fluctuations.  It is tempting to explore the possibility that this multiplicity of states may be useful for performing neural information processing tasks.  However, several works have noted that while these states are stable with respect to thermal fluctuations, they are not structurally stable with respect to perturbations either to the inverse temperature $\beta$, or the connectivity matrix $\J$  \cite{huse1999pure, bray1987chaotic, chatterjee2009disorder}.  Indeed very small changes to $\beta$ or $\J$ induce macroscopic changes in the location of energy minima in the space of neuronal activity patterns.   This sensitive dependence of low energy activity patterns to either $\beta$ or $\J$ was called temperature or disorder chaos respectively in \cite{bray1987chaotic}.   For neural information processing, it would be useful to instead have network connectivities whose noisy dynamics not only thermally stabilize a prescribed set of neuronal activity patterns, but do so in a manner that is structurally stable with respect to changes in either the connectivity or level of noise.  
 \par 
An early proposal to do just this was the Hopfield model \cite{hopfield1982neural}.   Suppose one wishes to find a network connectivity $\J$ that stabilizes a prescribed set of $P$ $N$-dimensional patterns \mbox{\boldmath{$\xi$}}$^\mu$, for $\mu=1,\dots,P$, where \mbox{\boldmath{$\xi$}}$^\mu_i = \pm 1$.  Hopfield's proposal was to choose
\begin{equation}
\J_{ij} = \frac{1}{N} \sum_{\mu=1}^P \xi^\mu_i \xi^\mu_j.
\label{eq:hopfieldconn}
\end{equation}
This choice reflects the outcome of a Hebbian learning rule \cite{hebb1949organization} in which each synapse from neuron $j$ to neuron $i$ changes its synaptic weight by an amount proportional to the correlation between the activity on its presynaptic and postsynaptic neuron.  When the activity pattern  \mbox{\boldmath{$\xi$}}$^\mu$ is imposed upon the network, this correlation is $\xi^\mu_i \xi^\mu_j$, and when all $P$ patterns are imposed upon the network in succession, the learned synaptic weights are given by \eqref{eq:hopfieldconn}. 
\par 
This synaptic connectivity $\J$ induces an equilibrium probability distribution over neuronal activity patterns $\s$ through \eqref{eq:spingibbs}.  Ideally, this distribution should have $2P$ free energy valleys, corresponding to lumps of probability mass located near the $P$ patterns \mbox{\boldmath{$\xi$}}$^\mu$ and their reflections
$-$\mbox{\boldmath{$\xi$}}$^\mu$.   If so, then when network activity $\s$ is initialized to either a corrupted or partial version of one of the learned patterns \mbox{\boldmath{$\xi$}}$^\mu$, the network will relax (under a dynamics whose stationary distribution is given by \eqref{eq:spingibbs}) to the free energy valley corresponding to \mbox{\boldmath{$\xi$}}$^\mu$.   This relaxation process is often called pattern completion.   Thus Hopfield's prescription provided a unifying framework for thinking about learning and memory:  the structure of past experience (i.e. the  patterns \mbox{\boldmath{$\xi$}}$^\mu$) are learned, or stored, in the network's synaptic weights (i.e. through \eqref{eq:hopfieldconn}), and subsequent network dynamics can be viewed as motion down a free energy landscape determined by the weights.   If learning is successful, the minima of this free energy landscape correspond to past experiences, and the process of recalling past experience corresponds to completing partial or corrupted initial network activity patterns induced by current stimuli.
\par
A key issue then is storage capacity:  how many patterns $P$ can a network of $N$ neurons store?  This issue was addressed in \cite{amit55storing, amit1987statistical} via the replica method in the situation where the stored patterns \mbox{\boldmath{$\xi$}}$^\mu$ are random and uncorrelated (each $\xi^\mu_i$ is chosen independently to be $+1$ or $-1$  with equal probability).   These works extensively analyzed the properties of free energy valleys in the Gibbs distribution \eqref{eq:spingibbs} with connectivity \eqref{eq:hopfieldconn}, as a function of the inverse temperature $\beta$ and the level of storage saturation $\alpha = \frac{P}{N}$.  This problem fits the classic mold of disordered statistical physics, where the patterns \mbox{\boldmath{$\xi$}}$^\mu$ play the role of quenched disorder, and neuronal activity patterns play the role of thermal degrees of freedom.   In particular, the structure of free energy minima can be described by a collection of self-averaging order parameters $m^\mu = \frac{1}{N}$ \mbox{\boldmath{$\xi$}}$^\mu \cdot \s$, denoting the overlap of neuronal activity with pattern $\mu$.  Successful pattern completion is possible if there are $2P$ free energy valleys such that the average of $m^\mu$ in each valley is large for one pattern $\mu$ and small for all the rest.  These free energy valleys can be thought of as recall states.  The replica method in \cite{amit55storing, amit1987statistical} yields a set of self-consistent equations for these averages.  Solutions to the replica equations, in which precisely one order parameter $m^\mu$ is large, are found at low temperature only when $\alpha < \alpha_c = 0.138$.  For $\alpha > \alpha_c$, the system is in a spin glass state with many free energy minima, none of which have a macroscopic overlap with any of the patterns (in the solutions to the replica equations, no average $m^\mu$ is O(1) as $P,N \rightarrow \infty$ with $\alpha > \alpha_c$).  At such high levels of storage, so many patterns ``confuse" the network, so that its low energy states do not look like any one pattern \mbox{\boldmath{$\xi$}}$^\mu$.  Indeed the free energy landscape of the Hopfield model as $\alpha$ becomes large behaves like the low temperature spin glass phase of the SK model discussed in the previous section.
\par Even for $\alpha < \alpha_c$,  at low enough temperatures, spurious, metastable free energy valleys corresponding to mixtures of patterns can also arise.    These mixture states are characterized by solutions to the replica equations in which the average $m^\mu$ is $O(1)$ for more than one $\mu$.   However, as the temperature is increased,  such mixture states melt away.  This phenomenon illustrates a beneficial role for noise in associative memory operation.  However, there is a tradeoff to melting away mixture states by increasing temperature, as  $\alpha_c$ decreases with increasing temperature.  Nevertheless, in summary there is a robust region in the $\alpha-\beta$ phase plane with $\alpha = O(0.1)$ and $\beta$ corresponding to low temperatures, in which the recall states dominate the free energy landscape over neural activity patterns, and the network can successfully operate as a pattern completion, or associative memory device.  Many important details about the phase diagram of free energy valleys as a function of $\alpha$ and $\beta$ and be found in \cite{amit55storing, amit1987statistical}.
 
\subsection{Cavity Method}
We now return to an analysis of the SK model through an alternate method that sheds light on the physical meaning of the saddle point equation for the replica symmetric order parameter $q$ in (\ref{eq:skrseq}), which may seem a bit obscure.  In particular, we give an alternate derivation of (\ref{eq:skrseq}) through the cavity method \cite{mezard1987spin, shamir2000thouless} which provides considerable physical intuition for (\ref{eq:skrseq}) by describing it as a self-consistency condition.  In general, the cavity method, while indirect, can often provide intuition for the final results derived via more direct replica methods.  
\par The starting point involves noting that the SK Hamiltonian (\ref{eq:skham}) governing the fluctuations of $N$ neurons can be written as
\begin{equation}
H_N(\s, \J) = - s_1 h_1 + \H1,
\label{eq:skdecomp}
\end{equation}
where
\begin{equation}
h_1 = \sum_{i=2}^N \J_{1i} s_i
\label{eq:localfield}
\end{equation}
is the local field acting on neuron $1$,
and 
\begin{equation}
\H1 = - \frac{1}{2}  \sum_{ij=2}^N \J_{ij} s_i s_j,
\label{eq:cavham}
\end{equation}
is the Hamiltonian of the rest of the neurons $s_2, \dots, s_N$. Since $h_1$ is a sum of many terms, it is tempting to approximate its thermal fluctuations in the full system of $N$ neurons in (\ref{eq:skdecomp}) by a Gaussian distribution.   However, such a Gaussian approximation is generally invalid because the individual terms are correlated with each other.   One source of correlation arises from a common coupling of all the neurons $s_2, \dots, s_N$ to $s_1$.  For example, because  $s_1$ interacts with $s_i$ through the symmetric coupling $\J_{1i} = \J_{i1}$, whenever $s_1= +1$ (or $s_1= -1$) this exerts a positive (or negative)  effect on the combination $\J_{1i} s_i$.  Thus all individual terms in (\ref{eq:localfield}) exhibit correlated fluctuations due to common coupling to the fluctuating neuron $s_1$.   
\par
\begin{figure}[htbp]
   \centering
   \includegraphics[width=5.5in]{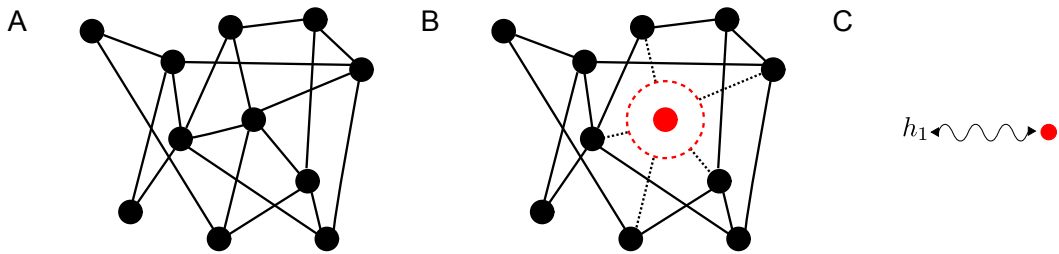} % requires the graphicx package
   \caption{The cavity Method.  (A) A network of neurons, or spins.  (B) A cavity surrounding a single neuron, $\s_1$, that has been removed from the system.  (C) In a replica symmetric approximation, the full distribution of the field $h_1$ exerted on the cavity (in the absence of $\s_1$) by all other neurons can be approximated by a Gaussian distribution, while the joint distribution of  $\s_1$ and $h_1$ takes the form in equation \eqref{eq:cavdecomp}.  }
   \label{FigCavity}
\end{figure}

The key idea behind the cavity method is to consider not the distribution of the local field $h_1$ acting on neuron $1$ in the full system of $N$ neurons in (\ref{eq:skdecomp}), but instead the distribution of $h_1$ in a ``cavity system" of $N-1$ neurons obtained by removing $s_1$ from the system, thereby leaving ``cavity" (see Fig. \ref{FigCavity}AB).   Then $h_1$ is known as the cavity field, or the field exerted on neuron $1$ by all the others in the {\em absence} of neuron $1$, and its distribution is given by that of $h_1$ (\ref{eq:localfield}) in a Gibbs distribution with respect to (\ref{eq:cavham}):
\begin{equation}
P_{\bk{1}}(h_1) = \frac{1}{Z_{\bk{1}}}\sum_{s_2,\dots,s_N} \delta(h_1 - \sum_{i=2}^N \J_{1i} s_i) \, e^{-\beta \H1} 
\end{equation}  
The joint distribution of $s_1$ and its local field $h_1$ in the full system of $N$ spins can be written in terms of the cavity field distribution as follows:
\begin{eqnarray}
P_N(s_1,h_1) & = \frac{1}{Z_N} \sum_{s_2,\dots,s_N} \delta(h_1 - \sum_{i=2}^N \J_{1i} s_i) \, e^{-\beta H_N} \nonumber \\
                       & = \frac{1}{Z} e^{-\beta V(s_1,h_1)} P_{\bk{1}}(h_1) \label{eq:cavdecomp}, 
\end{eqnarray} 
where $V(s_1,h_1) = -s_1 h_1$.  
\par The advantage of writing the joint distribution of $s_1$ and $h_1$ in terms of the cavity field distribution $P_{\bk{1}}(h_1)$ is that one can now plausibly make a Gaussian approximation to $P_{\bk{1}}(h_1)$, i.e. the distribution of (\ref{eq:localfield}) in the cavity system (\ref{eq:cavham}) of neurons $2,\dots,N$ in the absence of $1$.  Because the cavity system does not couple to neuron $1$, it does not know about the set of couplings $\J_{1i}$, and therefore the thermal fluctuations of cavity 
activity patterns $\s_2,\dots,s_N$, while of course correlated with each other,  must be uncorrelated with the couplings $\J_{1i}$, unlike the case of these same fluctuations in the presence of $s_1$.    Motivated by this lack of correlation, we can make a Gaussian approximation to the thermal fluctuations of $h_1$ in the cavity system $\H1$.  Note that this does not imply that the local field $h_1$ in the full system $H_N$ is Gaussian.  Indeed, if $P_{\bk{1}}(h_1)$ in (\ref{eq:cavdecomp}) is Gaussian, then $P_N(h_1)$ obtained by marginalizing out $s_1$ in $P_N(s_1,h_1)$ cannot be Gaussian; as discussed above, this non-gaussianity arises due to positive correlations between $s_1$ and $h_1$ induced by their coupling $V(s_1,h_1)$.   The simplification in replacing the network with a fluctuating field is shown in the transition from Fig.\ref{FigCavity}B to \ref{FigCavity}C.   
\par
Under a Gaussian approximation, $P_{\bk{1}}(h_1)$ is characterized by its mean 
\begin{equation}
\tav{h_1}_{\bk{1}} = \sum_{i=2}^{N} \J_{1i} \tav{s_i}_{\bk{1}},
\end{equation}
and variance
\begin{eqnarray}
\tav{(\delta h_1)^2}_{\bk{1}} & = \sum_{i,j=2}^{N} \J_{1i} \J_{1j} \tav{\delta s_i \delta s_j}_{\bk{1}} \label{eq:cavvar1} \\
                                             & = \sum_{i=2}^{N} \J_{1i}^2  \tav{(\delta s_i)^2}_{\bk{1}}  \label{eq:cavvar2}\\
                                             & = 1 - \sum_{i=2}^{N} \frac{1}{N}  \tav{s_i}^2_{\bk{1}} \nonumber \\
                                             & = 1 - q
\end{eqnarray}
where $q$ is the order parameter
\begin{equation}
q = \frac{1}{N} \sum_{i=1}^{N}   \tav{s_i}^2_N,
\end{equation}
and $\delta s_i = s_i - \tav{s_i}_{\bk{1}}$.
Here we have neglected various terms that vanish in the large $N$ limit, but most importantly, in going from (\ref{eq:cavvar1}) to (\ref{eq:cavvar2}), we have made a strong assumption that the connected correlation $\tav{\delta s_i \delta s_j}_{\bk{1}}$ vanishes in the large $N$ limit fast enough that we can neglect all off-diagonal terms in (\ref{eq:cavvar1}).   This can be true if the cavity system (and consequently the full system) is accurately described by a single free energy valley.   On the otherhand, if the system is described by multiple free energy valleys, the connected correlation will receive contributions from fluctuations across valleys, and we cannot neglect the off-diagonal terms \cite{mezard1987spin}.  Thus the validity of this cavity approximation is tantamount to an assumption of replica symmetry, or a single valley in the free energy landscape.  As discussed above, under the assumption of a single valley, we expect  $q$ to be self-averaging: it does not depend on the detailed realization of $\J_{ij}$ in the large $N$ limit.  Finally, we note that the cavity method can be extended to scenarios in which replica symmetry is broken and there are multiple valleys \cite{mezard1987spin}. 
\par
In the replica symmetric scenario, under the Gaussian approximation to $P_{\bk{1}}(h_1)$, (\ref{eq:cavdecomp}) becomes 
\begin{equation}
P_N(s_1,h_1)  = \frac{1}{Z[\tav{h_1}_{\bk{1}} , 1-q]} e^{-\beta V(s_1,h_1) - \frac{1}{2(1-q)} (h_1 - {\langle h_1 \rangle}_{\bk{1}})^2},
\label{eq:fulldist}
\end{equation}
allowing us to compute the mean activity of neuron $i$ in the full system of $N$ neurons, in terms of the mean $\tav{h_1}_{\bk{1}}$ and variance $1-q$ of its cavity field,
\begin{equation}
\tav{s_1 \, | \, \tav{h_1}_{\bk{1}} , 1-q \, }_N = \sum_{s_1} s_1 P_N(s_1,h_1).
\label{eq:tavs1}
\end{equation}  
But now we must compute $q$, which we can do by demanding self consistency of the cavity approximation.  First of all, we note that there was nothing special about neuron $1$; the above procedure of forming a cavity system by removing neuron $1$ could have been done with any neuron.  Thus (\ref{eq:fulldist}, \ref{eq:tavs1}) holds individually for all neurons $i$, and we can average these equations over all $i$ to obtain an expression for $q$:
\begin{equation}
q = \frac{1}{N} \sum_{i=1}^N\tav{s_i \, | \, \tav{h_i}_{\bk{i}} , 1-q \, }^2_N.
\label{eq:qselfcon1}
\end{equation}
However, we do not yet know $\tav{h_i}_{\bk{i}}$ for each $i$.   For each $i$,  $\tav{h_i}_{\bk{i}} = \sum_{k \neq i} \J_{ik} \tav{s_k}_{\bk{i}}$ is a random variable due to the randomness of the couplings $\J_{ik}$, which are uncorrelated with $\tav{s_k}_{\bk{i}}$ by virtue of the fact that this thermal average occurs in a cavity system in the absence of $i$.     
Thus we expect the distribution of  $\tav{h_i}_{\bk{i}}$ over random realizations of $\J_{ik}$ to be gaussian, with a mean and variance that are easily computed to be $0$ and $q$ respectively.   Furthermore, we expect this distribution to be self-averaging:  i.e. the distribution of $\tav{h_i}_{\bk{i}}$ for a fixed $i$ across different realizations of $\J$ should be the same as the distribution of  $\tav{h_i}_{\bk{i}}$ across different neurons $i$ for a fixed realization of $\J$, in the large $N$ limit.  Under this assumption, although we may not know each individual $\tav{h_i}_{\bk{i}}$, we can replace the average over neurons in (\ref{eq:qselfcon1}) with an average over a Gaussian distribution, yielding
\begin{equation}
q = \qav{ \, \tav{s_i \, | \, \sqrt{q}z , 1-q \, }_N^2 \, }_z.
\label{eq:selfconscav}
\end{equation}
Here $\qav{\cdot}_z$ denotes a ``quenched" average with respect to a zero mean unit variance Gaussian variable $z$, reflecting the heterogeneity of the mean cavity field across neurons, and the thermal average $\tav{\cdot}_N$ is computed via (\ref{eq:fulldist}, \ref{eq:tavs1}), and reflects the thermal fluctuations of a single neuron in the presence of a cavity field with mean and variance $\sqrt{q}z$ and $1-q$, respectively.  
\par 
 Equation (\ref{eq:selfconscav}) is a self-consistent equation for the order parameter $q$ which is itself a measure of the heterogeneity of mean activity across neurons.  So physically, (\ref{eq:selfconscav}) reflects a demand that the statistical properties of the cavity fields are consistent with the heterogeneity of mean neural activity. Now finally, we can specialize to the $SK$ model in which $V(s,h) = -sh$ in (\ref{eq:fulldist}), which yields $\tav{s_1 \, | \, \sqrt{q}z , 1-q \, }_N = \tanh \beta \sqrt{q}{z}$ in (\ref{eq:tavs1}), and when this is substituted into (\ref{eq:selfconscav}), we recover the self-consistent equation for $q$ in (\ref{eq:skrseq}), derived via the replica method.

\subsection{Message Passing}
\label{subsecmpass}
So far, we have seen two methods which allow us to calculate self-averaging quantities (for example $q = \frac{1}{N} \sum_{i} \tav{s_i}^2$) that do not depend on the detailed realization $\J$.  However, we may wish to understand the detailed pattern of mean neural activity, i.e. $\tav{s_i}$ for all $i$, for some fixed realization of $\J_{ij}$.  Mathematically, this corresponds to computing the marginal distribution of a single neuron in a full joint distribution given by (\ref{eq:spingibbs}).  Here we introduce efficient distributed message passing algorithms from computer science \cite{mezard2009information, pearl1988probabilistic, koller2009probabilistic} that have been developed to compute such marginals in probability distributions which obey certain factorization properties.
  
\par Consider for example a joint distribution over $N$ variables $x_1,\cdots,x_N$ that factorizes into a set of $P$ factors, or interactions, indexed by $a=1,\dots P$:
\begin{equation}
P(x_1, \dots, x_N) = \frac{1}{Z} \prod_{a=1}^P \psi_a(x_a).
\label{eq:factorize}
\end{equation} 
Here $x_i$ is any arbitrary variable that could be either continuous or discrete, and $x_a$ denotes the collection of variables that factor $a$ depends on.  Thus we systematically abuse notation and think of each factor index $a$ also as a subset of the $N$ variables, with variable $i \in a$ if and only if factor $\psi_a$ depends on $x_i$.    The factorization properties of $P$ can be visualized in a factor graph, which is a bipartite graph whose nodes correspond either to variables $i$ or factors $a$, and 
\begin{figure}[htbp]
   \centering
   \includegraphics[width=6in]{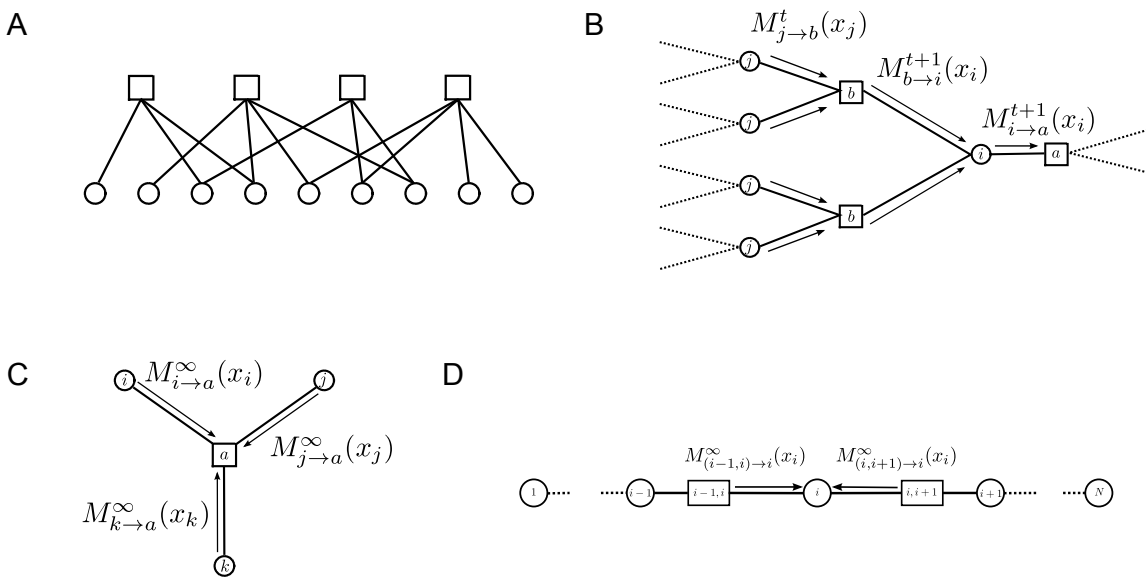} % requires the graphicx package
   \caption{Message Passing.  (A) A factor graph in which variable nodes are represented by circles, and factor nodes are represented by squares.  (B) The flow of messages involved in the update of the message $M^{t+1}_{i \rightarrow a}(x_i)$. (C) The message passing approximation to the joint distribution of $x_i, x_j$, and $x_k$.  Here the interaction $a$ is treated exactly, while the effect of all other interactions besides $a$ are approximated by a product of messages. (D) Exact message passing in a chain; the marginal on $x_i$ is computed exactly as a product of two messages.}
   \label{FigMessage}
\end{figure}
there is an edge between a factor $a$ and variable $i$ if and only if $i \in a$, or equivalently factor $\psi_a$ depends on $x_i$ (see Fig. \ref{FigMessage}A).   For example, the SK model, or more generally any neural system with an equilibrium distribution, corresponds to a factor graph in which the neurons $s_i$ are the variables $x_i$, and the factors correspond to nonzero synaptic weights connecting pairs of neurons.  Thus each $a$ corresponds to a neuron pair $a=(ij)$, and in the SK model of equation (\ref{eq:spingibbs}), $\psi_a(s_i,s_j) = e^{\beta \mathbf{J}_{ij} s_i s_j}$.
\par The utility of the factor graph representation is that an iterative algorithm to compute the marginals,
\begin{equation}
P(x_i) = \sum_{ \{ x_j \}, j \neq i} P(x_1,\dots,x_N)
\end{equation}
for all $i$ can be visualized as the flow of messages along the factor graph (\ref{FigMessage}B).  We first define this iterative algorithm and then later give justification for it.   Every message is a probability distribution over a single variable, and at any given time $t$ there are two types of messages, one from variables to factors, and the other from factors to variables.   We denote by $M^t_{j \rightarrow b}(x_j)$ the message from variable $j$ to factor $b$, and by $M^{t}_{b \rightarrow i}(x_i)$ the message from factor $b$ to variable $i$.  Intuitively, we can think of $M^t_{j \rightarrow b}(x_j)$ as an approximation to distribution on $x_j$ induced by all other interactions besides interaction $b$.    In contrast, we can think of $M^{t}_{b \rightarrow i}(x_i)$ as an approximation to the distribution on $x_i$ induced by the direct influence of interaction $b$ alone.   These messages will be used below to approximate the marginal of $x_i$ in the full joint distribution of all interactions (see e.g. \eqref{eq:marg1}).
\par The (unnormalized) update equation for a factor to variable message is given by 
\begin{equation}
M^{t+1}_{b \rightarrow i}(x_i) = \sum_{x_{b \backslash i}} \psi_b(x_b) \prod_{j \in b \backslash i} M^t_{j \rightarrow b}(x_j),
\label{eq:mpass1}
\end{equation}  
where $b \bk{i}$ denotes the set of all variables connected to factor node $b$ except $i$ (see Fig. \ref{FigMessage}b).
Intuitively, the direct influence of $b$ alone on $i$ (the left hand side of (\ref{eq:mpass1})) is obtained by marginalizing out all variables other than $i$ in the factor $\psi_b$, supplemented by accounting for the effect of all of the other interactions besides $b$ on variables $j \in b \bk{i}$ by the product of messages $M^t_{j \rightarrow b}(x_j)$ (see Fig. \ref{FigMessage}b).   
The (unnormalized) update equation for the variable to factor messages is then given by
\begin{equation}
M^{t+1}_{i \rightarrow a}(x_i) = \prod_{b \in i \backslash a} M^{t+1}_{b \rightarrow i}(x_i),
\label{eq:mpass2}
\end{equation}  
Intuitively, the distribution on $x_i$ induced by all other interactions besides interaction $a$ (the left hand side of (\ref{eq:mpass2})) is simply the product of the direct influences of all interactions $b$ that involve variable $i$, except for interaction $a$ (see Fig. \ref{FigMessage}b).  Message passing involves randomly initializing all the messages  and then iteratively running the update equations (\ref{eq:mpass1},\ref{eq:mpass2}) until convergence.  One exception to the random initialization is the situation where any variable $i$ is connected to only $1$ variable node $a$.  In this case, $M_{i  \rightarrow a}(x_i)$ is initialized to be a uniform distribution over $x_i$, since in the absence of $a$, variable $i$ feels no influence from the rest of the graph.  Under the message passing dynamics, $M_{i  \rightarrow a}(x_i)$ will remain a uniform distribution.  Now for general factor graphs, convergence is not guaranteed, but if the algorithm does converge, then the marginal distribution of a variable $x_i$ can be approximated via 
\begin{equation}
P(x_i) \propto \prod_{a \in i} M^{\infty}_{a \rightarrow i}(x_i).
\label{eq:marg1}
\end{equation}
and indeed the the joint distribution of all variables $i \in a$ can be approximated  via
\begin{equation}
P(x_a) \propto \psi_a(x_a) \prod_{i \in a} M^{\infty}_{i \rightarrow a}(x_i).
\label{eq:marg2}
\end{equation}
\par
The update equations (\ref{eq:mpass1},\ref{eq:mpass1}), while intuitive, lead to two natural questions:  for which factor graphs will they converge, and if they converge, how well will the fixed point messages $M^{\infty}_{a \rightarrow i}$ and $M^{\infty}_{i \rightarrow a}$ approximate the true marginals though equations 
(\ref{eq:marg1},\ref{eq:marg2})?  A key intuition arises from the structure of the approximation to the joint marginal of the variables $x_a$ in (\ref{eq:marg2}) (see also Fig. \ref{FigMessage}C).   This approximation treats the coupling of the variables $i \in a$ through interaction $a$ by explicitly including the factor $\psi_a$.  However, it approximates the effects of all other interactions $b$ on these variables by a simple product of messages $M^{\infty}_{i \rightarrow a}(x_i)$.  An exactly analogous approximation is made in the update equation (\ref{eq:mpass1}).   Such approximations might be expected to work well whenever removing the interaction $a$ leads to a factor graph in which all the variables $i$ that were previously connected to $a$ are now weakly coupled (ideally independent) under all the remaining interactions $b \neq a$. 
\par
This weak coupling assumption under the removal of a single interaction holds exactly whenever the factor graph is a tree, with no loops.  Indeed in such a case, removing any one interaction $a$ removes all paths through the factor graph between variables $i \in a$.   In the absence of any such paths, all pairs of variables $i \in a$ are independent, and their joint distribution factorizes, consistent with the approximations made in (\ref{eq:mpass1}) and (\ref{eq:marg2}).  In general, whenever the factor graph is a tree,  the message passing equations converge in finite time, and the fixed point messages yield the true marginals  \cite{pearl1988probabilistic}.  We will not give a general proof of this fact, but we will illustrate it in the case of a one dimensional Ising chain (see Fig. \ref{FigMessage}D).   Consider the marginal distribution of a spin at position $i$ in the chain.  This spin feels an interaction to its left and right, and so \eqref{eq:marg2} tells us the marginal is a product of two converged messages at time $t=\infty$:
\begin{equation}
P(s_i) \propto M^{\infty}_{(i-1,i) \rightarrow i}(s_i) \, M^{\infty}_{(i,i+1) \rightarrow i}(s_i), \label{eq:cmarg}
\end{equation} 
Each of these two messages can be computed by iterating messages from either end of the chain to position $i$.  For example, the rightward iteration for computing $M^{\infty}_{(i-1,i) \rightarrow i}(s_i)$ is
\begin{equation}
M^{t+1}_{k  \rightarrow (k,k+1)}(s_{k})  = M^{t+1}_{(k-1,k) \rightarrow k}(s_k)   = \sum_{s_{k-1}} e^{\beta \mathbf{J}_{k,k-1} s_k \, s_{k-1}} M^{t}_{k-1  \rightarrow (k-1,k)}(s_{k-1}) 
\end{equation}
where the first equality is a special case of \eqref{eq:mpass2} and the second is a special case of \eqref{eq:mpass1}.  The first message in this iteration, 
$M^{0}_{1  \rightarrow (1,2)}(s_{1})$ is initialized to be a uniform distribution, since spin $1$ is only connected to a single interaction $(1,2)$.  A similar leftward iteration leads to the calculation of $M^{\infty}_{(i,i+1) \rightarrow i}(s_i)$.  Each iteration converges in an amount of time given by the path length from each corresponding end to $i$, and after convergence, we have
\begin{eqnarray}
M^{\infty}_{(i-1,i) \rightarrow i}(s_i) & = \sum_{s_1,\dots, s_{i-1}} e^{\beta \sum_{k=1}^{i-1} \mathbf{J}_{k,k+1} s_k s_{k+1}}  \label{eq:cm1} \\
M^{\infty}_{(i,i+1) \rightarrow i}(s_i) & = \sum_{s_{i+1},\dots, s_{N}} e^{\beta \sum_{k=i}^{N-1} \mathbf{J}_{k,k+1} s_k s_{k+1}}. \label{eq:cm2} 
\end{eqnarray}
Inserting \eqref{eq:cm1} and \eqref{eq:cm2} into \eqref{eq:cmarg} yields the correct marginal for $s_i$, and the normalization factor can be fixed at the end by demanding $P(+1) + P(-1) = 1$.  
Note that whereas a naive sum over all spin configurations to compute the marginal over $s_i$ would require $O(2^N)$ operations, this iterative procedure for computing the marginal requires only $O(N)$ operations.  Moreover, two sweeps through the chain allow us to compute all the messages, and therefore all $N$ marginals simultaneously, as \eqref{eq:cmarg} holds for all $i$.  Overall, this method is essentially identical to the transfer matrix method for the $1D$ Ising chain, and is a generalization of the Bethe approximation \cite{bethe1935}.  
\par
Although message passing is only exact on trees, it can nevertheless be applied to graphical models with loops, and as discussed above, it should yield good approximate marginals whenever the variables adjacent to a factor node are weakly correlated upon removal of that factor node.  We will see successful examples of this in the contexts of learning in section \ref{sec:mpasslearn} and compressed sensing in \ref{sec:csmpass}.  An early theoretical advance in partially justifying the application of message passing to graphical models with loops was a variational connection:  each solution to the fixed point equations of message passing are in one to one correspondence with extrema of a certain Bethe free energy \cite{yedidia2005constructing}, an approximation to the Gibbs free energy in variational approaches to inference in graphical models that is exact on trees (see \cite{wainwright2008graphical} for a review).  However there are no known general and precise conditions under which message passing in graphical models with many loops is theoretically guaranteed to converge to messages that yield a good approximation to the marginals. 
Nevertheless, in practice, message passing seems to achieve empirical success in approximating marginals when correlations between variables adjacent to a factor node are indeed weak after removal of that factor. 
\par 
We conclude this section by connecting message passing back to the replica method.   In general, suitable averages of the messaging passing equations reduce to both the cavity equations and the replica equations \cite{mezard2009information}.  To illustrate this in the special case of the SK model, we outline the derivation of the replica saddle point equation  \eqref{eq:skrseq} from the perspective of message passing.  We first note that since every factor node in the SK model has degree $2$, the update of a message from a factor $(i,j)$ to variable $j$, i.e. $M_{(i,j) \rightarrow j}(s_j)$ depends only on the message $M_{i \rightarrow (i,j) }(s_i)$ through,
\begin{equation}
M^{t+1}_{(i,j) \rightarrow j}(s_j) = \sum_{s_i} \, e^{\beta \mathbf{J}_{ij} s_i s_j}  \,  M^{t}_{i \rightarrow (i,j)}(s_i),
\end{equation}
which is a special case of \eqref{eq:mpass1}.  Thus we can take one set of messages, for example the node to factor messages, $M^{t}_{i \rightarrow (i,j)}(s_i)$, as the essential degrees of freedom upon which the message passing dynamics operates.   We simplify notation a little by letting $M^{t}_{i \rightarrow j}(s_i) \equiv M^{t}_{i \rightarrow (i,j)}(s_i)$.  Then the remaining message passing update \eqref{eq:mpass2} yields the dynamics
\begin{equation}
M^{t+1}_{i \rightarrow j}(s_i) = \prod_{k \in i \backslash j} \sum_{s_k} \, e^{\beta \mathbf{J}_{ik} s_i s_k} \, M^{t}_{k \rightarrow i}(s_k),
\label{eq:SKmesspass}
\end{equation}
where $k \in i$ if and only if $s_k$ is coupled to $s_i$ through a nonzero $\J_{ik}$.  
\par Now each message is a distribution over a binary variable, and all such distributions can be usefully parameterized by a single scalar parameter:
\begin{equation}
M^{t}_{i \rightarrow j}(s_i) \propto e^{\beta h^t_{i \rightarrow j} s_i}. 
\end{equation} 
Here the scalar parameter $h^t_{i \rightarrow j}$ can be thought of as a type of cavity field;  as $t \rightarrow \infty$, if message passing is successful, $h^t_{i \rightarrow j}$ converges to the field exerted on spin $i$ by all the spins in a cavity system in which the interaction $\J_{ij}$ is removed.  In terms of this parameterization, the message passing updates \eqref{eq:SKmesspass} yield a dynamical system on the cavity fields \cite{mezard2001bethe},
\begin{equation}
h^{t+1}_{i \rightarrow j} = \sum_{k \in i \backslash j} u(\J_{ki}, h^{t}_{k \rightarrow i}). 
\label{eq:skcavdyn}
\end{equation}
Here $u(J,h)$ is defined implicitly through the relation
\begin{equation}
e^{\beta u(J,h) s} \propto \sum_{s'} e^{\beta J s s' + \beta h s'}.
\end{equation}
Physically $u(J,h)$ is the effective field on a binary spin $s$ coupled with strength $J$ to another spin $s'$ that experiences an external field of strength $h$, after marginalizing out $s'$. Explicitly, 
\begin{equation}
u(J,h)  = \frac{1}{\beta} \mathrm{arctanh} \left[ \tanh (\beta J) \tanh (\beta h)  \right].
\label{eq:ujh}
\end{equation}
In the weak coupling limit of small $J$, $u(J,h) \approx J \tanh (\beta h)$, which reflects the simple approximation that the average magnetization of $\s'$, due to the external field $h$ (which would be $\tanh (\beta h)$ if the back-reaction from $s$ can be neglected), exerts a field  $J \tanh (\beta h)$ on $s$.  The more complex form of $u(J,h)$ in \eqref{eq:ujh} reflects the back-reaction of $s$ on $s'$ that becomes non-negligible at larger values of the bi-directional coupling $J$.  In \eqref{eq:skcavdyn},
the updated cavity field $h^{t+1}_{i \rightarrow j}$ turns out to be a simple sum over all the spins $k$ (besides $j$) of this same effective field $u$ obtained by marginalizing out $s_k$ in the presence of its own cavity field $h^{t}_{k \rightarrow i}$.  
\par
Using \eqref{eq:skcavdyn}, we are now ready to derive \eqref{eq:skrseq}.  The key point is to consider self-consistency conditions for the distribution of cavity fields $h^{t}_{i \rightarrow j}$ as $t \rightarrow \infty$.   We can think of this distribution in two ways.  First, for a fixed $i$ and $j$, $h^{\infty}_{i \rightarrow j}$ is a random variable due to the random choice of couplings $\J$.  Second, for a fixed realization of $\J$, at a message passing fixed point, there is an empirical distribution of cavity fields $h^{\infty}_{i \rightarrow j}$ across all choices of pairs $i$ and $j$.   The assumption of self-averaging means that as $N \rightarrow \infty$, the latter empirical distribution converges to the distribution of the former random variable.  In any case, we would like to write down a self-consistent equation for this distribution, by observing that this distribution must be self-reproducing under the update equation \eqref{eq:skcavdyn}.  More precisely, in \eqref{eq:skcavdyn}, if the couplings $J_{ik}$ are drawn i.i.d from a distribution $P(J)$, and the cavity fields $h^{t}_{k \rightarrow i}$ are drawn i.i.d. from a distribution $Q(h)$, then the induced distribution on $h^{t+1}_{i \rightarrow j}$ should be identical to $Q(h)$.  This yields a recursive distributional equation characterizing the distribution of cavity fields $Q(h)$ at a message passing fixed point:
\begin{equation}
Q(h) =  \int \prod_k \, d \J_k \, P( \J_k ) \, \prod_k  \, dh_k \, Q( h_k ) \, \delta \left (h - \sum_{k} u(\J_k, h_k) \right ).
\label{eq:disteq}
\end{equation}
Here we have suppressed the arbitrary indices $i$ and $j$.  More generally, one can track the time-dependent evolution of the distribution of cavity fields, an algorithmic analysis technique known as density evolution \cite{mezard2009information}.
\par
In general, it can be difficult to solve the distributional equation for $Q(h)$.  However, in the SK model, one could make an approximation that the distribution of cavity fields is a zero mean Gaussian with variance $q$.  Then the distributional equation for $Q(h)$ reduces to a self-consistency condition for $q$ by taking the expectation of $h^2$ on each side of \eqref{eq:disteq}.  The left hand side is by definition $q$.  To simplify the right hand side, since the couplings $J_k$ have variance of $\frac{1}{N}$, we can use the small coupling approximation $u(J,h) \approx J \tanh (\beta h)$.  Then averaging $h^2$ on both sides of \eqref{eq:disteq}
yields
\begin{eqnarray}
q & =  \int \prod_k \, d \J_k \, P( \J_k ) \, \prod_k  \, dh_k \, Q( h_k ) \, \left ( \sum_{k} \J_k \tanh \beta h_k \right )^2 \\
   & =  \int \prod_k \, d \J_k \, P( \J_k ) \, \prod_k  \, dh_k \, Q( h_k ) \, \left ( \sum_{k} \J^2_k \tanh^2 \beta h_k  \right ) \\
   & =  \int  \prod_k  \, dh_k \, Q( h_k ) \, \left ( \frac{1}{N} \sum_{k} \tanh^2 \beta h_k  \right )\\
   & =  \int  dh \, Q( h ) \, \tanh^2 \beta h. 
\end{eqnarray}
Now, since we have assumed $Q(h)$ is zero mean Gaussian with variance $q$, this is equivalent to the replica symmetric saddle point equation \eqref{eq:skrseq}. 
\par
In summary, we have employed a toy model of a neural network, the SK spin glass model, to introduce the various replica, cavity and message passing approaches to analyzing disordered statistical mechanical systems.  In each case we have discussed in detail the simplest possible ansatz concerning the structure of free energy landscape, namely the replica symmetric ansatz, corresponding to a single valley with weak connected correlations between degrees of freedom.  While this assumption is not true for the SK model, it nevertheless provides a good example system in which to gain familiarity with the various methods.   And fortunately, for many of the applications discussed below, the assumption of a single free energy valley governing fluctuations will turn out to be correct.  Finally, we note that just as the replica and cavity methods can be extended \cite{mezard1987spin} to scenarios in which replica symmetry is broken, corresponding to many free energy valleys and long range correlations, so too can message passing approaches.  Indeed viewing optimization and inference problems through the lens of statistical physics has lead to a new algorithm, known as survey propagation \cite{mezard2002analytic, braunstein2005survey}, which can find good marginals, or minimize costs, in free energy landscapes characterized by many metastable minima that can confound more traditional, local algorithms.  
\section{Statistical Mechanics of Learning}
\label{sec:smoflearning}

In the above sections, we have reviewed powerful machinery designed to understand the statistical mechanics of fluctuating neural activity patterns in the presence of disordered synaptic connectivity matrices.   A key conceptual advance made by Gardner \cite{gardner1988space, gardner1999optimal} was that this same machinery could be applied to the analysis of learning, by performing statistical mechanics directly on the space of synaptic connectivities, with the training examples presented to the system playing the role of quenched disorder.   In this section we will explore this viewpoint and its applications to diverse phenomena in neural and unsupervised learning (see \cite{engel01} for an extensive review of this topic).   

\subsection{Perceptron Learning}
\label{sec:perceptlearn}

The perceptron is a simple neuronal model defined by a vector of $N$ synaptic weights $\w$, which linearly sums a pattern of incoming activity \mbox{\boldmath${\xi}$},
and fires depending on whether or not the summed input is above a threshold.  Mathematically, in the case of zero threshold, it computes the function $\sigma = {\rm sgn}(\w \cdot$\mbox{\boldmath{$\xi$}}$)$, where $\sigma=+1$ represents the firing state and $\sigma=-1$ represents the quiescent state.   Geometrically, it separates its input space into two classes, each on opposite sides of the $N-1$ dimensional hyperplane orthogonal to the weight vector $\w$.   Since the absolute scale of the weight vector $\w$ is not relevant to the problem, we will normalize the weights to satisfy $\w \cdot \w = N$, so that the set of perceptrons live on an $N-1$ dimensional sphere. 

\par Suppose we wish to train a perceptron to memorize a desired set of $P$ input-output associations, \mbox{\boldmath{$\xi$}}$^\mu \rightarrow \sigma^\mu$.    Doing so requires a learning rule (an algorithm for modifying the synaptic weights $\w$ based on the inputs and outputs) that finds a set of synaptic weights $\w$ that satisfies the $P$ inequalities 
\begin{equation}
\w \cdot \sigma^\mu $\mbox{\boldmath{$\xi$}}$^\mu \geq 0 \quad \forall \quad \mu=1,\dots,P.   
\label{eq:ineq}
\end{equation}
We will see below, that as long as there exists a simultaneous solution $\w$ to the $P$ inequalities, then remarkably, a learning rule, known as the perceptron learning rule \cite{rosenblatt1958perceptron}, can find the solution.  The main remaining question is then, under what conditions on the training data $\{$\mbox{\boldmath{$\xi$}}$^\mu, \sigma^\mu \}$ does a solution to the inequalities exist?

\par
\begin{figure}[htbp]
   \centering
   \includegraphics[width=5.5in]{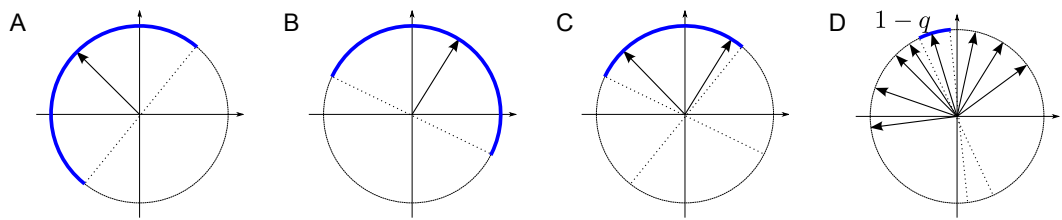} % requires the graphicx package
   \caption{Perceptron Learning. (A) The total sphere of all perceptron weights (grey circle) and a single example (black arrow).  The blue region is the set of perceptron weights that yield an output $+1$ on the example.  (B) The same as (A), but for a different example. (C) The set of weights that yield $+1$ on both examples in (A) and (B).  (D) As more examples are added, the space of correct weights shrinks, and its typical volume is given by $1-q$, where $q$ is the replica order parameter introduced in section \ref{sec:repllearn}}
   \label{FigPerceptron}
\end{figure}

 A statistical mechanics based approach to answering this question involves defining an energy function on the $N-1$ dimensional sphere of perceptrons as follows,
\begin{equation}
E(\w) = \sum_{\mu=1}^P V(\lambda^\mu),
\label{eq:percenergy}
\end{equation} 
where  $\lambda^\mu = \frac{1}{\sqrt{N}}\w\cdot\sigma^\mu$\mbox{\boldmath{$\xi$}}$^\mu$ is the alignment of example $\mu$ with the weight vector $\w$.  Successfully memorizing all the patterns requires all alignments to be positive, so $V$ should be a potential that penalizes negative alignments and favors positive ones.  Indeed a wide variety of learning algorithms for the perceptron architecture can be formulated as gradient descent on $E(\w)$ for various choices of potential functions $V(\lambda)$ in (\ref{eq:percenergy}) \cite{engel01}.  However, if we are interested in probing the space of solutions to the inequalities (\ref{eq:ineq}), it is useful to take $V(\lambda) = \theta(-\lambda)$, where $\theta(x)$ is the Heaviside function ($\theta(x)=1, x \geq 0$, and $0$ otherwise).    With this choice, the energy function in (\ref{eq:percenergy}) simply counts the number of misclassified examples, and so the Gibbs distribution
\begin{equation}
P(\w) = \frac{1}{Z} e^{-\beta E(\w)}
\label{eq:pergibbs}
\end{equation} 
in the zero temperature ($\beta \rightarrow \infty$) limit becomes a uniform distribution on the space of perceptrons satisfying (\ref{eq:ineq}) (see Fig. \ref{FigPerceptron}).
Thus the volume of the space of solutions to (\ref{eq:ineq}), and in particular, whether or not it is nonzero,  can be computed by analyzing the statistical mechanics of  (\ref{eq:pergibbs}) in the zero temperature limit.

\subsection{Unsupervised Learning}

This same statistical mechanics formulation can be extended to more general unsupervised learning scenarios.  In unsupervised learning, one often starts with a set of $P$ data vectors 
\mbox{\boldmath{$\xi$}}$^\mu$, for $\mu=1,\dots,P$, where each vector is of dimension $N$.  For example, each vector could be a pattern of expression of $N$ genes across $P$ experimental conditions, or a pattern of activity of $N$ neurons in response to $P$ stimuli.   The overall goal of unsupervised learning is to find simple hidden structures or patterns in the data.   The simplest approach is to find an interesting single dimension spanned by the vector $\w$, such that 
the projections
$\lambda^\mu = \frac{1}{\sqrt{N}}\w\cdot$\mbox{\boldmath{$\xi$}}$^\mu$ of the data onto this single dimension yield a useful one dimensional coordinate system for the data.  This interesting dimension can often be defined by minimizing the energy function (\ref{eq:percenergy}), with the choice of $V(\lambda)$ determining the particular unsupervised learning algorithm.  One choice, $V(\lambda) = -\lambda$ corresponds to Hebbian learning.  Upon minimization of $E(\w)$, this choice leads to $\w \propto \sum_\mu$\mbox{\boldmath{$\xi$}}$^\mu$; i.e. $\w$ points in the direction of the center of mass of the data.  In situations in which the data has its center of mass at the origin, a useful choice is $V(\lambda) = -\lambda^2$.   Under this choice, $\w$ points in the direction of the eigenvector of maximal eigenvalue of the data covariance matrix $\bo{C} = \sum_\mu$\mbox{\boldmath{$\xi$}}$^\mu$\mbox{\boldmath{$\xi$}}$^{\mu^T}$.   This is the direction of maximal variance in the data, also known as the first principal component of the data; i.e.  it is the direction which maximizes the variance of the distribution of $\lambda^\mu$ across data points $\mu$. 
\par
Beyond finding an interesting dimension in the data, another unsupervised learning task to find clusters in the data.  A popular algorithm for doing so is $K$-means clustering.  This is an iterative algorithm in which one maintains a guess about $K$ potential cluster centroids in the data, $\w_1,\dots, \w_K$.  At each iteration in the algorithm, each cluster $i$ is defined to be the set of data points closer to centroid $\w_i$ than to any other centroid.  Then all the cluster centroids $\w_i$ are optimized by minimizing the sum of the distances from $\w_i$ to those data points {\boldmath{$\xi$}}$^\mu$ assigned to cluster $i$.   In the case where the distance measure is euclidean distance, this step just sets each centroid $\w_i$ to be the center of mass of the data points assigned to cluster $i$.   The cluster assignments of the data are then recomputed with the new centroids, and the whole process repeats. The idea is that if there are $K$ well separated clusters in the data, this iterative procedure should converge so that each $\w_i$ is the center of mass of cluster $i$. 
\par  For general $K$, this iterative procedure can be viewed as an alternating minimization of a joint energy function over cluster centroids and cluster membership assignments.  For the special case of $K=2$, and when both the data and cluster centroids are normalized to have norm $N$, this energy function can be written as
\begin{equation}
E(\w_1,\w_2) =  \sum_{\mu=1}^P V(\lambda_1^\mu, \lambda_2^\mu),
\end{equation}
where
\begin{equation}
\lambda^\mu_i  = \frac{1}{\sqrt{N}} \w_i\cdot $\mbox{\boldmath{$\xi$}}$^\mu,
\end{equation}
and,
\begin{eqnarray}
V(\lambda_1, \lambda_2) & = - \lambda_1 \theta(\lambda_1-\lambda_2)  - \lambda_2 \theta(\lambda_2 - \lambda_1)  \label{eq:kmeans1}\\
                                         & = \frac{1}{2} (\lambda_1 + \lambda_2) - \frac{1}{2} |\lambda_1-\lambda_2|. \label{eq:kmeans2}
\end{eqnarray}
Gradient descent on this energy function forces each centroid $\w_i$ to perform Hebbian learning only on the data points that are currently closest to it.
\par

\subsection{Replica Analysis of Learning}
\label{sec:repllearn}
Both perceptron learning and unsupervised learning, when formulated as statistical mechanics problems as above, can be analyzed through the replica method. 
A natural question for perceptron learning is how many associations $P$ can a perceptron with $N$ synapses memorize?  One benchmark is the case of random associations where \mbox{\boldmath{$\xi$}}$^\mu$ is a random vector drawn from a uniform distribution on a sphere of radius $N$ and $\sigma^u = \pm 1$ each with probability half.   Similarly, a natural question for unsupervised learning, is how do we assess the statistical significance of any structure or pattern we find in a high dimensional dataset consisting of $P$ points in $N$ dimensions?  To address this question it is often useful to analyze what structure we may find in a null data distribution that itself has no structure, for example when the data points \mbox{\boldmath{$\xi$}}$^\mu$ are drawn from a uniform distribution on the $N-1$ sphere (or equivalently, in the large $N$ limit, from a Gaussian distribution with identity covariance matrix). 
\par
In both cases, the analysis simplifies in the "thermodynamic" limit $P,N \rightarrow \infty$ with the ratio $\alpha = P/N$ held constant.  Fortunately, this is the limit of relevance to neural models with many synaptic weights, and to high dimensional data.  The starting point of the analysis involves understanding the low energy configurations of the Gibbs distribution in (\ref{eq:pergibbs}).  In the thermodynamic limit, important observables, like the volume of low energy configurations or the distribution of data along the optimal direction(s) become self-averaging;  they do not depend on the detailed realization of  \mbox{\boldmath{$\xi$}}$^\mu$ or $\sigma^\mu$.  Therefore we can compute these observables by averaging $\log Z$ over these realizations.  This can be done by first averaging the replicated partition function
\begin{equation}
\qav{Z^n} = \bqav{  {\int} \, \prod_{a=1}^n \, d\w^a e^{- \sum_{a=1}^n \sum_{\mu=1}^P V(\lambda_a^\mu)} },
\label{eq:unsuprepl1}
\end{equation}
where 
$\lambda^\mu_a = \frac{1}{\sqrt{N}}\w_a\cdot$\mbox{\boldmath{$\xi$}}$^\mu$.  (For the case of perceptron learning, we can make the redefinition $\sigma^\mu$ \mbox{\boldmath{$\xi$}}$^\mu \rightarrow $\mbox{\boldmath{$\xi$}}$^\mu$, since both have the same distribution; in essence we absorb the sign of the desired output into the input yielding only positive examples.)  Averaging over \mbox{\boldmath{$\xi$}}$^\mu$ then reduces to averaging over $\lambda_a^\mu$.  These variables are jointly Gaussian distributed with zero mean and covariance matrix  $\qav{\lambda_a^\mu \lambda_b^\nu} = Q_{ab} \delta_{\mu\nu}$   where $Q_{ab} = \frac{1}{N} \w^a \cdot \w^b$ is the replica overlap matrix.  Thus after averaging over $\lambda_a^\mu$, the integrand depends on the configuration of replicated weights only through their overlap.  Therefore it is useful to separate the integral over $\w^a$ into an integral over all possible overlaps $Q_{ab}$, and an integral over all configurations with the same overlap.
Following section \ref{sec:repapp}, this yields
\begin{equation}
\qav{Z^n}  =  {\int} \,  \prod_{ab}  \,  d\Qab \,    e^{- N [ E(Q) - S(Q))]},
\end{equation}
where
\begin{equation} 
E(Q) = - \alpha \ln   {\int} \, \prod_{a=1}^n \, \frac{d\lambda_a}{\sqrt{2\pi}} \, \frac{1}{\sqrt{\det Q}} e^{-\frac{1}{2}  \lambda_a Q^{-1}_{ab}\lambda_b - \sum_a \beta V(\lambda_a)}
\label{eq:perceptlearnbef}
\end{equation}
and 
\begin{equation}
S(Q) = \frac{1}{2} \Tr \log Q
\end{equation}
is the entropy of the volume of weight vectors with overlap matrix $Q$.  For example, for perceptron learning when $V(\lambda) = \theta(-\lambda)$, in the zero temperature limit $\beta \rightarrow \infty$, $E(Q)$ is an energetic term that promotes the alignment of the replicated weights so that they all yield the correct answer on any given set of examples (i.e. $\lambda^a > 0$ for all $a$), while $S(Q)$ is an entropic term that promotes replicated weight configurations with small overlaps, since they have larger volume.  

\par At large $N$ the integral over $Q_{ab}$ can be done via the saddle point method, and the competition between entropy and energy selects a saddle point overlap matrix.   We make the ansatz that the saddle point has a replica symmetric form $Q_{ab} = (1-q) \delta_{ab} + q$.  Given the connection (explained in section \ref{sec:ovlapmeaning}) between replica overlap matrix elements, and the distribution of overlaps of pairs of random weights $\w$ drawn independently from (\ref{eq:pergibbs}), this choice suggests the existence of a single free energy valley.   This is reasonable to expect as most of the energy functions we will be analyzing for unsupervised learning are convex.  Also, in the zero temperature limit, this ansatz suggests that the space of ground state energy configurations, if degenerate, should form a convex, connected set.  This is indeed true for perceptron learning, since the space of ground states is the intersection of a set of $P$ half-spheres (see Fig. \ref{FigPerceptron}).  Thus unlike the SK model, we expect a replica symmetric assumption to be a good approximation.
\par
Taking the $n \rightarrow 0$ limit yields a saddle point equation for $q$ which, as explained in section \ref{sec:n0unsup} can be derived from extremizing a free energy
\begin{equation}
F(q) =  \alpha \qav{\ln \zeta}_z + \frac{1}{2} \bigg[ \frac{q}{1-q}  + \ln (1-q) \bigg],
\label{eq:feper}
\end{equation} 
where 
\begin{equation}
\zeta  = {\int} \,  \frac{d\lambda}{\sqrt{2 \pi (1-q)}} \,  e^{ -\frac{1}{2} \frac{(\lambda - \sqrt{q}z)^2}{1-q} -  \beta V(\lambda)}
\label{eq:zetaper}
\end{equation}
is the partition function of the distribution appearing inside the average over $z$ in \eqref{eq:unsupdistalign}.   Now in the case of perceptron learning, $1-q$ reflects the typical volume of the solution space to \eqref{eq:ineq} (see Fig. \ref{FigPerceptron}D), in that $q$ in the $\beta \rightarrow \infty$ limit is the typical overlap between two zero energy synaptic weight configurations (see section \ref{sec:ovlapmeaning}).   $q$ arises from a minimization of the sum of two terms in $F(q)$ in \eqref{eq:feper}.  The first term is an energetic term which is a decreasing function of $q$, reflecting a pressure for synaptic weights to agree on all examples (promoting larger $q$). The second term is an entropic term that is an increasing function of $q$, which thus promotes smaller values of $q$ which reflect larger volumes in weight space.  As $\alpha$ increases, placing greater weight on the first term in $F(q)$, $q$ increases as energy becomes more important than entropy.  As shown in \cite{gardner1999optimal}, $q \rightarrow 1$ as $\alpha \rightarrow \alpha_c = 2$.   Thus a perceptron with $N$ weights can store at most $2N$ random associations. 

\subsection{Perceptrons and Purkinje Cells in the Cerebellum}
\par 
Interestingly, in \cite{brunel2004optimal} the authors developed a replica based analysis of perceptron learning and applied it to make predictions about the distribution of synaptic weights of Purkinje cells in the cerebellum. Indeed an analogy between the Purkinje cell and the perceptron was first posited over 40 years ago
\cite{marr1969theory, albus1971theory}.   The Purkinje cell has one of the largest and most intricate dendritic arbors of all neuronal cell types; this arbor is capable of receiving excitatory synaptic inputs from about 100,000 granule cells which, in areas of the cerebellum devoted to motor control, convey a sparse representation of ongoing internal motor states, sensory feedback, and contextual states.  The Purkinje cell output in turn can exert an influence on outgoing motor control signals.  In addition to the granule cell input, each Purkinje cell receives input from on average one cell in the inferior olive through a climbing fiber input, whose firing induces large complex spikes in the Purkinje cell as well as plasticity in the granule cell to Purkinje cell synapses.  Since inferior olive firing is often correlated with errors in motor tasks, climbing fiber input is thought to convey an error signal that can guide plasticity.  Thus at a qualitative level, the Purkinje cell can be thought of as performing supervised learning in order to map ongoing task related inputs to desired motor outputs, where the desired mapping is learned over time using error corrective signals transmitted through the climbing fibers.  
\par
Now the actual distribution of synaptic weights between granule cells and Purkinje cells has been measured \cite{isope2002properties}, and a prominent feature of this distribution is that it has a delta function at $0$, while the rest of the nonzero weights follow a truncated Gaussian distribution.  In particular about $80$ percent of the synaptic weights are exactly $0$.  If the Purkinje cell is implementing an important sensorimotor mapping, why then are a majority of the synapses silent?  In general, the distribution of synaptic weights in a network should reflect the properties of the learning rule as well as the statistics of inputs and outputs.   Thus one might be able to quantitatively derive the distribution of weights by positing a particular learning rule and input-output statistics and then derive the weight distribution.  However, the authors of \cite{brunel2004optimal} took an even more elegant approach that did not depend on even positing any particular learning rule.  They simply modeled the Purkinje cell architecture as a perceptron, assumed that it operated optimally at capacity, and derived the distribution of synaptic weights of perceptrons operating at capacity via a replica based Gardner type analysis.  Remarkably, for a wide range of input-output statistics, whenever the perceptron implemented the maximal number of input-output associations at a given level of reliability (its capacity), its distribution of synaptic weights consisted of a delta-function at $0$ plus a truncated Gaussian for the nonzero weights.  Indeed, like the data, a majority of the synapses were silent.  This prediction only relies on the perceptron operating at (or near) capacity, and does not depend on the learning rule; any learning rule that can achieve capacity would necessarily yield such a weight distribution.  
\par
The key intuition for why a majority of the synapses are silent comes from the constraint that all the granule cell to Purkinje cell synapses are excitatory.   Thus the either the Purkinje cell or the perceptron faces a difficult computational task:  it must find a nonnegative synaptic weight vector that linearly combines nonnegative granule cell activity patterns and fires for some fraction of granule cell patterns while not firing for the rest.  It turns out that false positive errors dominate the weight structure of the optimal perceptron operating at or near capacity:  there are many granule cell activation patterns for which the perceptron must remain below threshold and not fire, and the only way to achieve this requirement with nonnegative weights is to set many synapses exactly to zero.  Indeed by quantitatively matching the parameters of the replica based perceptron learning theory to physiological data, the capacity of the generic Purkinje cell was estimated to be about 40,000 input-output associations, corresponding to 5 kilobytes of information stored in the weights of a single cell \cite{brunel2004optimal}.

\subsection{Illusions of Structure in High Dimensional Noise}
\label{secunsupill}
In contrast to perceptron learning, in the applications of the statistical mechanics formulation in \eqref{eq:percenergy} and \eqref{eq:pergibbs}
to unsupervised learning discussed here, $V(\lambda)$ has a unique minimum leading to a non-degenerate ground state.  Thus in the zero temperature $\beta \rightarrow \infty$ limit we expect thermal fluctuations in the synaptic weights, reflected by $1-q$, to vanish.   Indeed we can find self consistent solutions to the extremization of $F(q)$ in \eqref{eq:feper} by assuming $1-q = \frac{\Delta}{\beta}$ as $\beta \rightarrow \infty$ with $\Delta$ remaining $O(1)$.  In this limit, 
\eqref{eq:feper} and \eqref{eq:zetaper} become
\begin{equation}
\frac{1}{\beta}{F(\Delta)} =   - \alpha  \qav{ {\rm min}_\lambda \bigg[ \frac{(\lambda-z)^2}{2 \Delta} + V(\lambda) \bigg]}_z+ \frac{1}{2 \Delta}.
\label{eq:fbetaunsup}
\end{equation}
Extremization of \eqref{eq:fbetaunsup} over $\Delta$ determines the value of $\Delta$ as a function of $\alpha$.  
\par Furthermore, the interesting observable for unsupervised learning is the distribution of alignments across examples with the optimal weight vector,
\begin{equation}
P(\lambda) = \frac{1}{P} \sum_{\mu=1}^P \delta(\lambda - \lambda^\mu),
\end{equation}
where $\lambda^\mu = \frac{1}{\sqrt{N}}\w \cdot$\mbox{\boldmath{$\xi$}}$^\mu$, and 
$\w$ minimizes $E(\w)$ in \eqref{eq:percenergy}.  This is essentially the distribution of the data \mbox{\boldmath{$\xi$}}$^\mu$ along the dimension discovered by unsupervised learning.  This distribution is derived via the replica method in section \ref{sec:aligndist} at finite temperature, and is given by \eqref{eq:unsupdistalign}.
Its zero temperature limit yields      
\begin{equation}
P(\lambda) = \qav{\delta(\lambda - \lambda^*(z,\Delta)}_z,
\label{eq:plam}
\end{equation}
where 
\begin{equation}
\lambda^*(z,\Delta) = {\rm argmin}_\lambda \bigg[ \frac{(\lambda-z)^2}{2 \Delta}  + V(\lambda) \bigg]
\label{eq:lamstar}
\end{equation}
\par Equations \eqref{eq:fbetaunsup}, \eqref{eq:plam} and \eqref{eq:lamstar} have a simple interpretation within the zero temperature cavity method applied to unsupervised learning \cite{mezard1999space, griniasty1993cavity}.  Consider a cavity system in which one of the examples, say example $1$ in \eqref{eq:percenergy} is removed, and let $\w^{\backslash 1}$ be the``cavity" weight vector which optimizes $E(\w)$ in the presence of all other examples \mbox{\boldmath{$\xi$}}$^\mu$ for $\mu = 2,\dots,P$.   Since $\w^{\backslash 1}$ does not know about 
about the random example \mbox{\boldmath{$\xi$}}$^{\\1}$, its overlap with this example, $z = \frac{1}{\sqrt N} \w^{\backslash 1} \cdot$\mbox{\boldmath{$\xi$}}$^1$ is a zero mean unit variance random gaussian variable.   Now suppose example $1$ is then included in the unsupervised learning problem.  Then upon re-minimization of the total energy $E(\w)$ in the presence of \mbox{\boldmath{$\xi$}}$^{1}$, the weight vector $\w^{\backslash 1}$ will change to a new weight vector, and consequently its alignment with \mbox{\boldmath{$\xi$}}$^{1}$ will also change from $z$ to an optimal alignment $\lambda^*$.   It can be shown \cite{griniasty1993cavity} that for large $N$ and $P$, this new optimal alignment arises through the minimization in \eqref{eq:lamstar}.   This minimization reflects a competition between two effects;  the second term in \eqref{eq:lamstar} favors optimizing the alignment with respect to the new example, but the first term tries to prevent changes from the original alignment $z$.  This term arises because $\w$ was already optimal with respect to all the other examples, so any changes in $\w$ incur an energy penalty with respect to the old examples.  The 
\begin{figure}[htbp]
   \centering
   \includegraphics[width=5.5in]{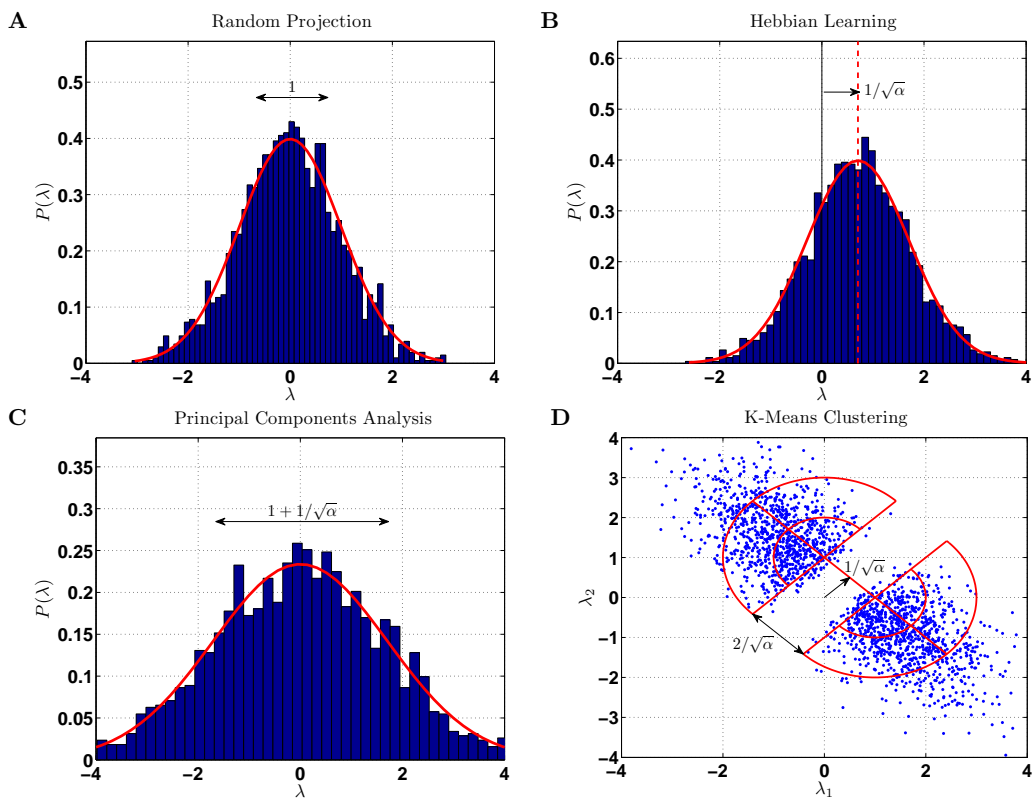} % requires the graphicx package
   \caption{Illusions of structure.  $P = 2000$ random points in $N=1000$ dimensional space (so $\alpha=2$) are drawn from a structureless zero mean, identity covariance Gaussian distribution.  These points are projected onto different directions.  (A) A histogram of the projection of these points onto a random direction; in the large $N,P$ limit this histogram is Gaussian with $0$ mean and unit variance. (B) A histogram of the same point cloud projected onto the Hebbian weight vector. (C) A projection onto the principal component vector. (D) The same point cloud projected onto two clusters directions found by $K$-means clustering with $K=2$. }
   \label{FigUnsup}
\end{figure}
parameter $\Delta$ plays the role of an inverse stiffness constant which determines the scale of a possible realignment of a weight vector with respect to a new example, and a 
self-consistency condition for $\Delta$ can be derived within the cavity approximation and is identical to the extremization of \eqref{eq:fbetaunsup}.   This extremization makes $\Delta$ implicitly a function of $\alpha$, and it is usually a decreasing function of $\alpha$.  Thus unsupervised learning becomes stiffer as $\alpha$, or the number of examples, increases and the weight vector responds less to the presentation of any new example.   Finally, example $1$ is not special in any way.  Thus repeating this analysis for each example, and averaging over the gaussian distribution of alignments $z$ before learning any example, yields the distribution of alignments across examples after learning in \eqref{eq:plam}. 
\par
We can now apply these results to an analysis of illusions of structure in high dimensional data.  Consider an unstructured dataset, i.e. a random gaussian point cloud consisting of $P$ points, \mbox{\boldmath{$\xi$}}$^{1},\dots,$\mbox{\boldmath{$\xi$}}$^{P}$ in $N$ dimensional space, where each point \mbox{\boldmath{$\xi$}}$^{\mu}$ is drawn i.i.d from a zero mean multivariate gaussian distribution whose covariance matrix is the identity matrix.   Thus if we project this data onto a random direction $\w$, the distribution of this projection $\lambda^{\mu}  = \frac{1}{\sqrt N} \w \cdot$\mbox{\boldmath{$\xi$}}$^\mu$ across examples $\mu$ will be a zero mean unit variance gaussian (see Fig. \ref{FigUnsup}A).  However, suppose we performed Hebbian learning to find the center of mass of the data.  This corresponds to the choice 
$V(\lambda) = -\lambda$, and leads to $\lambda^*(z,\Delta) = z + \Delta$ from \eqref{eq:lamstar} with $\Delta = \frac{1}{\sqrt{\alpha}}$ from \eqref{eq:fbetaunsup}.  Thus Hebbian learning yields an additive shift in the alignment to a new example whose magnitude decreases with the number of previous examples as $\frac{1}{\sqrt{\alpha}}$.  After learning, we find that the distribution of alignments in \eqref{eq:plam} is a unit variance gaussian with a {\it nonzero} mean given by $\frac{1}{\sqrt{\alpha}}$ (see Fig. \ref{FigUnsup}B).  Thus a high dimensional random gaussian point cloud typically has a nonzero center of mass when projected onto the optimal Hebbian weight vector.

\par
Similarly,  we could perform principal components analysis to find the direction of maximal variance in the data.  This corresponds to the choice $V(\lambda) = -\lambda^2$ and leads through \eqref{eq:lamstar} and \eqref{eq:fbetaunsup} to $\lambda^*(z,\Delta(\alpha)) = (1+\frac{1}{\sqrt{\alpha}})z$.  Thus PCA scales up the alignment to a new example, and \eqref{eq:plam} leads to a gaussian distribution of alignments along the principal component with zero mean, but a standard deviation equal to $1+\frac{1}{\sqrt{\alpha}}$ (see Fig \ref{FigUnsup}C).   This extra width is larger than any unity eigenvalue of the covariance matrix and leads to an illusion that the high dimensional gaussian point cloud has a large width along the principal component direction.

\par Finally, K-means clustering for $K=2$, defined by the energy function in \eqref{eq:kmeans1}, involves a projection of the data onto two dimensions, determined by the two cluster centroids.  However, the form of this energy function in \eqref{eq:kmeans2} reveals a lack of interaction between the projected coordinates $\lambda_+ = \lambda_1 + \lambda_2$ and $\lambda_- = \lambda_1 + \lambda_2$.  Along the direction $\lambda_+$, the algorithm behaves like Hebbian learning, so we should expected a gaussian distribution of the data along $\lambda_1 + \lambda_2$ with a mean of $\frac{1}{\sqrt{\alpha}}$.  However, along $\lambda_1 - \lambda_2$ the algorithm is maximizing the absolute value of the projection, so that $V(\lambda) = - | \lambda |$.  With this choice, \eqref{eq:lamstar} yields $\lambda^*(z,\Delta) = z + {\rm sgn}{(z)} \Delta$ with $\Delta = \frac{1}{\sqrt{\alpha}}$ determined by \eqref{eq:fbetaunsup}.   Note that this implies that the distribution of alignments in \eqref{eq:plam} has a gap of zero density in the region $-\frac{1}{\sqrt{\alpha}} \le \lambda \le \frac{1}{\sqrt{\alpha}}$ and outside this region, the distribution is a split gaussian.   The joint distribution of high dimensional data in K-means clustering factorizes along $\lambda_1 + \lambda_2$ and $\lambda_1-\lambda_2$ and does indeed have a gap of width $\frac{2}{\sqrt{\alpha}}$ along the $\lambda_1-\lambda_2$ direction (see Fig. \ref{FigUnsup}D).   So quite remarkably, K-means clustering (with $K=2$) of a random high dimensional gaussian point cloud reveals the illusion that there are $2$ well separated clusters in the cloud.   There is not a perfect match between the replica symmetric theory and numerical experiments because the discontinuity in the derivative of the energy in \eqref{eq:kmeans2} actually leads to replica symmetry breaking \cite{lootens2007analysing}. However the corrections to the replica symmetric result are relatively small, and replica symmetry is a good approximation in this case; in contrast it is exact for Hebbian learning and PCA (see e.g. Fig \ref{FigUnsup}B,C). 
\par
In summary, Fig. \ref{FigUnsup} reveals different types of illusions in high dimensional data whose effects diminish rather slowly as $O(\frac{1}{\sqrt{\alpha}})$ as the amount of data $\alpha$ increases.   Indeed it should be noted that the very ability of the perceptron to store random patterns also depends on a certain illusion of structure: $P$ random points in an $N$ dimensional space will typically lie on one side of some hyperplane as long as $\alpha = P/N \leq 2$.  
  
\subsection{From Message Passing to Synaptic Learning}
\label{sec:mpasslearn}

We have seen in section \ref{sec:perceptlearn}, that a perceptron with $N$ synapses has the capacity to learn $P$ random associations as long as $\alpha = P/N < \alpha_c = 2$.   But what learning algorithm can allow a perceptron to learn these associations, up to the critical capacity?  In the case of analog valued synaptic weights that we have been discussing, a simple algorithm, known as the perceptron learning algorithm \cite{rosenblatt1958perceptron, block1962perceptron} can be proven to learn any set of associations that can be implemented (i.e. those associations $\{ $\mbox{\boldmath{$\xi$}}$^\mu, \sigma^\mu \}$  for which a solution weight vector to \eqref{eq:ineq} exists).   The perceptron learning algorithm iteratively updates a set of randomly initialized weights as follows (for simplicity, we assume, without loss of generality, that  $\sigma^\mu = 1$, for all patterns).

\begin{itemize}
\item When presented pattern $\mu$, compute the current input $I = \w \cdot $\mbox{\boldmath{$\xi$}}$^\mu$. 
\item Rule 1: If $I \geq 0$, do nothing.
\item Rule 2: If $I < 0$, update all weights: $\w_i \rightarrow \w_i + $\mbox{\boldmath{$\xi$}}$^\mu_i$.
\item Iterate to the next pattern, until all patterns are learned correctly.  
\end{itemize}

\par
Such an algorithm will find realizable solutions to \eqref{eq:ineq} in finite time for analog synaptic weights.  However, what if synaptic weights cannot take arbitrary analog values?  Indeed evidence suggests that biological synapses behave like noisy binary switches \cite{petersen1998all, oconnor2005graded}, and thus can reliably code only two levels of synaptic weights, rather than a continuum.   The general problem of learning in networks with binary weights (or weights with a finite discrete set of values) is much more difficult than the analog case; it is in fact an NP-complete problem \cite{blum1992training, amaldi1991complexity}.   An exact enumeration and theoretical studies have revealed that when weights are binary (say $\w_i = \pm 1$), the perceptron capacity is
is reduced to $\alpha_c = 0.83$, i.e. the space of binary weight vector solutions to \eqref{eq:ineq} is nonempty only when $\alpha = P/N < \alpha_c = 0.83$ \cite{krauth1989critical, krauth1989storage}. Of course, below capacity, one can always find a solution through a brute-force search, but such a search will require a time that is exponential in $N$.   It is unlikely to expect to find a learning algorithm that provably finds solutions in a time that is polynomial in $N$, as this would imply $P = NP$.  However, is it possible to find a learning algorithm that can typically (but not provably) find solutions in polynomial time at large $\alpha < 0.83$, and moreover, can this algorithm be biologically plausible?  
\par
The work of \cite{braunstein2006learning, baldassi2007efficient} provided the first such algorithm.  Their approach was to consider message passing on the joint probability distribution over all synaptic weights consistent with the desired associations (again we assume $\sigma^\mu=1$):
\begin{equation}
P(\w) = \frac{1}{Z} \prod_{\mu=1}^P \theta ( \w \cdot $\mbox{\boldmath{$\xi$}}$^\mu ).
\end{equation}
Here the factors are indexed by examples. The messages from examples to synapses and synapses to examples are all distributions on binary variables, and therefore can be parameterized by real numbers, as $u_{\mu \rightarrow i} = M_{\mu \rightarrow i}(+1) -  M_{\mu \rightarrow i}(-1)$, and $M_{i \rightarrow \mu}(w_i) \propto e^{h_{i \rightarrow \mu} w_i}$.  The message passing equations \eqref{eq:mpass1}-\eqref{eq:mpass2} then yield a dynamical system on the variables $u_{\mu \rightarrow i}$ and $h_{i \rightarrow \mu}$.  This system drives the messages to approximate the marginal distribution of a synapse across all synaptic weight configurations which correctly learn all $P$ associations.  However, we would like to find a single synaptic weight configuration, not a distribution.  To do this, in \cite{baldassi2007efficient} the message passing equations are supplemented by a positive feedback term on the updates for $h_{i \rightarrow \mu}$.   This positive feedback amplifies the messages over time and forces the weights to polarize to a single configuration, so that the approximation to the marginals through \eqref{eq:marg1} becomes a delta function on $\pm 1$ for all synapses $i$.  Furthermore, in the large $N$ limit, one can approximate the dynamical system on the $2PN$ variables $u_{\mu \rightarrow i}$ and $h_{i \rightarrow \mu}$ via an approximate message passing dynamical system on the time dependent variable $h_i^t = \sum_{t' < t} \sum_{\mu = 1}^P u^{t'}_{\mu \rightarrow i}$ \cite{braunstein2006learning, baldassi2007efficient}.  Thus one obtains a learning rule in which each synapse maintains a single analog hidden variable $h_i$.   
\par 
This rule was further simplified by allowing $h_i$ to only take a finite number of discrete values, and the actual value of the synaptic weight was related to the hidden variable via $w_i = \mathrm{sgn}(h_i)$ \cite{baldassi2007efficient}.    After this simplification, the (amplified) message passing equations can be written in an online form in terms of the following algorithm (for convenience, $h_i$ is allowed to take only odd integer values to avoid the ambiguous state $h_i = 0$):
\begin{itemize}
\item For pattern $\mu$, compute the current input $I = \w \cdot $\mbox{\boldmath{$\xi$}}$^\mu$, where $w_i = \mathrm{sgn}(h_i)$. 
\item Rule 1: If $I \geq 1$, do nothing.
\item Rule 2: If $I < 0$, update all internal states : $h_i \rightarrow h_i + 2$\mbox{\boldmath{$\xi$}}$_i^\mu$.
\item Rule 3: If $I = 1$, then update each internal state $h_i \rightarrow h_i + 2$\mbox{\boldmath{$\xi$}}$_i^\mu$, but only if $h_i$\mbox{\boldmath{$\xi$}}$_i^\mu  \geq 1$. 
\item Iterate to the next pattern, until all patterns are learned correctly.  
\end{itemize}

The resulting rule is quite similar to the perceptron learning rule above, except for the modification of rule $1$ and the addition of rule $3$.  Rule $3$ concerns situations in which pattern $\mu$ is barely correct, i.e. a change in a single synaptic weight, or a single pattern component would cause $I$ to be below threshold (which is $0$), resulting in an error.  For barely learned patterns, rule $3$ reinforces those internal variables that are already pointing in the right direction, i.e. contributing positively to the input current $I$ on pattern $\mu$, by making them larger in absolute value.   Note that rule $3$ cannot change any synaptic weight $w_i$;  it is thus a {\it metaplastic} rule \cite{montgomery2004discrete}, or a rule that changes an internal state of the synapse without changing its synaptic efficacy.     

\par Remarkably, the addition of rule $3$, while seeming to be an innocuous modification of the perceptron learning rule, turns out to have a large impact on the learning capabilities of the discrete perceptron.   For example, for a neuron with $N = O(10^5)$ synapses, when $\alpha \in [0.3 \dots 0.6]$, the message passing derived algorithm finds a solution with a few tens of presentations per pattern, whereas a similar clipped perceptron algorithm obtained by removing rule $3$ is unable to find such a solution in $O(10^4)$ presentations per pattern \cite{baldassi2007efficient}. Given the remarkable performance of message passing, it is intriguing to speculate whether some signature of message passing may exist within synapses.   The key prediction is that in neurons that learn via error signals, metaplastic changes should occur whenever an error signal is absent, but the neuron is close to threshold.

\section{Random Matrix Theory}
\label{sec:rmt}

The eigenvalue distributions of large random matrices play a central role in a variety of fields \cite{mehta2004random, akemann2011oxford}.  For example, within the context of neuroscience, these distributions determine the stability of linear neural networks, the transition to chaos in nonlinear networks \cite{sompolinsky1988chaos}, and they are relevant to the statistical analysis of high dimensional data.   Replica theory provides a powerful method to compute the eigenvalue spectrum of many different classical random matrix ensembles, including random symmetric \cite{dhesi1999asymptotic} and asymmetric \cite{sommers1988spectrum} matrices.   More recently, it has been applied to matrices whose connectivity obeys Dale's law, which stipulates that all the outgoing synaptic weights of any neuron have the same sign \cite{rajan2006eigenvalue}.  Here we will introduce the replica formalism for symmetric matrices, focusing on the Wishart matrix ensemble \cite{sengupta1999distributions, hoyle2004principal} because of its applications to high dimensional statistics  discussed in section \ref{sec:corrextreme}.     

\subsection{Replica Formalism for  Random Matrices}

Suppose $\W$ is an $N$ by $N$ random matrix whose elements are drawn from some probability distribution.   For any specific realization of $\W$, its eigenvalue distribution is
\begin{equation}
\rho_{\mathbf W}(z) = \frac{1}{N} \sum_{i=1}^N \delta(z - z_i),
\label{eq:eigdist}
\end{equation}
where $z_i$ are the eigenvalues of $\W$.  Now for large $N$, and for many distributions on the matrix elements of $\W$, this eigenvalue distribution is self-averaging;  for any realization of $\W$, it converges as $N \rightarrow \infty$, to its average over $\W$, which we denote by $\qav{\rho_{\mathbf W}(\lambda)}_{\mathbf W}$.   We would like to theoretically compute this average, but it is difficult to average (\ref{eq:eigdist}) directly, since the eigenvalues $z_i$ are complicated functions of the matrix elements of $\W$ (i.e. the roots of the characteristic polynomial $\det (z - \W)$). 
\par To perform this average, it is useful to physically think of the eigenvalues $z_i$ as a collection of coulomb charges in the complex plane.  In two dimensions, such charges repel each other with a force that decays inversely with distance.   Then the resolvent,
\begin{equation}
R_{\mathbf W}(z) = \frac{1}{N}  \sum_{i=1}^N \frac{1}{z_i-z} =  \int dz'  \, \frac{\rho_{\mathbf W}(z')}{z'-z},
\label{eq:resovlentdef}
\end{equation} 
can be thought of as (the negative of) the electric force field on a test charge placed at a point $z$ in the complex plane, due to the presence of all the other charges $z_1,\dots,z_N$.  In mathematics, the transformation from $\rho_{\mathbf W}(z)$ to $R_{\mathbf W}(z)$ in \eqref{eq:resovlentdef} is known as the Stieltjes transform.  For the case of symmetric $\W$, the charge density is confined to the real axis, and one can recover the charge density from its force field via the relation
\begin{equation}
\rho_{\mathbf W}(z) = \lim_{\epsilon \rightarrow 0^+} \frac{1}{\pi} \, Im \, R_{\mathbf W}(z + i \epsilon).
\label{eq:chargetoforce}
\end{equation} 
See section \ref{sec:contour} for a derivation of this relation. Now the force on a test charge at a point $z$ is the derivative of the electrostatic potential $\Phi_{\mathbf W}(z)$, and it turns out this potential, as opposed to either the charge density $\rho_{\mathbf W}(z)$ or the electric force $R_{\mathbf W}(z)$,  will be easy to average over $\W$.  We can derive a simple expression for the potential via the following sequence:
\begin{eqnarray}
R_{\mathbf W}(z) & = \frac{1}{N} \Tr \frac{1}{{\mathbf W}-z} \nonumber \\
                               & = - \frac{\partial}{\partial z} \, \frac{1}{N} \Tr \log \big( {\mathbf W}-z\big) \nonumber \\
                               & =  \frac{\partial}{\partial z} \, \frac{2}{N} \log \bigg[ \det \big( {\mathbf W}-z  \big) \bigg]^{-\frac{1}{2}} \nonumber \\
                               & =  \frac{\partial}{\partial z}  \, \Phi_{\mathbf W}(z),  \label{eq:restopot}                     
\end{eqnarray}
where, 
\begin{equation}
\Phi_{\mathbf W}(z) =  \frac{2}{N} \log Z_{\mathbf W}(z)
\label{eq:pottopart}
\end{equation} 
and 
\begin{equation}
Z_{\mathbf W}(z) =  \int d\u  \, e^{-\frac{i}{2} {\mathbf u} ^T ( {\mathbf W} - z ) {\mathbf u}}.
\label{eq:gintdet}
\end{equation} 
Here we have used a Gaussian integral representation of $[ \det ( z - {\mathbf W} ) ]^{-\frac{1}{2}}$ in (\ref{eq:gintdet}) and neglected factors which do not survive differentiation by $z$ in (\ref{eq:restopot}).  
\par
Now the electrostatic potential $\Phi_{\mathbf W}(z)$ is expressed in (\ref{eq:pottopart}) as the free energy of a partition function $Z_{\mathbf W}(z)$ given by (\ref{eq:gintdet}).  We can use this representation to average the potential over $\W$,  via the replica method to appropriately take care of the logarithm:
\begin{eqnarray}
\qav{\Phi_{\mathbf W}(z)}_{\mathbf W}  &=   \frac{2}{N} \, \qav{\log Z_{\mathbf W}(z)}_{\mathbf W} ,\nonumber \\ 
                                                                      &=   \frac{2}{N} \, \lim_{n \rightarrow 0}  \, \frac{\partial}{\partial n} \, \qav{Z^n_{\mathbf W}(z)}_{\mathbf W}.
\label{eq:avpot}                                                                      
\end{eqnarray}
This yields a general procedure for computing the average eigenvalue spectrum (i.e. charge density) of random Hermitian matrices.   We first average a replicated version of the partition function in (\ref{eq:gintdet}) (see \eqref{eq:replrmt} below).   This allows us to recover the average electrostatic potential through (\ref{eq:avpot}), which then leads to the the average electric field through (\ref{eq:restopot}), which in turn leads to the average charge density through  (\ref{eq:chargetoforce}).  We note that although we have focused on the case of hermitian matrices, this analogy between eigenvalues and coulomb charges extends to non-hermitian matrices in which the  eigenvalue density is not confined to the real axis.

\subsection{The Wishart Ensemble and the Marcenko-Pastur Distribution}
\label{sec:wishartmp}
\par As seen in the previous section, the first step in computing the eigenvalue spectrum of a Hermitian random matrix involves computing the average
\begin{equation}
\bqav{Z^n_{\mathbf W}(z)}_{\mathbf W} =  \bqav{\int \prod_{a=1}^n d\u^a  \, e^{-\frac{i}{2} \sum_a {\mathbf u^a} ^T ( {\mathbf W} - z ) {\mathbf u^a}}}_{\mathbf W}.
\label{eq:replrmt}
\end{equation}
At this point we must choose a probability distribution over $\W$.   When the matrix elements of $\W$ are chosen i.i.d. from a gaussian distribution, one obtains Wigner's semicircular law \cite{wigner1958distribution} for the eigenvalue distribution, which was derived via the replica method in \cite{dhesi1990asymptotic}.   
\par Here we will focus on the Wishart random matrix ensemble in which
\begin{equation}
\W = \frac{1}{P} \, \A^T \A,
\label{eq:wishartdef}
\end{equation}
where $\A$ is a $P$ by $N$ matrix whose elements are chosen i.i.d. from a zero mean, unit variance Gaussian.  This matrix has a simple interpretation in terms of high dimensional data analysis.   We can think of each row of the matrix $\A$ as a data vector in an $N$ dimensional feature space.   Each data vector, or row of $\A$ is then a single draw from a multivariate Gaussian distribution in $N$ dimensions, whose covariance matrix is the identity matrix.   $\W$ is then the empirical covariance matrix of the $P$ samples in the data set $\A$.   In the low dimensional limit where the amount of data $P \rightarrow \infty$ and $N$ remains $O(1)$, the empirical covariance matrix $\W$ will converge to the identity, and its spectrum will be a delta-function at $1$.   However, in the high dimensional limit in which $P,N \rightarrow \infty$ and $\alpha = P/N = O(1)$, then even though on average $\W$ will be the identity matrix, fluctuations in its elements are strong enough that its eigenvalue spectrum for typical realizations of the data $\A$ will not converge to that of the identity matrix.   Even when $\alpha > 1$, the case of interest here, there will be some spread in the density around $1$, and this spread can be thought of as another illusion of structure in high dimensional data, which we now compute via the replica method. 
\par   
Inserting \eqref{eq:wishartdef} into \eqref{eq:replrmt} we obtain
\begin{equation}
\bqav{Z^n_{\mathbf W}(z)}_{\mathbf W} =  \int \prod_{a=1}^n d\u^a  \, \bigg[ \qav{e^{-\frac{i}{2} \sum_a {\frac{1}{P} \mathbf u^a} ^T ( {\mathbf A}^T{\mathbf A}) {\mathbf u^a}}}_{\mathbf A}e^{\frac{iz}{2} \sum_a {\mathbf u^a} ^T {\mathbf u^a}} \bigg].
\label{eq:replrmtwis}
\end{equation}
Now the integrand depends on the quenched disorder $\A$ only through the variables $\lambda^a_\mu = \frac{1}{\sqrt{N}} \a_\mu \cdot \u^a$, where $\a_\mu$ is row $\mu$ of the matrix $\A$.  These variables are jointly gaussian distributed with zero mean and covariance $\qav{\lambda^a_\mu \lambda^b_\nu} = Q_{ab} \delta_{\mu \nu}$ where $Q_{ab} = \frac{1}{N} \u^a \cdot \u^b$.   Thus the average over $\A$ can be done by a gaussian integral over the variables $\lambda^a_\mu$:
\begin{eqnarray}
 \qav{e^{-\frac{i}{2} \sum_a {\frac{1}{P} \mathbf u^a} ^T ( {\mathbf A}^T{\mathbf A}) {\mathbf u^a}}}_{\mathbf A} 
& = \qav{e^{-\frac{1}{2} \frac{i}{\alpha}  \sum_a \sum_\mu (\lambda_a^\mu)^2}}_{\{\lambda^a_\mu\}} \label{t1}\\
& = \bigg[ \qav{e^{-\frac{1}{2} \frac{i}{\alpha}  \sum_a (\lambda_a)^2}}_{\{\lambda^a\}} \bigg]^P \label{t2}\\
& = \bigg[ \det \big(I + \frac{i}{\alpha} Q\big)^{-\frac{1}{2}} \bigg]^P \label{t3} \\
& = e^{-N\frac{\alpha}{2} \Tr \log (I + \frac{i}{\alpha} Q)} \label{t3}
\end{eqnarray}
Here in going from \eqref{t1} to \eqref{t2}, we have exploited the fact that the variables $\lambda^a_\mu$ are uncorrelated for different $\mu$, yielding a single average over variables $\lambda_a$ with covariance $\qav{\lambda_a  \lambda_b} = Q_{ab}$, raised to the power $P$.  In going from \eqref{t2} to \eqref{t3} we performed the gaussian integral over $\lambda_a$.   
\par Thus consistent with the general framework in section \ref{sec:repframe}, averaging over the disorder introduces interactions between the replicated degrees of freedom $\u^a$ which depend only on the overlap matrix $Q_{ab}$.  Therefore we can compute the remaining integral over $\u^a$ in \eqref{eq:replrmtwis} by integrating over all overlaps $Q_{ab}$, and integrating over all configurations of $\u^a$ with a given overlap $Q$.  This latter integral yields an entropic factor that depends on the overlap.  In the end \eqref{eq:replrmtwis} becomes
\begin{equation}
\bqav{Z^n_{\mathbf W}(z)}_{\mathbf W} =  {\int} \,  \prod_{ab}  \,  d\Qab \, e^{-N(E(Q) - S(Q))},
\label{eq:replmatzn}
\end{equation}
where 
\begin{equation}
E(Q) = \frac{\alpha}{2} \Tr \log (I + \frac{i}{\alpha} Q) - \frac{i z}{2} \Tr Q,
\label{eq:replmaten}
\end{equation}
and
\begin{equation}
S(Q) = \frac{1}{2} \Tr \log Q, 
\end{equation}
is the usual entropic factor.  The first term in \eqref{eq:replmaten} comes from \eqref{t3} while the second term in \eqref{eq:replmaten} comes from the part outside the average over $\A$ in \eqref{eq:replrmtwis}.    
\par
Now the final integral over $Q_{ab}$ can be done via the saddle point method, and the integral can be approximated by the value of the integrand at the saddle point matrix $Q$ which extremizes $F(Q) = E(Q) - S(Q)$.   We can make a decoupled replica symmetric ansatz for this saddle point, $Q_{ab} = q \delta_{ab}$.   With this choice, \eqref{eq:avpot} leads to the electrostatic potential
\begin{equation}
\qav{\Phi_{\mathbf W}(z)}_{\mathbf W} = - \alpha \log (1 + \frac{i}{\alpha} q) + i z q + \log q
\label{eq:avpotavw}
\end{equation}
and \eqref{eq:restopot} leads to the electric field
\begin{equation}
\qav{R_{\mathbf W}(z)}_{\mathbf W} = iq.
\end{equation}
Here $q$ satisfies the saddle point equation obtained by extremizing $F(q)$, or equivalently the right hand side of  \eqref{eq:avpotavw},
\begin{equation}
- \frac{\alpha}{\alpha + iq} + z + \frac{1}{i q} = 0.
\end{equation}
This is a $z$ dependent quadratic equation for $iq$, and due to the relation between the electric field and charge density in \eqref{eq:chargetoforce}, we are interested in those real values of $z$ for which the solution $iq$ has a nonzero imaginary part.  It is in these regions of $z$ that charges (eigenvalues) will accumulate, and their density will be proportional to this imaginary part.  In the regime in which $\alpha > 1$ (so we have more data points than dimensions), a little algebra shows that $iq$ has an imaginary part only when $z_- < z < z_+$ where $z_\pm = (1 \pm \frac{1}{\sqrt{\alpha}})^2$.  In this region the charge density is 
\begin{equation}
\qav{\rho_{\mathbf W}(z)}_{\mathbf W} = \frac{\alpha \sqrt{(z-z_-)(z_+-z)}}{2\pi z},
\label{eq:rangemp}
\end{equation}  
which is the Marcenko-Pastur (MP) distribution (see Fig. \ref{mptw}A below).   Thus due to the high dimensionality of the data, the eigenvalues of the sample covariance matrix spread out around $1$ over a range of $O(\pm \frac{1}{\alpha})$.  This illusory spread becomes smaller as we obtain more data (increased $\alpha$).

\subsection{Coulomb Gas Formalism}
\label{sec:cg}
In the previous section we found the marginal density of eigenvalues for a Wishart random matrix, but what about the entire joint distribution of all $N$ eigenvalues?  This distribution has a physically appealing interpretation that provides intuition for applications in high dimensional statistics discussed below.  Consider  the distribution of  $\mathbf{W} = \A^T \A$, i.e. the matrix in \eqref{eq:wishartdef} without the $\frac{1}{P}$ scaling.  Because the $P$ by $N$ matrix $\A$ has i.i.d zero mean unit variance Gaussian elements, the distribution of $\A$ is given by
\begin{equation}
P(\A) \propto  e^{-\frac{1}{2} \Tr {\mathbf A}^T {\mathbf A}},
\label{eq:gaussA}
\end{equation}
Now each matrix $\A$ has a unique singular value decomposition (SVD), $\A = {\mathbf {U \Sigma V}^T}$, where \bo{U} and \bo{V} are unitary matrices and ${\mathbf \Sigma}$ is a $P$ by $N$ matrix whose only $N$ nonzero elements are on the diagonal: ${\mathbf \Sigma}_{ii} = \sigma_i$.  The $\sigma_i$'s are the singular values of $\A$, and the eigenvalues $\lambda_i$ of $\W$ are simply the square of these singular values.  Thus to obtain the joint distribution for $\lambda_i$, we first perform the change of variables $\A = {\mathbf {U \Sigma V}^T}$ in the measure \eqref{eq:gaussA}.
\par Fortunately, $P(\A)$ is independent of \bo{U} and \bo{V}, and depends only on ${\mathbf \Sigma}$.   However, we need to transform the full measure $P(\A) \, d\A$, and therefore we must include the Jacobian of the change of variables, given by (see e.g. \cite{edelman2005random})
\begin{equation}
d\A = \prod_{i<j}{\left(\sigma_i^2-\sigma_j^2\right)}\prod_{i=1}^N{\sigma_i^{P-N}}({\mathbf U}^T d{\mathbf U})(d {\mathbf \Sigma})({\mathbf V}^T d{\mathbf V}).
\label{eq:measurech}
\end{equation}
Now the angular variables \bo{U} and \bo{V} decouple from the singular values $\sigma_i$, so we can integrate them out, yielding a constant.  Furthermore, we can perform the change of variables $\lambda_i = \sigma_i^2$ to obtain
\begin{equation}
P(\lambda_1,\dots,\lambda_N) \propto e^{-\frac{1}{2} \sum_{i=1}^N{\lambda_i}}\prod_{i=1}^N{\lambda_i^{\frac{1}{2}(P-N-1)}}\prod_{j<k}{|\lambda_j-\lambda_k|}.
\label{eq:jointD}
\end{equation}
Here the first factor in the product arises from $P(\A)$ in \eqref{eq:gaussA} while the second two factors arise from the Jacobian incurred by the change of measure in \eqref{eq:measurech}.
This joint distribution can be written as a Gibbs distribution at unit temperature, $P(\{\lambda_i\}) \propto e^{- E(\{\lambda_i\})}$, where the energy is 
\begin{equation}
E = \frac{1}{2} \sum_{i=1}^N \left( \lambda_i - (P-N-1) \ln \lambda_i \right) -\sum_{j\ne k}{\ln|\lambda_j-\lambda_k|}.
\label{eq:mpenergy}
\end{equation}
This energy function has a simple interpretation in which each eigenvalue is a Coulomb charge on confined to the real axis on the 2D complex plane.  Each charge moves in a linear plus logarithmic potential which confines the charges, and there is a pairwise repulsion between all charges governed by a logarithmic potential (the Coulomb interaction in two dimensions).   The Coulomb repulsion balances the confinement due to the external potential when the charges, or eigenvalues spread out
over a typical range of $O(N)$.   More precisely, this range is given by $(1 \pm \sqrt{\alpha})^2 N$ where $\alpha = P/N$ (note this is consistent with $z_{\pm}$ defined above \eqref{eq:rangemp} after rescaling by $\frac{1}{P}$), and within this range, the charge density in the $N \rightarrow \infty$ limit is given by the MP distribution.

\subsection{Tracy-Widom Fluctuations}
\label{sec:tw}

In the previous sections we have seen that full joint distribution of eigenvalues behaves like a Coulomb gas and its typical density at large $N$ is given by the MP distribution. However, what do typical as well as large fluctuations of the maximal eigenvalue, or right most charge behave like?  The distribution of the maximal eigenvalue forms a null distribution to test for the statistical significance of outcomes in PCA, and also plays role in random dimensionality reduction, so its fluctuations are of great interest.  The mean of the maximal eigenvalue $\lambda_{MAX}$ of course lies at the end of the MP charge density and is given, to leading order in $N$, by $\langle \lambda_{MAX} \rangle  = (1 + \sqrt{\alpha})^2 N$.  
Typical fluctuations about this mean have been found to scale as $O(N^{\frac{1}{3}})$ \cite{tracy1994level}.  More precisely, for large $N$ they have the limiting form
\begin{equation}
\lambda_{MAX}=(1 + \sqrt{\alpha})^2 N + \alpha^{-\frac{1}{6}} (1 + \sqrt{\alpha})^{\frac{4}{3}} N^{\frac{1}{3}} \chi,
\label{eq:twfinN}
\end{equation}
where $\chi$ is the Tracy-Widom distribution that has a range of $O(1)$ \cite{tracy1994level}.
\par
The computation of these typical fluctuations is involved, but often we are interested in the probability of large deviations in which $| \lambda_{MAX} - \langle \lambda_{MAX} \rangle | = O(N)$.  These large deviations were computed in \cite{vivo2007large, majumdar2009large} in a very simple way using the Coulomb gas picture. 
Suppose for example the largest eigenvalue occurs at distance that is $O(N)$ to the right of the typical edge of the MP density, $(1 + \sqrt{\alpha})^2 N$.  The most likely 
\begin{figure}[htbp]
   \centering
   \includegraphics[width=5.5in]{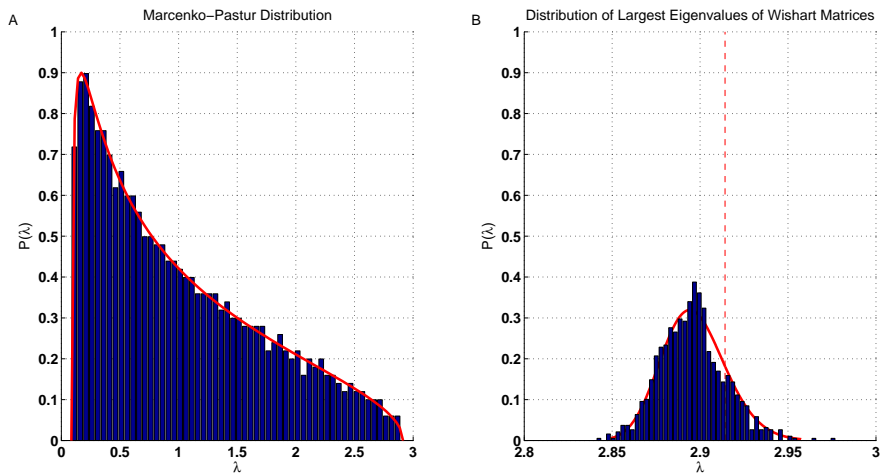} % requires the graphicx package
   \caption{Spectral distributions of empirical noise covariance matrices. (A) $P = 2000$ random points in $N=1000$ dimensional space ($\alpha=2$) are drawn from a $0$ mean identity covariance normal distribution. The blue histogram is the distribution of eigenvalues of the empirical covariance matrix $\W = \frac{1}{P} \, \A^T \A$, where $\A$ is a $P$ by $N$ data matrix whose rows correspond to the points (see Eq. \eqref{eq:wishartdef}).  The red curve is the Marcenko-Pastur distribution (see Eq. \eqref{eq:rangemp}) for $\alpha=2$.   (B) A histogram, in blue, of the maximal eigenvalues of $1000$ random covariances matrices $\W$, each constructed exactly as in (A). The red curve is the Tracy-Widom distribution in \eqref{eq:twfinN} for $\alpha=2$, rescaled by $\frac{1}{P}$. The dashed red line marks the edge of the Marcenko-Pastur distribution in (A). The discrepancy between this edge and the mean of the maximal eigenvalue distribution is a finite size effect; this discrepancy, like the fluctuations in the maximal eigenvalue, vanishes as $O(N^{-2/3})$. }
\label{mptw}
\end{figure}
way this could happen (i.e. the saddle point configuration of charges in \eqref{eq:jointD} \cite{majumdar2009large}), is that a {\it single} eigenvalue pops out of the MP density, while the remaining eigenvalues are unperturbed, and preserve the shape of the MP density.  The energy paid by a single eigenvalue popping out of the MP density is dominated by the linear confining term in \eqref{eq:mpenergy}, and is therefore proportional to the distance it pops out.  Since the probability of a fluctuation is exponentially suppressed by its energy cost (here entropy plays no role because the MP density is unperturbed), we obtain 
\begin{equation}
{\mathrm{Prob}}\left(\lambda_{MAX} = \langle \lambda_{MAX} \rangle + cN \right) \propto e^{-N \Phi_+(c)} \quad {\mathrm{for}}  \quad cN \gg \langle \lambda_{MAX}\rangle .
\end{equation}
Thus large deviations of $O(N)$ are exponentially suppressed by $N$, and the $O(1)$ large deviation constant $\Phi_+(c)$ can be computed explicitly by quantitatively working out this Coulomb gas argument \cite{vivo2007large, majumdar2009large}. 
\par On the other hand, suppose that the maximal eigenvalue $\lambda_{MAX}$ occurs at a distance $cN$ to the left of right edge of the $MP$ density.  In order for this fluctuation to occur, the entire MP density must become compressed, incurring a much larger energy cost compared to a positive or rightward fluctuation of $\lambda_{MAX}$.  Indeed because of the Coulomb repulsion between all pairs of charges in \eqref{eq:mpenergy}, the energy cost of compression is $O(N^2)$ leading to the stronger suppression,
\begin{equation}
{\mathrm{Prob}}\left(\lambda_{MAX} = \langle \lambda_{MAX} \rangle - cN \right) \propto e^{-N^2 \Phi_-(c)} \quad {\mathrm{for}}  \quad cN \gg \langle \lambda_{MAX}\rangle,
\end{equation}
where $\Phi_-(c)$ is an $O(1)$ large deviation function computed in \cite{vivo2007large, majumdar2009large}.  Thus the physics of Coulomb gases gives a nice explanation for the asymmetry in the large deviations of the Tracy-Widom distribution. 
\par
For the reader's convenience, we summarize the implications of the Coulomb gas formalism for high dimensional statistics by reintroducing the $\frac{1}{P}$ scaling in the definition \eqref{eq:wishartdef} of the empirical covariance matrix.  The above results then tell us that the maximal eigenvalue of the empirical covariance matrix of $P$ random Gaussian points in $N$ dimensional space, in the limit $N,P \rightarrow \infty$ with $\alpha = P/N$ remaining $O(1)$ (but strictly greater than $1$),  has a mean $\langle \lambda_{MAX} \rangle = (1 + \frac{1}{\sqrt{\alpha}})^2$ with typical fluctuations about this mean that are $O(N^{-\frac{2}{3}})$ (see Fig. \ref{mptw}B).  Moreover, the probability of large $O(1)$ positive deviations of $\lambda_{MAX}$ are $O(e^{-N})$, while the probability of large $O(1)$ negative deviations of $\lambda_{MAX}$ are $O(e^{-N^2})$.

\section{Random Dimensionality Reduction}
\label{rdimred}

We have seen in section \ref{secunsupill} that we may need to be careful when we perform dimensionality reduction of high dimensional data by looking for optimal directions along which to project the data, as this process can potentially lead to illusions of structure.  An alternate approach might be to skip the optimization step responsible for the illusion, and simply project our data onto randomly chosen directions.  However, it is not at all obvious that such random dimensionality reduction would preserve the true or interesting structure that is present in the data.  Remarkably, a collection of theoretical results reveal that random projections (RP's) preserve much more structure than one might expect. 
 
\subsection{Point Clouds}

\par
\begin{figure}[htbp]
   \centering
   \includegraphics[width=5.5in]{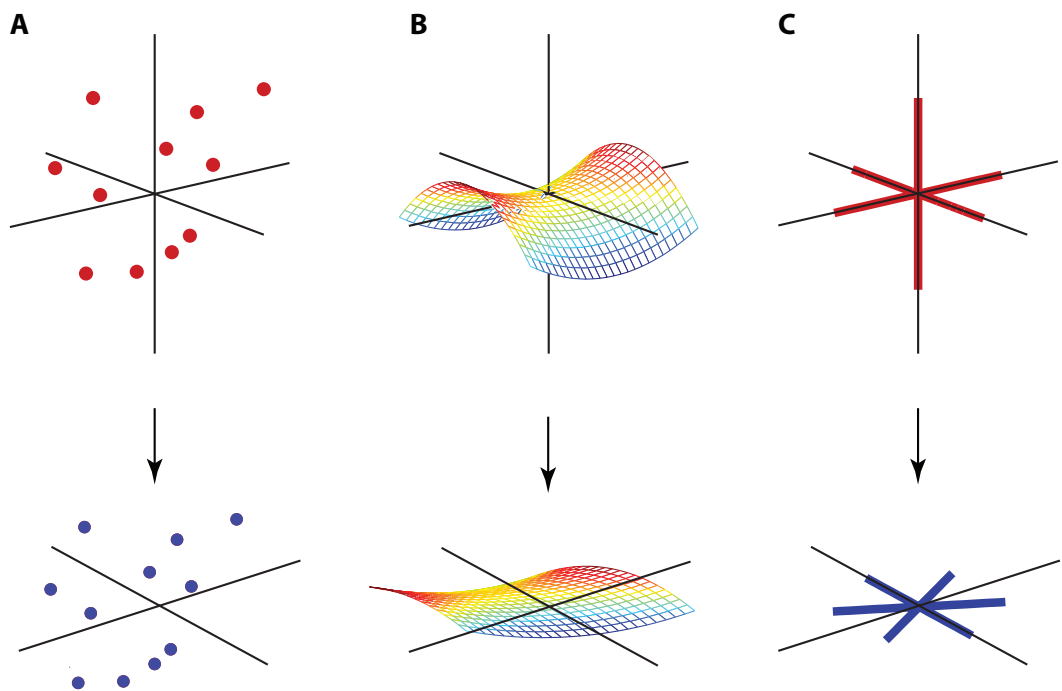} % requires the graphicx package
   \caption{Random Projections. (A,B) Projection of a point cloud, and a nonlinear manifold respectively. (C) A manifold of $K$-sparse signals (red) in $N$ dimensional space is randomly projected down to an $M$ dimensional space (here $K=1$, $N=3$, $M=2$). }
   \label{rproj}
\end{figure}

  A very generic situation is that data often lies along a low dimensional manifold embedded in a high dimensional space.   An extremely simple manifold is a point cloud consisting of a finite set of points, as in Fig. \ref{rproj}A.  Suppose this cloud consists of $P$ points  $\s^\alpha$, for $\alpha=1,\dots,P$,  embedded in an $N$ dimensional space, and we project them down to the points $\x^\alpha = \A \s^\alpha$ in a low $M$ dimensional space through an appropriately normalized random $M \times N$ random projection matrix $\A$.  The squared euclidean distances between pairs of points in the high dimensional space are given by $ ||\s^\alpha-\s^\beta||^2$ and in the low dimensional space by ${\| \x^\alpha-\x^\beta\|}^2$. The fractional distortion in the squared distance incurred by the projection is given by
\begin{equation}
D_{\alpha\beta} =  \frac{\|\x^\alpha - \x^\beta \|^2  -  \| \s^\alpha-\s^\beta \|^2}{\| \s^\alpha-\s^\beta \|^2}.
\label{eq:distortion}
\end{equation}  
How small can we make $M$ before the point cloud becomes distorted in the low dimensional space, so that the low and high dimensional distances are no longer similar?

\par The celebrated Johnson-Lindenstrauss (JL) lemma \cite{johnson1984extensions}  (see \cite{indyk1998approximate, dasgupta2003elementary} for more recent and simpler proofs) and provides a striking answer.  It states that for any distortion level $0 < \delta < 1$, as long as $M > O(\frac{\ln P}{\delta^2})$, with high probability, one can find a projection such that
\begin{equation}
-\delta \le D_{\alpha\beta} \le \delta
\end{equation}  
for all pairs of points $\alpha$ and $\beta$.  Thus the distortion between any pair of points rarely exceeds $\delta$.  This is striking because the number of projected dimensions $M$ need only be {\it logarithmic} in the number of points $P$  {\it independent} of the embedding dimension of the source data, $N$.  Of course, with so few projections, one cannot reconstruct the original data from its projections. Nevertheless, surprisingly, with so few random projections the geometry of the entire point cloud is preserved.  We will discuss a statistical mechanics based approach for understanding the JL lemma in section \ref{sec:corrextreme}.

 \subsection{Manifold Reduction}
 \label{sec:manifoldred}

\par Consider data distributed along a nonlinear $K$ dimensional manifold embedded in $N$ dimensional space, as in Fig \ref{rproj}B.  An example might  be a set of images of a single object observed under different lighting conditions, perspectives, rotations and scales. Another example would be the set of neural firing rate vectors in a brain region in response to a continuous family of stimuli.  \cite{baraniuk2009random,  baraniuk2010low, yap2011stable} show that $M > O(\frac{K}{\delta^2} \log{NC})$ random projections preserve the geometry of the manifold up to distortion $\delta$.  Here $C$ is a number related to the curvature of the manifold, so that highly curved manifolds require more projections.  Overall, the interesting result is that the required number of  projections depends linearly on the intrinsic dimensionality of the manifold, and only logarithmically on its ambient embedding dimension.

\par The simplest finite dimensional manifold is a $K$ dimensional linear subspace in an $N$ dimensional space.   It can be shown \cite{baraniuk2008simple} that  $M > O(\frac{K}{\delta^2})$ RP's are sufficient to preserve all pairwise distances between data points within a distortion level $\delta$.  We will give an alternate proof of this result in section \ref{sec:corrextreme} below using the results of sections \ref{sec:wishartmp} and \ref{sec:cg}.  Of course, for such a simple manifold, there exists a nonrandom, optimal geometry preserving projection, namely the PCA basis consisting of $M=K$ orthogonal vectors spanning the manifold.  Thus we pay a price in the number of projections for choosing random projections rather than the optimal ones.  Of course, for data that is not distributed along a hyperplane, a PCA based projection will no longer be optimal, and will not generically preserve geometry.  

\par Sparsity is another example of an interesting low dimensional structure.   Consider for example, the (nonsmooth) manifold of $N$ dimensional signals with only $K$ nonzero components.  This is a manifold of $N \choose K$ coordinate hyperplanes in $N$ dimensional space, as in Fig. \ref{rproj}C.    The geometry of this manifold can also be preserved by random projections.  In particular \cite{baraniuk2008simple} shows that random projections down to an $M > O(\frac{K}{\delta^2} \log \frac{N}{K})$ dimensional space preserve the distance between any pair of $K$-sparse signals, with distortion less than $\delta$. 

\par Beyond the issue of preserving geometry under an RP, one might be interested in situations in which one can invert the random projection; i.e. given the projection of a data vector, or signal in the low $M$ dimensional space, how might one recover the original signal in the high $N$ dimensional space?  For the case of point clouds  (Fig. \ref{rproj}A) and general nonlinear manifolds (Fig. \ref{rproj}B), there are no general computationally tractable algorithms capable of achieving this high dimensional signal recovery.   However, for the case of $K$-sparse signals (Fig. \ref{rproj}C), there exists a simple, computationally tractable algorithm, known as $L_1$ minimization, reviewed below, that can provably, under various assumptions on the RP matrix $\A$, recover the high dimensional signal from its projection.  It turns out that geometry preservation is a sufficient condition for signal recovery; in particular, \cite{candes2005decoding} shows that any projection which preserves the geometry of all $K$-sparse vectors, allows one to reconstruct these vectors from the low dimensional projection, efficiently and robustly using $L_1$ minimization.   This is one of the observations underlying the field of compressed sensing, reviewed below. 

\par However, even in situations where one cannot accurately reconstruct the high dimensional signal from its projection, RP's can still be very useful by allowing for compressed computation directly in the low dimensional projection space, without the need for signal recovery.   This can be done because many interesting machine learning and signal processing algorithms depend only on pairwise distances between data points.  For example,  regression \cite{zhou2009compressed}, signal detection \cite{duarte2006sparse}, classification \cite{davenport2007smashed, duarte2007multiscale, haupt2006compressive, blum2006random}, manifold learning \cite{hegde2007random}, and nearest neighbor finding \cite{indyk1998approximate} can all be accomplished in a low dimensional space given a relatively small number of RP's.  Moreover, task performance is often comparable to what can be obtained by performing the task directly in the original high dimensional space. The reason for this remarkable performance is that these computations rely only on the distances between data points, which are preserved by RP's. 

\subsection{Correlated Extreme Value Theory and Dimensionality Reduction}
\label{sec:corrextreme}

The proofs \cite{johnson1984extensions, indyk1998approximate, dasgupta2003elementary, baraniuk2008simple, baraniuk2009random} behind the remarkable sufficient conditions for low distortion of submanifolds under RP's rely on sequences of potentially loose inequalities.  Thus they leave open the question of whether these sufficient conditions are actually necessary, and how the typical, as opposed to worst case, distortion behaves.  Is it possible to use a direct approach, more in the spirit of statistical mechanics, to simply compute the probability distribution of the typical distortion of random manifolds under random projections?  Here the random choice of manifold plays the role of quenched disorder, the random choice of projection plays the role of thermal degrees of freedom, and the observable of interest is the distribution, across choices of RP's, of the maximal distortion over all pairs of points in the fixed manifold.  The hope is that this distribution is self-averaging, in that it does not depend on the choice of a particular manifold from a suitable ensemble of manifolds.  In general, this approach is challenging, but here we show that it can be carried out for two very simple classes of manifolds: random point clouds and hyperplanes.   Our main goal in this section is simply to obtain intuition for the scaling behavior of some of the inequalities discussed in the previous section. 
\par 
Consider a fixed realization of a Gaussian point cloud consisting of $P$ points $\s^\alpha$, $\alpha =1,\dots,P$ in $N$ dimensional space.  Let $\A$ be an $M$ by $N$ random projection operator whose matrix elements are i.i.d Gaussian with zero mean and variance $\frac{1}{M}$, and let $\x^\alpha = \A \s^\alpha$ be the low dimensional image of the cloud.  With this choice of scaling for the variance of the projection operator, it is straightforward to show that any one distortion $D_{\alpha \beta}$ in \eqref{eq:distortion} is in the large $N$ limit, for a fixed (i.e. quenched) point cloud, a Gaussian random variable with zero mean and variance $O(\frac{1}{M})$, due to the random choice of $\A$.   Now there are ${P \choose 2} = O(P^2)$ pairs of points, or possible distortions, and we are interested in the maximum distortion, whose behavior could in principle depend on the correlations between pairs of distortions.  For random Gaussian point clouds, the correlation coefficient between two pairs of distortions are weak, in fact $O(\frac{1}{N})$,  and can be neglected.  In this manner, the ambient dimensionality of the point cloud disappears from the problem.  Thus the maximum distortion over all pairs of points can be well approximated by the maximum of $O(P^2)$ independent Gaussian variables each with variance $O(\frac{1}{M})$.   In general the maximum of $R$ independent Gaussian variables with variance
$\sigma^2$ approaches a Gumbel distribution in the large $R$ limit, and takes typical values of $O(\sigma \sqrt{\ln R})$.  Indeed the Gumbel distribution is a universal distribution governing extreme values of any random variables whose tails vanish faster than exponentially \cite{coles2001introduction};  this strong suppression of extreme values in any single variable directly leads to an extremely slow $\sqrt{\ln R}$ growth of the maximum over $R$ realizations of such variables.  Applying this general result with 
$\sigma^2 = O(\frac{1}{M})$ and $R = O(P^2)$ yields the conclusion that the maximal distortion over all pairs of points obeys a Gumbel distribution, and its typical values scale with $P$ and $M$ as $O( \sqrt{\frac{\ln P}{M}})$.   Thus the origin of the extremely slow $\sqrt{\ln P}$ growth of the maximal distortion with the number of points $P$ is due to the strong, Gaussian suppression of any individual distortion.  This effect is directly responsible for the remarkable $JL$ lemma.  For example, if we desire our maximal distortion to be less than $\delta$, we must have
\begin{equation}
O \left( \sqrt{\frac{\ln P}{M}} \right) < \delta,
\end{equation}
or equivalently $M > O \left( \frac{\ln P}{\delta^2} \right )$, which, up to constants, is the JL result.  Thus extreme value theory for uncorrelated Gaussian variables provides a natural intuition for why the number of random projections $M$ need only be logarithmic in the number of points $P$, and independent of the ambient dimension $N$, in order to achieve an $O(1)$ distortion $\delta$.   
\par For random Gaussian point clouds, we were able to neglect correlations in the distortion between different pairs of points.  For more general manifold ensembles, we will no longer be able to do this.  However, for the ensemble of random hyperplanes, an exact analysis is still possible despite the presence of these correlations.
Let $\bo{U}$ be an $N$ by $K$ random matrix whose $K$ orthonormal columns form a basis for a random $K$ dimensional subspace of $N$ dimensional space (drawn uniformly from the space of such subspaces).     What is the distribution of the maximal distortion in \eqref{eq:distortion} where $\s^\alpha$ and $\s^\beta$ range over all pairs of points in this subspace?  First, by exploiting rotational invariance of the ensemble of $\A$ and $\bo{U}$, we can always perform a change of basis in $N$ dimensional space so that the columns of $\bo{U}$ are mapped to the first $K$ coordinate axes.  Thus points in the hyperplane can be parameterized by $N$ dimensional vectors whose only nonzero components are the first $K$ coordinates, and the statistics of their projection to $M$ dimensional space can be determined simply by the $M$ by $K$ submatrix of $\A$ consisting of its first $K$ columns.   In this manner, the dimensionality $N$ of the ambient subspace again disappears from the problem.  Second, by exploiting the linearity of the projection operator, to compute the maximal distortion over all pairs of points in the plane, it suffices to compute the maximal distortion over all points on the unit sphere in $K$ dimensional space.   Thus if we let $\bar{\A}$ denote the $M$ by $K$ submatrix of $\A$, and let $\s$ denote a $K$-dimensional coordinate vector for the hyperplane, then we have
\begin{equation}
{\mathrm{max}}_{\alpha \beta} \, D_{\alpha \beta} = {\mathrm{max}}_{\{ {\mathbf s} \, , \,  || {\mathbf s}||^2 =1 \}}  \, \sqrt{\s^T \bar{\A}^T \bar{\A}   \s} - 1.
\label{eq:planedist}
\end{equation}
Here, to obtain a slightly cleaner final result, we are now measuring the distortion $D_{\alpha \beta}$ in terms of fractional change in Euclidean distance as opposed to the squared Euclidean distance used in \eqref{eq:distortion}, hence the square root in \eqref{eq:planedist}. The constrained maximum over $\s$ of $\s^T \bar{\A}^T \bar{\A} \s$  in \eqref{eq:planedist} is simply the maximum eigenvalue of the matrix $\bar{\A}^T \bar{\A}$, and its distribution over the random choice of $\A$ has been characterized in sections \ref{sec:wishartmp} and \ref{sec:cg}.  In fact the results in these sections carry over with the replacements $P \rightarrow M$ and $N \rightarrow K$.  The maximal eigenvalue is with high probability equal to $(1 + \frac{1}{\sqrt{\alpha}})^2$, with $\alpha = \frac{M}{K}$.  Its typical fluctuations are $O(M^{-\frac{2}{3}})$, while its large positive deviations of $O(1)$ are exponentially suppressed in $M$, i.e. are $O(e^{-M})$.  So the maximal distortion in \eqref{eq:planedist} is close to $\frac{1}{\sqrt{\alpha}}$.  As long as $M > K$, a similar argument holds for the minimum distortion, which will be close to  $-\frac{1}{\sqrt{\alpha}}$.  Indeed, if $M<K$, then $\bar{\A}^T \bar{\A}$ will have zero eigenvalues, which correspond geometrically to vectors in the random hyperplane $\bo{U}$ that lie in the kernel of the random projection $\A$.   So as long as $\alpha > 1$, the distribution of distance distortions $D_{\alpha \beta}$ will with high probability lie in the range $-\frac{1}{\sqrt{\alpha}}$ to $+\frac{1}{\sqrt{\alpha}}$.   This means of course, that if one wants all distortions $D_{\alpha \beta}$ to obey $-\delta < D_{\alpha \beta} < +\delta$, then one can achieve this with high probability as long as $\delta < \frac{1}{\sqrt{\alpha}}$, or equivalently, the number of random projections obeys $M > \frac{K}{\delta ^2}$, which proves the claim about RP's of hyperplanes made in section \ref{sec:manifoldred}.  Overall, this argument shows how the extremal fluctuations of correlated random variables (i.e. the charges of a Coulomb gas described in section \ref{sec:cg}) can be used to understand geometric distortions induced by RP's of simple manifolds, namely hyperplanes.

 \section{Compressed Sensing}
 \label{sec:compsense}

\par We have seen in section \ref{rdimred} that random projections can preserve the geometric structure of low dimensional signal manifolds.   Furthermore, in the case in which the manifold is the space of $K$-sparse signals (Fig. \ref{rproj}C), as discussed above, one can actually recover the high dimensional signal from its projection using a computationally tractable algorithm, known as $L_1$ minimization.   Here we review this algorithm and its analysis based on statistical mechanics and message passing.   As discussed in the introduction, many applications of the ideas in this section and the previous one are described in \cite{ganguli2012annrevs}.

\subsection{$L_1$ Minimization}
\label{sec:csintro}
\par Suppose  $\s^0$ is an unknown sparse $N$ dimensional vector which has only a fraction $f=K/N$ of its elements nonzero.   Thus $\s^0$ is a point in the top manifold of Fig. \ref{rproj}C.  Suppose we are given a vector $\x$ of $M < N$ measurements, which is linearly related to $\s^0$ by an $M$ by $N$ measurement matrix $\A$, i.e. $\x = \A\s^0$.   $\x$ is then the projection of $\s^0$ in the bottom manifold of Fig. \ref{rproj}C.  Each measurement $x_\mu$, for $\mu=1,\dots,M$ is a linear function $\a_\mu \cdot \s^0$ of the unknown signal $\s^0$, where $\a_\mu$ is the $\mu$'th row of $\A$.   In the context of signal processing, $\s^0$ could be a temporal signal, and the $\a_\mu$ could be a set of $N$ temporal filters.  In the context of network reconstruction, $\s^0$ could be a vector of presynaptic weights governing the linear response of a single postsynaptic neuron $x_u$ to a pattern of presynaptic stimulation $\a_\mu$ on a trial $\mu$ \cite{hu2009reconstruction}.  

\par 
Now how might one recover $\s^0$ from $\x$?  In general, this is an underdetermined problem; there is an $N-M$ dimensional space of candidate signals $\s$ that satisfy the measurement constraint $\x = \A \s$.   The true signal $\s^0$ is but one point in this large space.   However, we can try to exploit our prior knowledge that $\s^0$ is sparse by searching for sparse solutions to the measurement constraints.  For example, one could solve the optimization problem
\begin{equation}
\hat{\s} =  {\rm arg min}_{\s} \, \sum_{i=1}^{N} V(s_i) \quad {\rm subject \ to} \,\, \x=\A\s,
\label{eq:opt1}
\end{equation}
to obtain an estimate $\hat{\s}$ of $\s^0$.   Here $V(x)$ is any sparsity promoting function.  A natural choice is $V(x) = 1$ if $x =0$ and $V(x) = 0$ otherwise, so that (\ref{eq:opt1}) yields the the signal consistent with the measurements $\x$ with the minimum number of nonzero elements.  However, this is in general a hard combinatorial optimization problem.  One could relax this optimization problem by choosing $V(s) = |s|^p$, so that (\ref{eq:opt1}) finds a solution to the measurement constraints with minimal $L_p$ norm.  However, this optimization problem is nonconvex for $p < 1$.   Thus a natural choice is $p=1$, the lowest value of $p$ for which the recovery algorithm (\ref{eq:opt1}) becomes a convex optimization problem, known as $L_1$ minimization.

\subsection{Replica Analysis}
\label{sec:csrepl}

\par Much of the seminal theoretical work in CS \cite{donoho2003optsparse,emmanuel2006candes,candes2005decoding} has focused on sufficient conditions on $\A$ to guarantee perfect signal recovery, so that $\hat{\s} = \s^0$ in (\ref{eq:opt1}),  in the case of $L_1$ minimization. But often, large $\it random$ measurement matrices $\A$ which violate these sufficient conditions nevertheless typically yield good signal reconstruction performance.    Thus these sufficient conditions are not necessary.  Here we review a statistical mechanics approach to CS based on the replica method \cite{kabashima09Lp, DBLP:journals/corr/abs-0906-3234, ganguli2010statistical}, which allows one to directly compute the typical performance of $L_1$ minimization.    Some of these results have also been derived using message passing \cite{donoho2009message} and polyhedral geometry \cite{donoho2005neighborliness}.

\par To understand the properties of the solution $\hat{\s}$ to the optimization problem in (\ref{eq:opt1}), we define an energy function on the residual $\u = \s-\s^0$ given by
\begin{equation}
E(\u) = \frac{\lambda}{2N} \u^T \A^T \A \u + \sum_{i=1}^{N} | u_i + s^o_i |,
\label{eq:en}
\end{equation} 
and analyze the statistical mechanics of the Gibbs distibution 
\begin{equation}
P_G(\u) = \frac{1}{Z} e^{-\beta E(\u)}.
\label{eq:csgibbs}
\end{equation}
By taking the limit $\lambda \rightarrow \infty$ we enforce the constraint $\x = \A\s$.  Then taking the low temperature $\beta \rightarrow \infty$ limit condenses the Gibbs distribution onto the vicinity of the global minimum of (\ref{eq:opt1}).  Then we can compute the average error
\begin{equation}
Q_0 = \frac{1}{N} \sum_{i=1}^N \tav{\u_i}_{P_G}^2,
\label{eq:gord0}
\end{equation} 
and, if needed, the thermal fluctuations
\begin{equation}
\Delta Q  =  \frac{1}{N} \sum_{i=1}^N  \tav{(\delta \u_i)^2}_{P_G}^2 
\label{eq:gord1}
\end{equation}
\par Now, $P_G$, and therefore its free energy $-\beta F = \log Z$, average error $Q_0$, and fluctuations $\Delta Q$ all depend on the measurement matrix $\A$ and on the signal $\s^0$.  We take these to be random variables; the matrix elements $A_{\mu i}$ are drawn independently from a standard normal distribution,  while $\s^0$ has $fN$ randomly chosen nonzero elements each drawn independently from a distribution $\pnz(s^0)$.  Thus $\A$ and $\s^0$ play the role of quenched disorder in the thermal distribution $P_G$.  In the limit $M,N \rightarrow \infty$ with $\alpha = M/N$ held fixed, we expect interesting observables, including the free energy, $Q_0$ and $\Delta Q$ to be self-averaging;  i.e. the thermal average of these observables over $P_G$ for any typical realization of $\A$ and $\s^0$ coincides with their thermal averages over $P_G$, further averaged over $\A$ and $\s^0$.  Thus the typical error $Q_0$ does not depend on the detailed realization of $\A$ and $\s^0$.   We can therefore compute $Q_0$ by computing the average free energy $-\beta \bar F \equiv \qav {\log \, Z}_{\A,\s^0}$ using the replica method.
\par Details of the replica calculation can be found in \cite{ganguli2010statistical}.  Basically, averaging over $\A$ in the replicated Gibbs distribution corresponding to the energy (\ref{eq:en}) reduces to averaging over the variables $\bamu = \frac{1}{\sqrt{N}} \a_\mu \cdot \u^a$, where $\u^a$, $a=1,\dots,n$ are the replicated residuals.  These variables are are jointly gaussian distributed with zero mean and covariance $\qav{\delta \bamu \, \delta b^{b}_{\nu}} = \Qab \delta_{\mu\nu},$ where $\Qab \equiv \frac{1}{T} \sum_{i=1}^T \uai u^{b}_{i}$.  The replica method yields a set of saddle point equations for the overlap matrix $Q$.  Given the convexity of the energy function (\ref{eq:en}), it is reasonable to choose a replica symmetric ansatz for the saddle point, $Q_{ab} = \Delta Q \delta_{ab} + Q_0$.   Under this replica symmetric ansatz, further averaging over $\s^0$, and taking the $\lambda \rightarrow \infty$ limit, yields a set of self-consistent equations
\begin{eqnarray} % requires amsmath; align* for no eq. number
  Q_0                                 &=  \sqav{ \tav{u}_{H^{MF}}^2 }_{z,s^0}             \label{eq:ordq0} \\  
   \Delta Q                         &=  \sqav{ \tav{\delta u^2}_{H^{MF}} }_{z,s^0}. \label{eq:ordq1}
\end{eqnarray}
Here the thermal average $\tav{\cdot}_{H^{MF}}$ is performed with respect to a Gibbs distribution 
\begin{equation}
P^{MF}(s \, | \, s^0) = \frac{1}{Z} e^{-H^{MF}},
\label{eq:pmf}
\end{equation} 
with an effective mean-field Hamiltonian
\begin{equation}
H^{MF}=\frac{\alpha}{2\Dq} \left (s - s^0 -z\sqrt{Q_0/\alpha}\right)^2 + \beta |s|,
\label{eq:hmf}
\end{equation}
where we make the substitution $s - s^0  = u$.   Furthermore the quenched average $\sqav{ \cdot }_{z,s^0}$ denotes an average over a standard normal variable $z$ and the full distribution of the signal component $s^0$ (given by $f \delta(s^0) + (1-f) P(s^0)$).  

\par The relationship between the mean field theory $P^{MF}$ in (\ref{eq:pmf}) and the original Gibbs distribution $P_G$ in (\ref{eq:csgibbs}) is as follows.  The replica parameters $Q_0$ and $\Delta Q$ in (\ref{eq:ordq0}) and (\ref{eq:ordq1}) are identified with the order parameters (\ref{eq:gord0}) and (\ref{eq:gord1}).   Thus solving (\ref{eq:ordq0}) and (\ref{eq:ordq1}) in the zero temperature $\beta \rightarrow \infty$ limit allows us to compute the typical error $Q_0$ of CS as a function of $\alpha$ and $f$.  Furthermore, consider the marginal distribution of a single signal component $\s_k$ in $P_G(\u) = P_G(\s - \s^0)$, given the true signal component is $\s^0_k$.   According to replica theory, the mean field theory prediction for the distribution of this marginal is given by
\begin{equation}
P_G(\s_k = s \, | \, \s^0_k = s^0) = \qav{P^{MF}(s \, | \, s^0)}_{z},
\label{eq:cscondist1}
\end{equation} 
where $P^{MF}(s \, | \, s^0)$ is defined by (\ref{eq:pmf}) and (\ref{eq:hmf}), and $Q_0$ and $\Delta Q$ are the solutions to (\ref{eq:ordq0}) and (\ref{eq:ordq1}). 

 \par
\begin{figure}[htbp]
   \centering
   \includegraphics[width=4.3in]{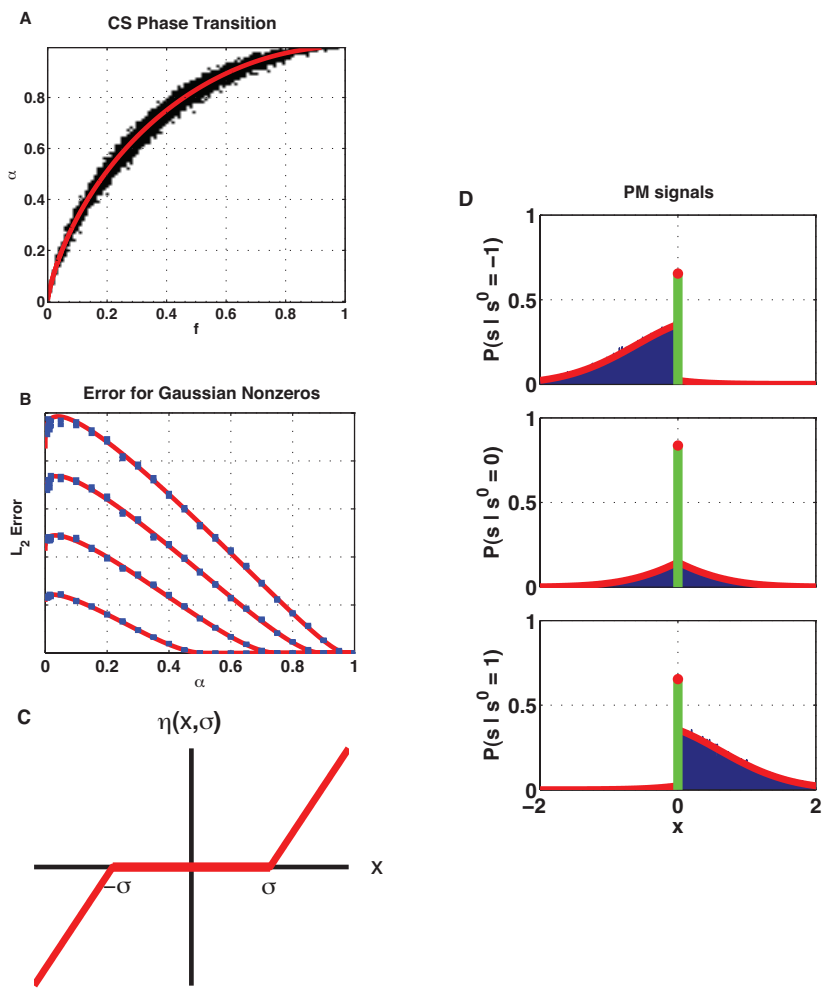} % requires the graphicx package
   \caption{Compressed sensing analysis. (A) The red curve is the theoretical phase boundary $\alpha_c(f)$ obtained by solving (\ref{eq:ordq0}) and (\ref{eq:ordq1}). We also use linear programming to solve \eqref{eq:opt1} $50$ times for each value of 
   $\alpha$ and $f$ in increments of $0.01$, with $N=500$.    The black transition region shows when the fraction of times perfect recovery occurs is neither $0$ nor $1$. For all other $\alpha > \alpha_c(f)$, we obtained perfect recovery all $50$ times, and for all other $\alpha < \alpha_c(f)$ we never once obtained perfect recovery.  The width of this transition region narrows as $N$ is increased (not shown), yielding a sharp transition in the $N \rightarrow \infty$ limit at the phase boundary $\alpha_c(f)$.  (B) Blue points are the average $L_2$ reconstruction error obtained by solving \eqref{eq:opt1} $100$ times for each for $4$ values of  $f = 0.2,0.4,0.6,0.8$, and various $\alpha$, with $N=500$.  Error bars reflect the standard error.  Red curves are plots of $Q_0$ obtained by solving (\ref{eq:ordq0}) and (\ref{eq:ordq1}) in the error phase. (C) The soft thresholding function defined in (\ref{eq:eta}). (D) The blue histograms are the conditional distribution of nonzero signal reconstruction components ${\hat{\mathbf s}}_k$ obtained from solving \eqref{eq:opt1} $2000$ times, while the height of the green bar represents the average fraction of components  ${\hat{\mathbf s}}_k$ that are zero,  all conditioned on the value of the true signal component $s^0_k$.  Here $N=500$, $\alpha = f = 0.2$, and $P(s^0) = \frac{1}{2} \delta(s^0 - 1) + \frac{1}{2} \delta(s^0 + 1)$.   For these values of $\alpha$ and $f$, the order parameters were numerically found to take the values $q_0 = 1.06$ and $\Dqb= 1.43$.   The red curves are the theoretically predicted distribution of nonzero reconstruction components in \eqref{eq:cscondist2}, while the red dot is the theoretically predicted height of the delta function at $s=0$ predicted in \eqref{eq:cscondist2}, all conditioned on the three possible values of the true signal $s^0$, $-1$ (top), $0$ (middle) and $+1$, bottom.  Each distribution can be thought of as arising from a Gaussian distribution with mean $s^0$ and variance $q_0$, fed through the soft threshold function in (C), with a noise threshold $\sigma = \Delta q$.}
   \label{figcs}
\end{figure}

 Now in solving (\ref{eq:ordq0}) and (\ref{eq:ordq1}) in the $\beta \rightarrow \infty$ limit, one finds two distinct classes of solutions \cite{ganguli2010statistical} depending on the values of $\alpha$ and $f$.  For $\alpha > \alpha_c(f)$ one finds solutions in which both $\Delta Q$ and $Q_0$ vanish as $O(\frac{1}{\beta^2})$.   It is expected that thermal fluctuations captured by $\Delta Q$ should always vanish in the low temperature limit, but the fact that $Q_0$, which captures the typical error of CS, also vanishes, suggests that for $\alpha > \alpha_c(f)$, $L_1$ minimization should exactly recover the true signal, so that $\hat{{\bf s}} = \s^0$ in (\ref{eq:opt1}).  On the otherhand, for $\alpha < \alpha_c(f)$, this class of solutions no longer exists, and instead a new class of solutions occurs in which $\Delta Q$ is $O(\frac{1}{\beta})$ but $Q_0$ remains $O(1)$ as $\beta \rightarrow \infty$.  This class of solutions predicts an error regime in which $\hat{{\bf s}} \neq \s^0$ due to too few measurements.   Thus replica theory predicts phase a transition between a perfect and an imperfect reconstruction regime in the $\alpha$-$f$ plane, as verified in Fig. \ref{figcs}A.   This phase boundary was first derived in  \cite{donoho2005sparse, donoho2005neighborliness} using very different methods of convex geometry.  
 
 \par The phase boundary simplifies in the $f\rightarrow0$ limit of high sparsity.  In this limit, $\alpha_c(f) = f \log 1/f$.  This result can be understood from an information theoretic perspective.   First, the entropy of a sparse signal of dimension $N$ is $O(N f \log 1/f)$.   Second, assuming that each of our measurements carries $O(1)$ bits of entropy, and that they are not redundant or highly correlated, then the entropy of our measurements is $O(M)$.   It will not be possible to perfectly reconstruct the signal 
using {\it any} reconstruction algorithm whatsoever, if the entropy of our measurements is less than the entropy of our signal.   The  requirement that the measurement entropy exceed the signal entropy then yields the inequality $\alpha = M/N > O(f \log 1/f)$.   Thus from the perspective of information theory, it is not surprising that we can reconstruct the signal when $\alpha > \alpha_c(f)$.   What is surprising is that a very simple, polynomial time algorithm, $L_1$ minimization, is capable of performing the reconstruction, down to a number of measurements that approaches the information theoretic limit at small $f$, up to constant factors.    

\par What is the nature of this phase transition?  For example if we decrease $\alpha$ from above $\alpha_c(f)$ to below, do we see a catastrophic rise in the error, or does performance gracefully degrade?  In the language of statistical physics, does $Q_0(\alpha, f)$ undergo a first or second order phase transition in $\alpha$?  Fortunately, it is a second order phase transition, so that $Q_0$ rises continuously from $0$.   The exponent governing the rise depends on the distribution of non-zeros $P(s^0)$; namely the more confined this distribution is to the origin, the shallower the rise (see Fig. \ref{figcs}B).  Note that the phase boundary $\alpha_c(f)$ in contrast is universal, in that it does not depend on the distribution of non-zeros in the signal.

\par Finally, we can understand the nature of the errors made by CS by looking at the distribution the signal reconstruction components conditioned on the true signal component.  This is of course interesting only in the error regime.  To take the zero temperature limit we can make the change of variables $\Dq = \frac{\alpha}{\beta} \Delta q$, and $Q_0 = \alpha q_0$ where $\Delta q$ and $q_0$ remain $O(1)$ as $\beta \rightarrow \infty$.  Then the mean field Hamiltonian in (\ref{eq:hmf}) becomes
\begin{equation}
H^{MF}=\beta \bigg[ \frac{1}{2 \Delta q} \left (s - s^0 -z\sqrt{q_0}\right)^2 +  |s| \bigg].
\label{eq:hmfkb}
\end{equation}  
Since the entire Hamiltonian is proportional to $\beta$, in the large $\beta$ limit, the statistics of $s$ are dominated by the global minimum of (\ref{eq:hmfkb}).
In particular, we have 
\begin{equation}
\tav{s}_{H^{MF}} = \eta \big( \, s^0 + z\sqrt{q_0} \, ,  \, \Delta q \big),
\label{eq:avs}
\end{equation}
where
\begin{equation}
\eta(x \, , \sigma) = {\rm arg min}_s  \bigg( \frac{1}{2}\frac{(s-x)^2}{\sigma}  + |s| \bigg)= \mathrm{sgn}(x) ( |x|-\sigma )_+,
\label{eq:eta}
\end{equation}
is a soft thresholding function (see Fig. \ref{figcs}D) which also arises in message passing approaches \cite{donoho2009message} to solving the CS problem in \eqref{eq:opt1}, and
$(y)_+ = y$ if $y>0$ and is otherwise $0$.   
\par The optimization in (\ref{eq:eta}) can be understood intuitively as follows:  suppose one measures a scalar value $x$ which is a true signal $s^0$ corrupted by additive gaussian noise with variance $\sigma$.  Under a Laplace prior $e^{-|s^0|}$ on the true signal, $\eta(x,\sigma)$ is simply the MAP estimate of $s^0$ given the data $x$, which basically chooses the estimate $s=0$ unless the data exceeds the noise level $\sigma$.  Thus we see that in \eqref{eq:hmfkb} and \eqref{eq:avs}, $s^0 + z\sqrt{q_0}$ plays the role of the observed, corrupted data $x$ and $\Delta q$ plays the role of an effective noise level $\sigma$.   
\par This optimization has an interpretation within the cavity method  \cite{kabashima09Lp}.  It is the optimization that a new signal component $s$ added to a cavity system in the absence of that component must perform to minimize its total energy in  (\ref{eq:csgibbs}).  This minimization reflects a compromise between minimizing its own absolute value, and satisfying all the measurement constraints, whose sum total effect is encapsulated by the quadratic term in the mean field theory of (\ref{eq:hmfkb}).  The cavity field encapsulating the effect of all other measurements will vary from component to component, and the average over components can be approximated by the average over the Gaussian variable $z$, in analogy to the SK model in going from (\ref{eq:qselfcon1}) to (\ref{eq:selfconscav}).  

\par The distribution of the signal reconstruction components, conditioned on the true signal component $s^0$ in (\ref{eq:cscondist1}) reduces to 
\begin{equation}
P_G(\s_k = s \, | \, \s^0_k = s^0) = \qav{\eta \big( s^0 + z\sqrt{q_0},  \Delta q \big)}_{z},
\label{eq:cscondist2}
\end{equation} 
and it reflects the Gaussian distribution of the zero-temperature cavity fields across components, fed through the soft-thresholding function which arises from the scalar $L_1$ minimization problem in (\ref{eq:eta}).  Here $q_0$ and $\Delta q$ are determined through (\ref{eq:ordq0}) and (\ref{eq:ordq1}), which can be thought of as a self-consistency condition within the cavity approximation demanding that the distribution across components of the cavity field is consistent with the distribution across components of the signal reconstruction, in analogy to the corresponding self-consistency condition (\ref{eq:selfconscav}) in the SK model.  An example of the match between replica theory and simulations for the signal reconstruction distribution is shown in Fig. \ref{figcs}D.

 \subsection{From Message Passing to Network Dynamics}
 \label{sec:csmpass}
 
 The $L_1$ minimization problem in \eqref{eq:opt1} can also be formulated as a message passing problem \cite{donoho2009message, montanari2010graphical}, and approximate formulations of the message passing dynamical system yield neural network-like dynamics which provide a fast iterative way to solve the $L_1$ problem.  The graphical model consists of $N$ variable nodes, one for each component of the unknown signal $\s_i$, and $M$ degree $N$ factor nodes, one for each measurement, plus $N$ more degree $1$ factor nodes to implement the $L_1$ norm.  For example, the Gibbs distribution defined by \eqref{eq:en} and \eqref{eq:csgibbs} decomposes as
\begin{equation}
P(\s) = \prod_{\mu=1}^M \psi_\mu(\s) \prod_{\i=1}^N e^{-\beta | \s_i |},
\end{equation}
where the factor $\psi_\mu(\s)$ is given by
\begin{equation}
\psi_\mu(\s)  = e^{-\frac{1}{2N} \beta \lambda (x_\mu - \mathbf{a}_\mu \cdot \mathbf{s})^2},
\end{equation}
and $\a_\mu$ is row $\mu$ of the measurement matrix $\A$. This decomposition is in a form suitable for the application of the message passing equations \eqref{eq:mpass1} and \eqref{eq:mpass2}.  However, a straightforward application of these equations is computationally complex.  First, since each unknown component $\s_i$ is a real number, every message becomes a distribution over the real numbers, and one must keep track of $MN$ such messages (the degree $1$ factors do not require associated messages and can be incorporated into the updates of the other messages).  
\par The first approximation made in \cite{donoho2009message} is to restrict the messages to be Gaussian, which is reasonable because the density of the random measurement matrix $\A$ implies that in each update, each message receives contributions from a large number of messages, allowing one to invoke the central limit theorem.  Thus one need keep track of only $2$ numbers for each message, leading to a dynamical system on $2MN$ variables.  This system can be further simplified by noting \cite{donoho2009message, montanari2010graphical} that messages from the same variable $i$ to different factors $\mu$ are all quite similar to each other; they differ only in excluding the effects of one factor $\mu$ out of $M$ possible factors.  Thus one might assume that $M_{i \rightarrow \mu} = M_i + O(\frac{1}{\sqrt{M}})$.   A similar argument holds for the factor to variable messages, suggesting  $M_{\mu \rightarrow i} = M_\mu + O(\frac{1}{\sqrt{N}})$.  By doing a careful expansion in $\frac{1}{N}$, one can then reduce the message passing equations to a dynamical system on $M+N$ variables.  
\par A readable account of this reduction can be found in \cite{montanari2010graphical}.  Here we simply quote the main result.  The dynamical system on $M$ + $N$ variables can be interpreted as an iterative update on two pairs of variables, a current estimate $\s^t$ for the unknown $N$ dimensional signal, and the resulting residual $\r^t$ in $M$ dimensional measurement space.  The update equations are  \cite{montanari2010graphical},
\begin{eqnarray}
 \s^{t+1} & = \eta(\s^t + \A^T \r^t , \theta) \\
 \r^t         & = \x - \A \s^t + b \r^{t-1}, \label{eq:brt}
 \end{eqnarray} 
where $\eta$ is the soft-thresholding function defined in \eqref{eq:eta}, and here is applied component wise to its vector inputs.  In \cite{montanari2010graphical}, it was shown that if these equations converge, then the resulting $\s^{\infty}$ is a global minimum of the $L_1$ energy function in \eqref{eq:en} with $\frac{1}{\lambda} = \theta(1-b)$.

\par The crucial term is the message passing derived term involving $b$ in \eqref{eq:brt} that endows the temporal evolution of the residual with a history dependence.  Without this term, reconstruction performance is severely impaired.  This dynamics can loosely be interpreted as that of a two layer interacting neuronal network, where the residual $\r^t$ is stored in the activity of $M$ neurons in the first layer, while the current estimate of the sparse representation $\s^{t}$ is stored in $N$ neurons in a second layer.  The first layer receives feedforward external input $\x$ and a top down inhibitory prediction of that input through synaptic connectivity $\A$.  The second layer neurons receive a feedforward drive from the residuals through a synaptic connectivity $\A^T$ and have a nonlinear transfer function $\eta$.  Interestingly, this network dynamics is different from other proposals for the implementation of $L_1$ minimization and sparse coding \cite{olshausen1996emergence, rozell2008sparse}.  Given the potential role of $L_1$ minimization as a computational description of early visual \cite{hu11early} and olfactory \cite{koulakov11rindberg} processing, it would be interesting to explore more fully the space of neuronal architectures and dynamics capable of implementing $L_1$ minimization type computations. 

\section{Discussion}

We have reviewed the applications of replicas, cavities, and message passing to basic models of network dynamics, memory storage, machine learning algorithms, and statistical models of data.  While the ideas and models reviewed yield a rich and sometimes surprisingly striking picture, it is natural to ask how the above results are modified when assumptions in the models are relaxed, and what theoretical progress can be made through a statistical mechanics based analysis of these more complex scenarios.  Here we will briefly discuss a few more prominent examples.  However, we warn the reader that in this discussion we will barely be scratching the surface of a deep literature lying at the intersection of statistical mechanics, computer science and neuroscience. 

\subsection{Network Dynamics}

\par First, in section \ref{sec:spinglassmodel}, when we introduced the SK model and the Hopfield model, we were considering models of neuronal networks that had many simplifying assumptions, including: symmetric connectivity, binary neurons, and lack of external inputs.  What happens for example, when the connectivity becomes asymmetric?  Then there is no simple form for the stationary distribution of neuronal activity analogous to the equilibrium Gibbs distribution in \eqref{eq:spingibbs}.  One must then posit a dynamical model for the network dynamics, and in many situations, one can use dynamic mean field theory methods  \cite{martin1973statistical, de1978dynamics} to understand time averaged asymptotic statistical properties of neuronal activity.  This was done in for example in \cite{crisanti1987dynamics, crisanti1988dynamics, derrida2007exactly} for asymmetric networks.  Interestingly, while fully asymmetric networks become ergodic, partially asymmetric networks retain fixed points, but the time it takes for transients to reach these fixed points can diverge exponentially with network size.

\par
Moreover dynamical versions of the SK and Hopfield models have a Lyapunov function that is bounded from below, implying that the long time asymptotic behavior at zero temperature consists of fixed points only.  In asymmetric networks, two more dynamical possibilities arise: oscillations and chaos.  Seminal work showed analytically (via dynamic mean field theory) and through simulations that neuronal networks can exhibit deterministic high dimensional chaos \cite{sompolinsky1988chaos, van1996chaos, vreeswijk1998chaotic}.  Even when driven by a constant current input, such networks exhibit chaos by dynamically achieving a balance between excitatory and inhibitory inputs to individual neurons.  This balance leads to spontaneous irregular neural activity characteristic of cortical states, in which neurons spike due to fluctuations in their input, as opposed to a mean superthreshold input current.  Interestingly, when the inputs have more nontrivial temporal structure, such networks exhibit a sharp phase transition from a chaotic state to an ordered state, that is entrained by the input, as the input strength increases.  This happens for example when the external input is either noisy \cite{molgedey1992suppressing} or oscillatory \cite{rajan2010stimulus}.  In the case of oscillatory input there is an interesting non-monotonic dependence in the input strength at which this phase transition occurs, as a function of oscillation frequency \cite{rajan2010stimulus}.

\par
In section  \ref{sec:spinglassmodel}, we discussed binary models of neurons.   However, biological neurons are characterized by analog internal states describing membrane voltage and ion channel conductance states, and their dynamics exhibits large spiking events in their membrane voltage.   Dynamic mean field theory methods  can be extended to spiking networks and were used to characterize the phase diagram of networks of excitatory and inhibitory leaky integrate and fire neurons \cite{brunel2000dynamics}.  For such neurons, whose internal state is characterized solely by a membrane voltage, one can derive an appropriate mean field theory by maintaining a distribution of membrane voltages across neurons in the network, and self-consistently solving for this distribution using Fokker-Planck methods (see \cite{renart2004mean, hertz2004mean} for reviews).  This work \cite{brunel2000dynamics} lead to four possible macroscopic phases of network dynamics, characterized by two possibilities for the temporal statistics of single neurons (regular periodic spike trains, or irregular aperiodic spike trains) times two possibilities for the population average firing rates (synchronous or temporally structured rates, or asynchronous or constant rates).   Varying strengths of excitation, inhibition, and single neuron properties allow all four combinations to occur.  

\par  
More recently, in \cite{monteforte2010dynamical, monteforte2012dynamic} the authors went beyond mean field theory to track entire microstate trajectories in spiking neural networks consisting of neurons in which it is possible to analytically compute the time of the first neuron to spike next, given the internal state of all neurons in the network.  This allowed the authors to perform numerically exact computations of the entire spectrum of Lyapunov exponents by computing products of Jacobians associated with every future spike starting from an initial condition.  They found classes of networks that exhibit extensive chaos, in which a finite fraction of all Lyapunov exponents were positive.  Moreover, they showed that the Lyapunov spectrum is highly sensitive to the details of the action potential shape, as positive feedback effects associated with the rise of the action potential contribute most heavily to the divergence of microstate trajectories.  Even more interestingly, the authors found ``flux" tubes of stability surrounding trajectories:  small perturbations to the network state decayed quickly, whereas larger perturbations lead to an exponential divergence between trajectories.  Thus each trajectory is surrounded by a stability tube.   However, the radius of this tube shrinks with the number of neurons, $N$.   This reveals that the calculation of Lyapunov exponents in spiking networks in the thermodynamic ($N \rightarrow \infty$) limit is extremely subtle, due to the non-commutation of limits.  
Computing Lyapunov exponents requires taking a small perturbation limit, which if taken before the thermodynamic limit would yield negative exponents, but if taken after the thermodynamic limit, would yield positive exponents.  In any case, injecting extra spikes into the network constitutes a large perturbation even at finite $N$, which leads to a divergence in trajectories.  This picture is consistent with recent experimental results suggesting that the injection of extra spikes into a cortical network leads to 
a completely different spiking trajectory, without changing the overall population statistics of neural activity \cite{london2010sensitivity}.  More generally, for reviews on network dynamics in neuroscience, see \cite{vogels2005neural, rabinovich2006dynamical}.

\subsection{Learning and Generalization}

\par In the beginning of section \ref{sec:smoflearning}, we considered the capacity of simple network architectures to store, or memorize a set of input-output mappings.  While memory is certainly important, the goal of most organisms is not simply to memorize past responses to past inputs, but rather to generalize from past experience in order to learn rules that can yield appropriate responses to novel inputs the organism has never seen before.  This idea has been formalized in a statistical mechanics framework for the perceptron in \cite{seung1992statistical, sompolinsky1990learning}.  Here the $P$ training inputs and outputs are no longer random, but are generated from a ``teacher'' perceptron.  The observable of interest then becomes the generalization error $\epsilon_g(\alpha)$, which is by definition the probability that the trained perceptron disagrees with the teacher perceptron's correct answer on a novel input, not present in the training set.  Here $\alpha = \frac{P}{N}$ is the ratio of the number of training examples to the number of synapses $N$.   For a wide variety of training procedures, or learning algorithms, statistical mechanics approaches found that $\epsilon_g(\alpha)$ decays as $O(\frac{1}{\alpha})$ for large $\alpha$, indicating that the number of examples should be proportional to the number synapses in order for good generalization to occur. 

\par  The perceptron, while acting in some sense as the {\it Drosophila} of statistical learning theory, is a very limited architecture in that it can only learn linearly separable classifications in which the two classes fall on opposite sides of a hyperplane.  Statistical mechanics approaches have been used to analyze memory  
\cite{barkai1990statistical, barkai1992broken, engel1992storage} and generalization \cite{opper1994learning, monasson1995weight, schwarze1999learning} in more sophisticated multilayered networks.  In a multilayered network, only the input and output layers are constrained to implement a desired mapping, while the internal, hidden layer activities remain unspecified.   This feature generically leads to replica symmetry breaking, where the space of solutions to a desired input-output mapping breaks into multiple disconnected components, where each component corresponds to a different internal representation of hidden layer activities capable of implementing the desired mapping.  Statistical mechanics has also had success in the analysis of learning in other architectures and machine learning algorithms, including support vector machines \cite{dietrich1999statistical, opper2001universal, malzahn2005statistical}, and Gaussian processes \cite{urry2012replica}.

\par Another generalization of the perceptron is the tempotron, an architecture and learning rule capable of learning to classify spatiotemporal patterns of incoming spikes \cite{gutig2006tempotron}.   The tempotron can be trained to fire a spike for one class of input spike time patterns, and no spikes for another class, while the precise timing of the output spike can be left unspecified.  A statistical mechanics analysis of a simplified binary tempotron was carried out in \cite{rubin2010theory}.  Interestingly, the space of solutions in synaptic weight space to any given spike time classification problem can be well described by a one-step replica symmetry broken phase shown schematically in Fig. \ref{FigLandscape}C.  Each component corresponds to a different output spike time for the positive classifications, in direct analogy to the replica symmetry broken phase of multilayered networks in which each component corresponds to a different internal representation.  The various solution components are both small (implying that very similar weights can yield very different classifications) and far apart (implying that very different weights can yield an identical classification).  The authors verified that these properties persist even in a more biologically realistic Hodgkin-Huxley model of a single neuron \cite{rubin2010theory}.   Overall this reveals a striking double dissociation between structure (synaptic connectivity) and function (implemented classification) even at the level of single neurons.  This double dissociation has important implications for the interpretation of incoming connectomics data \cite{lichtman2008ome}.  More generally, for reviews on applications of statistical mechanics to to memory, learning and generalization, see  \cite{engel01, watkin1993statistical}.

\subsection{Machine Learning and Data Analysis}

\par
Starting in the latter part of section \ref{sec:smoflearning}, we turned our attention to the statistical mechanics based analysis of machine learning algorithms designed to extract structured patterns from data, focusing on illusions of structure returned by such algorithms when applied to high dimensional noise.  In real data analysis problems, we have to protect ourselves from such illusions, and so understanding these illusions present in pure noise is an important first step.  However, we would ideally like to analyze the performance of machine learning algorithms when data contain both structured patterns as well as random noise.  A key question in the design and analysis of experiments is then how much data do we need to reliably uncover structure buried within noise?  Since, in the statistical mechanics based analysis of learning algorithms, the data plays the role of quenched disorder, we must analyze statistical mechanics problems in which the quenched disorder is no longer simply random, but itself has structure.     

\par This has been done for example in \cite{hoyle2010statistical, sengupta1999distributions} for PCA applied to signals confined to a low dimensional linear space, but corrupted by high dimensional noise.  The data is then drawn from a covariance matrix consisting of the identity plus a low rank part.   A replica based computation of the typical eigenvalue spectrum of empirical covariance matrices for data of this type revealed the presence of a series of phase transitions as the ratio between the amount of data and its ambient dimensionality increases.  As this ratio increases, signal eigenvalues associated with the low rank part pop out of a Marcenko-Pasteur sea (i.e. Fig.\ref{mptw}A)  associated with the high dimensional noise.  Thus this work reveals sharp thresholds in the amount of data required to resolve signal from noise.  Also, interesting work has been done on the statistical mechanics based analysis of typical learning outcomes for other structured data settings, including finding a direction separating two Gaussian clouds \cite{biehl1999statistical, biehl1994exactly}, supervised learning from clustered input examples \cite{marangi2007supervised}, phase transitions in clustering as a function of cluster scale \cite{rose1990statistical}, and learning Gaussian mixture models \cite{barkai1993scaling, barkai1994statistical}.
Moreover, statistical mechanics approaches to clustering have yielded interesting new algorithms and practical results, including superparamagnetic clustering \cite{blatt1996superparamagnetic, wiseman1998superparamagnetic}, based on an isomorphism between cluster assignments and Potts model ground states, and a method for computing p-values for cluster significance \cite{luksza2010significance}, using extreme value theory to compute a null distribution for the maximal number of data points over all regions in feature space of a given size.  
 
\par In section \ref{sec:corrextreme} we initiated a statistical mechanics based analysis of random dimensionality reduction by connecting the maximally incurred geometric distortion to correlated extreme value theory.  For the simple case of point clouds, correlations could be neglected, while for hyperplanes, the correlations arose from fluctuations in the Coulomb gas interactions of eigenvalues of random matrices, and could be treated exactly. It would be interesting to study more complex manifolds.  For example, rigorous upper bounds on the maximal distortion were proven in  \cite{baraniuk2009random} by surrounding arbitrary manifolds and their tangent planes by a scaffold of points, and then showing that if the geometry of this scaffold remains undistorted under any projection, then so does the geometry of the manifold.  An application of the JL lemma to the scaffold then suffices to obtain an upperbound on the distortion incurred by the manifold under a random projection.  To understand how tight or loose this upperbound is, it would useful to compute the typical distortion incurred by more complex manifold ensembles.   For example, for manifolds consisting of unions of planes, one would be interested in the fluctuations of the maximal eigenvalue of multiple correlated matrices, corresponding to the restriction of the same random projection to each plane.  Thus results from the eigenvalue spectra of random correlated matrices \cite{dyson1962brownian, schehr2008exact, tracy2007nonintersecting} could become relevant.    

\par Finally, we note that throughout most of this review we have focused on situations in which replica symmetry holds, though we have noted that several neuronal and machine learning problems, including multilayer networks, tempotrons, and clustering, are described by replica symmetry broken phases in which the solution space breaks up into many clusters, as well as suboptimal, higher energy metastable states.  As noted at the end of section \ref{subsecmpass}, statistical mechanics based approaches have inspired a new algorithm, known as survey propagation \cite{mezard2002analytic, braunstein2005survey} that can find good solutions despite the proliferation of metastable states whose presence can confound simpler optimization and inference algorithms.  Despite the power of survey propagation, its applications to neuroscience remain relatively unexplored.    

\par In summary, decades of interactions between statistical physics, computer science, and neuroscience, have lead to beautiful insights, both into how neuronal dynamics leads to computation, as well as how our brains might create machine learning algorithms to analyze themselves.  We suspect that further interactions between these fields are likely to provide exciting and insightful intellectual adventures for many years to come.

\subsection{Acknowledgements}
We thank DARPA, the Swartz foundation, Burroughs-Wellcome Foundation, Genentech Foundation, and Stanford Bio-X Neuroventures  for support.  S.G. thanks Haim Sompolinsky for many interesting discussions about replicas, cavities, and messages.

% survey propagation:  would be intriguing to speculate that neural circuits in our brains might themselves be implementing through their dynamics some version of survey propagation to solve complex computational tasks that, when formulated as optimization problems, elude solution by current, human engineered hardware.

\section{Appendix:  Replica Theory}
\label{sec:repapp}
\subsection{Overall Framework}
\label{sec:repframe}
Suppose we wish to do statistical mechanics on a set of $N$ thermal degrees of freedom encoded in the $N$ dimensional vector $\x$, where the components are coupled to each other through some quenched disorder $\D$, in a Hamiltonian $H(\x,\D)$. In the above applications, $\x$ could be the spins $\s$ in a spin glass (then $\D$ is the connectivity matrix $\J$), the synaptic weights $\w$ of a perceptron (then $\D$ is the set of examples to be stored),  the variables $\u$ in the Stieltjes transform of an eigenvalue spectrum (then $\D$ is a random matrix), or the residuals $\u$ in a compressed sensing problem (then $\D$ is the measurement matrix).  
As discussed above, to properly average over the quenched disorder $\D$, we must average the replicated partition function
\begin{equation}
\qav{Z^n} _{{\mathbf D}} = \bqav{  {\int} \, \prod_{a=1}^n \, d\x^a e^{- \sum_{a=1}^n H(\x^a, {\mathbf D})} }_{{\mathbf D}}.
\label{eq:repl1}
\end{equation}
Conditioned on any particular realization of the quenched disorder \D, the different replicated degrees of freedom $\x^a$ are independent.  However, integrating out the quenched disorder introduces interactions among the replicated variables.  In all of the above applications, the resulting interactions depend only on the overlap matrix  between replicas, defined as $\Qab = \frac{1}{N} \x^a \cdot \x^b$.  More precisely, the following identity holds,
\begin{equation}
\bqav{  e^{- \sum_{a=1}^n H(\x^a, {\mathbf D}) } }_{{\mathbf D}} = e^{-N E(Q)},
\label{eq:Eqabdef}
\end{equation}
for some function $E$ over the overlap matrix $Q$.
Therefore it is useful to separate the remaining integral over $\x^a$ in (\ref{eq:repl1}) into an integral over all possible overlaps $\Qab$, and then all possible $\x^a$ configurations with a prescribed set of overlaps by introducing a $\delta$-function:
\begin{equation}
\qav{Z^n} _{\D} = \int \, \prod_{ab}  \,  d\Qab \, e^{-N E(Q)} \int{ \,  \prod_{a=1}^n \, d\x^a \prod_{ab} \delta \big[ \x^a \cdot \x^b - N \Qab \big]. }
\label{eq:repl2}
\end{equation}
The integral over $\x^a$ with a fixed set of overlaps $\Qab$ can be carried out by introducing the exponential representation of the $\delta$ function,
\begin{equation}
\delta \big[ \x^a \cdot \x^b - N \Qab \big]  = \int \, d\Qhab \, e^{-\Qhab({\mathbf x}^a \cdot {\mathbf x}^b - N \Qab) } ,
\label{eq:deltafunc}
\end{equation}
where the integral over $\Qhab$ is understood to be along the imaginary axis.  Inserting (\ref{eq:deltafunc}) into (\ref{eq:repl2}) decouples the components of the vectors $\x^a$, yielding an integral over $n$ scalar variables $x^a$ raised to the $N$'th power.  This final result can be written as
\begin{equation}
\qav{Z^n} _{\D} =  {\int} \,  \prod_{ab}  \,  d\Qab \,   \, d\Qhab \, e^{- N [ E(Q) - \sum_{ab} \Qab \Qhab + G(\Qhab)]},
\label{eq:repl3}
\end{equation}  
where 
\begin{equation}
G(\Qhab) =  - \ln { \int \,  \prod_{a}  \,  dx^a  e^{ - H_{\rm eff}(x^1,\dots,x^n)}},
\label{eq:effpart}
\end{equation}  
is the partition function of an effective Hamiltonian
\begin{equation}
H_{\rm eff} =  \sum_{ab} x^a \, \Qhab \, x^b.
\label{eq:effham}
\end{equation}  
Now in the large $N$ limit, the final integrals over $\Qab$ and $\Qhab$ can be done via the saddle point method, yielding a set of self-consistent equations for the saddle point by extremizing the exponent in (\ref{eq:repl3}):
\begin{equation}
\Qhab = \frac{\partial E}{\partial \Qab}
\label{eq:repsadd1}
\end{equation}  
\begin{equation}
\Qab = \langle x^a x^b \rangle_n,
\label{eq:repsadd2}
\end{equation}  
where $\langle \cdot \rangle_n$ denotes an average with respect to a Gibbs distribution with effective Hamiltonian $H_{\rm eff}$ in (\ref{eq:effham}).
\par
In general both these equations must be solved in the $n \rightarrow 0$ limit.  Now in the case where $\x^a_i$ are real valued-variables (as opposed to binary variables in the SK model), these equations can be further simplified because the integral over $x^a$ in (\ref{eq:effpart}) can be done exactly, since it is  Gaussian, and furthermore, the extremum over $\Qhab$ in (\ref{eq:effpart}) can be performed.  Together, this yields an entropic factor (up to a multiplicative constant involving $n$ and $N$),
\begin{equation}
\int \,  \prod_{a=1}^n \, d\x^a \prod_{ab} \delta \big[ \x^a \cdot \x^b - N \Qab \big] = \e^{N S(Q)},
\end{equation}
where 
\begin{equation}
S(Q) = \frac{1}{2} \Tr \log Q
\label{eq:entropicfactor}   
\end{equation}
represents (up to an additive constant) the entropy of replicated configurations $\x^a$ with a prescribed overlap matrix $Q$.  (\ref{eq:repl3}) reduces to
\begin{equation}
\qav{Z^n} _{\mathbf D} =  {\int} \,  \prod_{ab}  \,  d\Qab \,    e^{- N [ E(Q) - S(Q))]},
\label{eq:repl4}
\end{equation}  
and the saddle point overlap configuration $Q$ represents a compromise between energy and entropy extremization in the exponent of (\ref{eq:repl4}).    

\subsection{Physical meaning of overlaps}

\label{sec:ovlapmeaning}
Here we make the connection between the replica overlap matrix $\Qab$ and the disorder averaged distribution of overlaps, $P(q)$, of two states $\x^1$ and $\x^2$ both drawn from a Gibbs distribution with Hamiltonian $H(\x,\mathbf{D})$.   For a given realization of the disorder, the overlap distribution is
\begin{equation}
P_{\mathbf{D}}(q) =  \frac{1}{Z({\mathbf D})^2} \int \,  d\x^1 \, d\x^2 \, \delta \left(q-\frac{1}{N} \x^1 \cdot \x^2 \right) e^{-H(\x^1,\mathbf{D})-H^(\x^2,\mathbf{D})}
\label{eq:pdq}
\end{equation}
where $Z(\D) = \int \,  d\x \, e^{-H(\x,\mathbf{D})}$. Averaging $P_\mathbf{D}(q)$ over the disorder  is difficult because $\D$ appears both in the numerator and denominator of \eqref{eq:pdq}.  To circumvent this, one can introduce replicas via the simple identity $Z^{-2} = \lim_{n \rightarrow 0} Z^{n-2}$ .  Using this, one can perform the easier average at integer $n>2$, and then take the limit $n \rightarrow 0$ at the end.  Thus
\begin{eqnarray}
P(q) & =   \qav{P_{\mathbf{D}}(q)}_{\mathbf{D}} \\ 
        & =   \lim_{n \rightarrow 0} \bqav{  {\int} \, \prod_{a=1}^n \, d\x^a e^{- \sum_{a=1}^n H(\x^a, \mathbf{D})} \, \delta\big(q-\frac{1}{N} \x^1 \cdot \x^2 \big)}_{\mathbf{D}}
\label{eq:pdqav}
\end{eqnarray}
Here $\x^1$ and $\x^2$ are the original degrees of freedom with $n-2$ additional replicas added to yield $Z^{n-2}$.   One can then average the right hand side of (\ref{eq:pdqav}) over $\D$ using a sequence of steps very similar to section \ref{sec:repframe}. The final answer yields
\begin{equation}
P(q) = \lim_{n \rightarrow 0} \frac{1}{n(n-1)} \sum_{a \neq b} \delta ( q - Q_{ab} )
\label{eq:pdqav2}
\end{equation}
where $Q_{ab}$ is the saddle point replica overlap matrix.   In situations where replica symmetry is broken, there will be multiple equivalent saddle points related to each other by the action of the permutation group on the replica indices $a,b$.   The sum over these saddle points yields the sum in (\ref{eq:pdqav2}).  In summary, the probability that two states have overlap $q$ is, according to replica theory, equal to the fraction of off-diagonal matrix elements $Q_{ab}$ that take the value $q$.

\subsection{Replica symmetric equations}

Here we show how to take the $n \rightarrow 0$ limit for various problems, in the replica symmetric approximation.   We use Einstein summation convention in which repeated indices are meant to be summed over.

\subsubsection{SK Model}

\par  We will now apply \eqref{eq:effpart}-\eqref{eq:repsadd2}
 to the SK model from section \ref{sec:spinglassmodel}.
As we saw in \eqref{eq:disorderavsk}, we have
\begin{equation}\label{eq:SKoverlapE}
  E(Q) = -\left(\frac{\beta}{2}\right)^2 Q_{ab}^2.
\end{equation}
Then, \eqref{eq:repsadd1} gives
\begin{eqnarray}\label{eq:SKhatQ}
  \hat{Q}_{ab} &=& -\frac{\beta^2}{2} Q_{ab}, \\
\label{eq:SKE-QhQ}
  E(Q)-Q_{ab}\hat{Q}_{ab} &=& \left(\frac{\beta}{2}\right)^2 Q_{ab}^2.
\end{eqnarray}

We make the following replica-symmetric ansatz for the saddle point:
\begin{equation}\label{eq:SKansatz}
  Q_{ab} = q +(1-q)\delta_{ab},
\end{equation}
where we used the fact that \eqref{eq:ovlapspin} guarantees that $Q_{aa}=1$. We will determine $q$ by minimizing the free energy. This leads to
\begin{eqnarray}\label{eq:SK-RS-E}
  E(Q)-Q_{ab}\hat{Q}_{ab} &=& \left(\frac{\beta}{2}\right)^2 \left[(1-q^2)n + q^2n^2\right] \\
\label{eq:SK-RS-Heff}
  H_{\mathrm{eff}} &=& -\frac{\beta^2}{2} \left[(1-q) n + q \left(\sum_a s^a\right)^2\right].
\end{eqnarray}

We can now evaluate \eqref{eq:effpart} using the identity \eqref{eq:hubstrat} with $\sigma=1$:
\begin{eqnarray*}
  G(Q) &=& -\ln\left[\sum_{\{s^a\}} e^{\frac{\beta^2}{2} \left[(1-q) n + q \left(\sum_a s^a\right)^2\right]}\right]\\
   &=& -\frac{\beta^2}{2} (1-q) n - \ln\left[ \sum_{\{s^a\}}
   \qav{ e^{\beta \sqrt{q}z \sum_{a} s^a} }_{z}\right]\\
   &=& -\frac{\beta^2}{2} (1-q) n - \ln\left[
   \qav{ \left[2 \cosh (\beta \sqrt{q}z)\right]^n }_{z}\right].
\end{eqnarray*}
This gives us the free energy density:
\begin{eqnarray}
  \qav{\frac{\beta F}{N}}_\mathbf{J}
   &=& -\frac{1}{N}\left.\frac{\partial}{\partial n} \qav{Z^n}_\mathbf{J}\right|_{n=0}
\nonumber\\\label{SKfree}
   &=& -\left(\frac{\beta J}{2}\right)^2 (1-q)^2
    - \qav{\ln [2\cosh(\beta \sqrt{q}z)]}_z.
\end{eqnarray}
As mentioned above, we determine $q$ by minimizing this. We will need the identity
\[
  \qav{z\,f(z)}_z = \qav{f'(z)}_z,
\]
which can be derived with integration by parts. We find
\begin{eqnarray}
 \frac{\partial}{\partial q} \qav{\frac{\beta F}{N}}_\mathbf{J}
   &=& \frac{\beta^2}{2}(1-q) - \frac{\beta}{2\sqrt{q}} \qav{z\tanh(\beta \sqrt{q}z)}_z
\nonumber\\
   &=& \frac{\beta^2}{2}\left(1-q-\qav{\mathrm{sech}^2(\beta \sqrt{q}z)}_z\right)
\nonumber\\\label{eq:SKsaddle}
   &=& \frac{\beta^2}{2}\left(\qav{\tanh^2(\beta \sqrt{q}z)}_z-q\right), \label{eq:saddlesk}
\end{eqnarray}
therefore, the minimum satisfies \eqref{eq:skrseq}.

\subsubsection{Perceptron and Unsupervised Learning}
\label{sec:n0unsup}
The starting point for the learning applications discussed here is the energy \eqref{eq:perceptlearnbef} and entropy \eqref{eq:entropicfactor} in the replicated partition function \eqref{eq:repl4}.  These can be derived by following sections \ref{sec:repllearn} and \ref{sec:repframe}.  Here we take the $n \rightarrow 0$ limit. For the energy, we obtain
\begin{eqnarray}
\lim_{n \rightarrow 0} E(Q) & = - \alpha \ln   {\int} \, \prod_{a=1}^n \, \frac{d\lambda_a}{\sqrt{2\pi}} \, \frac{1}{\sqrt{\det Q}} e^{-\frac{1}{2}  \lambda_a Q^{-1}_{ab}\lambda_b - \sum_a \beta V(\lambda_a)} \label{eq:unsuprs1}\\
         & = -  \alpha \ln   {\int} \, \prod_{a=1}^n \, \frac{d\lambda_a \, d\hat{\lambda_a }}{2 \pi} \,  e^{ i \lambda_a \hat{\lambda_a } -\frac{1}{2}  \hat{\lambda_a} Q_{ab}\hat{\lambda_b} - \sum_a \beta V(\lambda_a)} \nonumber \\
         & = -  \alpha  \ln   {\int} \, \prod_{a=1}^n \, \frac{d\lambda_a \, d\hat{\lambda_a }}{2 \pi} \,  e^{ i \lambda_a \hat{\lambda_a }  - \frac{1}{2}(1-q)\sum_a (\hat{\lambda_a} )^2 - \frac {1}{2}( \sqrt{q} \sum_a  \hat{\lambda_a} )^2  - \sum_a \beta V(\lambda_a)} \label{eq:unsuprs3} \\
         & = -  \alpha \ln   \qav{ \zeta^n }_z \label{eq:unsuprs4} \\
         & = -n  \alpha \qav{ \ln \zeta }_z \label{eq:unsuprs5}, 
\end{eqnarray}
where
\begin{eqnarray}
\zeta  & = {\int} \,  \frac{d\lambda \, d\hat{\lambda}}{2 \pi} \,  e^{ i \hat{\lambda}  (\lambda - \sqrt{q}z)  - \frac{1}{2}(1-q)\hat{\lambda}^2 -  \beta V(\lambda)} \nonumber \\
           & = {\int} \,  \frac{d\lambda}{\sqrt{2 \pi (1-q)}} \,  e^{ -\frac{1}{2} \frac{(\lambda - \sqrt{q}z)^2}{1-q} -  \beta V(\lambda)}
\end{eqnarray}
is the partition function of a distribution whose interpretation will be given in section \ref{sec:aligndist}.
\par
In going from (\ref{eq:unsuprs1}) to (\ref{eq:unsuprs3}) we used the identity
\begin{equation} 
\int \, \frac{d\lambda_a}{\sqrt{2\pi}} \, \frac{1}{\sqrt{\det Q}} \, e^{-\frac{1}{2}  \lambda_a Q^{-1}_{ab}\lambda_b} = {\int} \, \prod_{a=1}^n \, \frac{d\lambda_a \, d\hat{\lambda_a }}{2 \pi} \,  e^{ i \lambda_a \hat{\lambda_a } -\frac{1}{2}  \hat{\lambda_a} Q_{ab}\hat{\lambda_b}},
\end{equation}
and inserted the replica symmetric ansatz $\Qab = (1-q) \delta_{ab} + q$.   Then the only coupling between the various $\lambda_a$'s in (\ref{eq:unsuprs3}) occurs through the term 
$\frac {1}{2}( \sqrt{q} \sum_a  \hat{\lambda_a} )^2$.  We can thus decouple the $\lambda_a$ variables at the expense of introducing a Gaussian integral via the identity 
$ e^{-\frac{1}{2} b^2} = \qav{e^{i b z}}_z$, where $z$ is a zero mean, unit variance Gaussian variable and $\qav{\cdot}_z$ denotes an average with respect to $z$.  This transformation yields (\ref{eq:unsuprs4}), and, as $n\rightarrow0$, (\ref{eq:unsuprs5}). 
\par Now for the entropy, we obtain
\begin{eqnarray}
\lim_{n \rightarrow 0} S(Q) & = \lim_{n \rightarrow 0} \frac{1}{2} \Tr \log Q \\
                                           & =  \frac{n}{2} \bigg[ \frac{q}{1-q}  + \ln (1-q) \bigg], \label{eq:replsymment}
\end{eqnarray}
Here we have used the fact that the replica symmetric $Q$ has 1 eigenvalue equal to  $1+(n-1)q$ and $n-1$ eigenvalues equal to $1-q$.  Finally, inserting 
\eqref{eq:unsuprs5} and \eqref{eq:replsymment} into \eqref{eq:repl4} and performing the integration over $q$ via a saddle point yields a saddle point equation for $q$ corresponding to extremizing $F(q)$ in \eqref{eq:feper}.

\subsection{Distribution of Alignments}
\label{sec:aligndist}
Suppose we wish to compute the probability distribution across examples $\mu$ of the alignment of each example \mbox{\boldmath{$\xi$}}$^\mu$ with an optimal weight vector $\w$ derived from an unsupervised learning problem.  Alternatively, one can think of this as the distribution of the data projected onto the optimal dimension. This distribution is 
\begin{equation}
P(\lambda) = \frac{1}{P} \sum_{\mu=1}^P \delta(\lambda - \lambda^\mu),
\end{equation}
where $\lambda^\mu = \frac{1}{\sqrt{N}}\w \cdot$\mbox{\boldmath{$\xi$}}$^\mu$, and $\w$ is drawn from the distribution (\ref{eq:pergibbs}).  For large $N$ and $P$ we expect this distribution to be self-averaging, so for any fixed realization of the examples, it will be close to
\begin{equation}
P(\lambda) = \bqav{\frac{1}{Z} \int \, d\w \, \delta(\lambda - \lambda^1) \, e^{-\beta \sum_{\mu} V(\lambda^\mu)}},
\label{eq:lamdist}
\end{equation}
where 
\begin{equation}
Z =  \int \, d\w  \, e^{-\beta \sum_{\mu} V(\lambda^\mu)},
\end{equation}
and $\qav{\cdot}$ denotes an average over the examples \mbox{\boldmath{$\xi$}}$^\mu$.  This average is hard to perform because the examples occur both in the numerator and denominator.  This difficulty can be circumvented by introducing replicas via the simple identity $\frac{1}{Z} = \lim_{n \rightarrow 0} Z^{n-1}$.  Thus
\begin{equation}
P(\lambda) = \lim_{n\rightarrow 0} \bqav{\int \, \prod_{a=1}^n d\w_a \, \delta(\lambda - \lambda^1_1) \, e^{-\beta \sum_{a=1}^n \sum_{\mu} V(\lambda^\mu_a)}},
\end{equation}
where $\lambda^\mu_a = \frac{1}{\sqrt{N}}\w_a\cdot$\mbox{\boldmath{$\xi$}}$^\mu$.  Here the first replica plays the role of the numerator in (\ref{eq:lamdist}) and replicas $2,\dots,n$ play the role of $\frac{1}{Z}$ in the $n \rightarrow 0$ limit.  Now we can introduce an integral representation of $\delta(\lambda - \lambda^1_1)$, perform the Gaussian average over $\lambda^\mu_a$ and take the $n \rightarrow 0$ limit using a sequence of steps very similar to those in sections \ref{sec:repframe} and \ref{sec:n0unsup}.  This yields
\begin{equation}
P(\lambda) = \bqav{\frac{1}{\zeta}  \frac{1}{\sqrt{2 \pi (1-q)}} \,  e^{ -\frac{1}{2} \frac{(\lambda - \sqrt{q}z)^2}{1-q} -  \beta V(\lambda)}}_z,
\label{eq:unsupdistalign}
\end{equation}
where $\zeta$ is the partition function given by (\ref{eq:zetaper}) and $q$ extremizes the free energy (\ref{eq:feper}).

\subsection{Inverting the Stieltjies Transform}
\label{sec:contour}

It is helpful to think of \eqref{eq:resovlentdef} as a complex contour integral, with the contour running along the real axis.
We can't simply set $\epsilon=0$ in \eqref{eq:chargetoforce}, as the pole at $z'=z+i\epsilon$ would hit the contour.
However, Cauchy's theorem tells us that we can deform the contour without changing the integral, provided that it doesn't cross any singularities.
We will use the following contour, which takes a semicircle detour below the singularity:
\begin{eqnarray*}
  {}\qquad\qquad\quad z' &=& x, \qquad\qquad\quad x \,\in\, (-\infty,-\delta],\\
  C(\delta): \qquad
  z' &=& z+\delta\, e^{i\theta}, \qquad \theta \,\in\, [-\pi,0],\\
  {}\qquad\qquad\quad z' &=& x, \qquad\qquad\quad x \,\in\, [\delta,\infty).
\end{eqnarray*}
It will help to take the limit $\delta \rightarrow 0$.

We can write
\begin{eqnarray*}
   \lim_{\epsilon \rightarrow 0^+} \frac{1}{\pi} \, \mathrm{Im} \, R_{\mathbf W}(z + i \epsilon)
    &=& \lim_{\delta \rightarrow 0^+} \frac{1}{\pi} \, \mathrm{Im} \, \int_{C(\delta)}\!\! \mathrm{d}z'\, \frac{\rho_\mathbf{W}(z')}{z'-z} \\
    &=& \lim_{\delta \rightarrow 0^+} \frac{1}{\pi} \, \mathrm{Im} \left [
     \int_{-\infty}^{-\delta}\!\! \mathrm{d}x\, \frac{\rho_\mathbf{W}(x)}{x-z}
     \right. \\ && \left. \qquad
     + \int_{-\pi}^{0}\!\! \mathrm{d}\theta\, \left(i\delta\, e^{i\theta}\right) \frac{\rho_\mathbf{W}\left(z+\delta\, e^{i\theta}\right)}{\delta\, e^{i\theta}}
     + \int_{\delta}^{\infty}\!\! \mathrm{d}x\, \frac{\rho_\mathbf{W}(x)}{x-z}
     \right ].
\end{eqnarray*}
The first and third terms diverge as $\delta \rightarrow 0$.
However, their sum is finite.
It is referred to as the Cauchy principal value of the integral.
It also happens to be real, and we are only interested in the imaginary part.
This leaves the second term:
\[
   \lim_{\epsilon \rightarrow 0^+} \frac{1}{\pi} \, \mathrm{Im} \, R_{\mathbf W}(z + i \epsilon)
    = \lim_{\delta \rightarrow 0^+} \frac{1}{\pi} \, \mathrm{Im} \,
      \int_{-\pi}^{0}\!\! \mathrm{d}\theta\, i \rho_\mathbf{W}\left(z+\delta\, e^{i\theta}\right)
    = \rho_\mathbf{W}(z).
\]

 \section*{References}
 
\bibliography{csrev}
\bibliographystyle{unsrt}

\end{document}